\documentclass[twocolumn]{aastex631}

\usepackage{gensymb}
\usepackage{booktabs}
\usepackage{multirow}
\usepackage{graphicx}
\usepackage{xspace}
\usepackage{chngcntr}

\let\tablenum\relax

\usepackage{siunitx}

\usepackage{amsmath}
\usepackage[T2A]{fontenc}

\newcommand*{\thead}[1]{\multicolumn{1}{c}{\bfseries #1}}
\newcommand*{\tcenter}[1]{\multicolumn{1}{c}{#1}}

\shorttitle{KM3-230213A Blazar Candidate Counterparts}
\shortauthors{KM3NeT Collaboration et al.}

\graphicspath{{./}{figs/}}

\newcommand{\uhe}{KM3-230213A\xspace}
\newcommand{\uhevent}{KM3-230213A\xspace}

\newcommand{\numberOfCandidates}{seventeen\xspace}

\newcommand{\gammarays}{gamma rays\xspace}
\newcommand{\gammaray}{gamma-ray\xspace}

\newcommand{\Swift}{\emph{Swift}\xspace}
\newcommand{\SwiftXRT}{\emph{Swift}-XRT\xspace}

\newcommand{\Fermi}{\emph{Fermi}}
\newcommand{\FermiLAT}{\emph{Fermi}-LAT\xspace}
\newcommand{\eROSITA}{\emph{eROSITA}\xspace}
\newcommand{\WISE}{\emph{WISE}\xspace}
\newcommand{\NEOWISE}{\emph{NEOWISE}\xspace}
\newcommand{\Chandra}{\emph{Chandra}\xspace}
\newcommand{\ROSAT}{\emph{ROSAT}\xspace}
\newcommand{\SVOM}{\emph{SVOM}\xspace}
\newcommand{\GAIA}{\emph{Gaia}\xspace}

\newcommand{\sourcename}[1]{%
  \ifcase#1\relax
  \or MRC\,0614-083\xspace
  \or 4FGL\,J0616.2-0653\xspace
  \or PMN\,J0622-0657\xspace
  \or NVSS\,J062455-073536\xspace
  \or PMN\,J0618-0954\xspace
  \or 
  0605-085\xspace
  \or PMN\,J0609-0615\xspace
  \or PMN\,J0606-0724\xspace
  \or PMN\,J0616-1040\xspace
  \or NVSS\,J060639-063421\xspace
  \or PMN\,J0605-0759\xspace
  \or NVSS\,J060509-075747\xspace
  \or PMN\,J0612-0700\xspace
  \or PMN\,J0614-0918\xspace
  \or PMN\,J0622-0846\xspace
  \or NVSS\,J061237-055244\xspace
  \or NVSS\,J061050-095934\xspace
  \else UNDEFINED SOURCE\xspace
  \fi
}

\begin{document}

\title{Characterizing Candidate Blazar Counterparts of the Ultra-High-Energy Event KM3-230213A}


\author{O.~Adriani}
\affiliation{INFN, Sezione di Firenze, via Sansone 1, Sesto Fiorentino, 50019 Italy}
\affiliation{Universit{\`a} di Firenze, Dipartimento di Fisica e Astronomia, via Sansone 1, Sesto Fiorentino, 50019 Italy}
\author{S.~Aiello}
\affiliation{INFN, Sezione di Catania, (INFN-CT) Via Santa Sofia 64, Catania, 95123 Italy}
\author{A.~Albert}
\affiliation{Universit{\'e}~de~Strasbourg,~CNRS,~IPHC~UMR~7178,~F-67000~Strasbourg,~France}
\affiliation{Universit{\'e} de Haute Alsace, rue des Fr{\`e}res Lumi{\`e}re, 68093 Mulhouse Cedex, France}
\author{A.\,R.~Alhebsi}
\affiliation{Khalifa University of Science and Technology, Department of Physics, PO Box 127788, Abu Dhabi,   United Arab Emirates}
\author{M.~Alshamsi}
\affiliation{Aix~Marseille~Univ,~CNRS/IN2P3,~CPPM,~Marseille,~France}
\author{S. Alves Garre}
\affiliation{IFIC - Instituto de F{\'\i}sica Corpuscular (CSIC - Universitat de Val{\`e}ncia), c/Catedr{\'a}tico Jos{\'e} Beltr{\'a}n, 2, 46980 Paterna, Valencia, Spain}
\author{A. Ambrosone}
\affiliation{Universit{\`a} di Napoli ``Federico II'', Dip. Scienze Fisiche ``E. Pancini'', Complesso Universitario di Monte S. Angelo, Via Cintia ed. G, Napoli, 80126 Italy}
\affiliation{INFN, Sezione di Napoli, Complesso Universitario di Monte S. Angelo, Via Cintia ed. G, Napoli, 80126 Italy}
\author{F.~Ameli}
\affiliation{INFN, Sezione di Roma, Piazzale Aldo Moro 2, Roma, 00185 Italy}
\author{M.~Andre}
\affiliation{Universitat Polit{\`e}cnica de Catalunya, Laboratori d'Aplicacions Bioac{\'u}stiques, Centre Tecnol{\`o}gic de Vilanova i la Geltr{\'u}, Avda. Rambla Exposici{\'o}, s/n, Vilanova i la Geltr{\'u}, 08800 Spain}
\author{L.~Aphecetche}
\affiliation{Subatech, IMT Atlantique, IN2P3-CNRS, Nantes Universit{\'e}, 4 rue Alfred Kastler - La Chantrerie, Nantes, BP 20722 44307 France}
% ...

\collaboration{1000}{KM3NeT Collaboration}

\author[0000-0002-1460-3369]{M.~Lincetto}
\affiliation{Julius-Maximilians-Universit{\"a}t W{\"u}rzburg, Fakult{\"a}t f{\"u}r Physik und Astronomie, Institut f{\"u}r Theoretische Physik und Astrophysik, Lehrstuhl f{\"u}r Astronomie, Emil-Fischer-Stra{\ss}e 31, 97074 W{\"u}rzburg, Germany}

\author[0000-0003-2497-6836]{L.~Pfeiffer}
\affiliation{Julius-Maximilians-Universit{\"a}t W{\"u}rzburg, Fakult{\"a}t f{\"u}r Physik und Astronomie, Institut f{\"u}r Theoretische Physik und Astrophysik, Lehrstuhl f{\"u}r Astronomie, Emil-Fischer-Stra{\ss}e 31, 97074 W{\"u}rzburg, Germany}

\author[0000-0002-3308-324X]{S. Buson}
\affiliation{Deutsches~Elektronen-Synchrotron~DESY,~Platanenallee~6,~15738~Zeuthen,~Germany}
\affiliation{Julius-Maximilians-Universit{\"a}t W{\"u}rzburg, Fakult{\"a}t f{\"u}r Physik und Astronomie, Institut f{\"u}r Theoretische Physik und Astrophysik, Lehrstuhl f{\"u}r Astronomie, Emil-Fischer-Stra{\ss}e 31, 97074 W{\"u}rzburg, Germany}

\author[0009-0001-9486-1252]{Jose Maria Sanchez Zaballa}
\affiliation{Julius-Maximilians-Universit{\"a}t W{\"u}rzburg, Fakult{\"a}t f{\"u}r Physik und Astronomie, Institut f{\"u}r Theoretische Physik und Astrophysik, Lehrstuhl f{\"u}r Astronomie, Emil-Fischer-Stra{\ss}e 31, 97074 W{\"u}rzburg, Germany}

\author[0000-0002-2515-1353]{A. Azzollini}
\affiliation{Julius-Maximilians-Universit{\"a}t W{\"u}rzburg, Fakult{\"a}t f{\"u}r Physik und Astronomie, Institut f{\"u}r Theoretische Physik und Astrophysik, Lehrstuhl f{\"u}r Astronomie, Emil-Fischer-Stra{\ss}e 31, 97074 W{\"u}rzburg, Germany}

\author[0009-0002-6743-7808]{A. Bremer}
\affiliation{Julius-Maximilians-Universit{\"a}t W{\"u}rzburg, Fakult{\"a}t f{\"u}r Physik und Astronomie, Institut f{\"u}r Theoretische Physik und Astrophysik, Lehrstuhl f{\"u}r Astronomie, Emil-Fischer-Stra{\ss}e 31, 97074 W{\"u}rzburg, Germany}

\collaboration{1000}{\emph{MessMapp} Group}

\author[0000-0002-2028-9230]{A.~Acharyya}
\affiliation{Center for Cosmology and Particle Physics Phenomenology, University of Southern Denmark, Campusvej 55, DK-5230 Odense M, Denmark}
\author{A.~Adelfio}
\affiliation{Istituto Nazionale di Fisica Nucleare, Sezione di Perugia, I-06123 Perugia, Italy}
\author[0000-0002-6584-1703]{M.~Ajello}
\affiliation{Department of Physics and Astronomy, Clemson University, Kinard Lab of Physics, Clemson, SC 29634-0978, USA}
\author[0000-0002-9785-7726]{L.~Baldini}
\affiliation{Universit\`a di Pisa and Istituto Nazionale di Fisica Nucleare, Sezione di Pisa I-56127 Pisa, Italy}
\author[0000-0001-7233-9546]{C.~Bartolini}
\affiliation{Istituto Nazionale di Fisica Nucleare, Sezione di Bari, I-70126 Bari, Italy}
\affiliation{Universit\`a degli studi di Trento, via Calepina 14, 38122 Trento, Italy}
\author[0000-0002-6729-9022]{J.~Becerra~Gonzalez}
\affiliation{Instituto de Astrof\'isica de Canarias and Universidad de La Laguna, Dpto. Astrof\'isica, 38200 La Laguna, Tenerife, Spain}
\author[0000-0002-2469-7063]{R.~Bellazzini}
\affiliation{Istituto Nazionale di Fisica Nucleare, Sezione di Pisa, I-56127 Pisa, Italy}
\author[0000-0001-9935-8106]{E.~Bissaldi}
\affiliation{Dipartimento di Fisica ``M. Merlin" dell'Universit\`a e del Politecnico di Bari, via Amendola 173, I-70126 Bari, Italy}
\affiliation{Istituto Nazionale di Fisica Nucleare, Sezione di Bari, I-70126 Bari, Italy}
\author[0000-0002-1854-5506]{R.~D.~Blandford}
\affiliation{W. W. Hansen Experimental Physics Laboratory, Kavli Institute for Particle Astrophysics and Cosmology, Department of Physics and SLAC National Accelerator Laboratory, Stanford University, Stanford, CA 94305, USA}
\author[0000-0002-4264-1215]{R.~Bonino}
\affiliation{Istituto Nazionale di Fisica Nucleare, Sezione di Torino, I-10125 Torino, Italy}
\affiliation{Dipartimento di Fisica, Universit\`a degli Studi di Torino, I-10125 Torino, Italy}
% ...

\collaboration{1000}{\emph{Fermi}-LAT Collaboration}
\author[0000-0001-9152-961X]{A.~C.~S.~Readhead}
\affiliation{Owens Valley Radio Observatory, California Institute of Technology,  Pasadena, CA 91125, USA}
\author[0000-0002-0870-1368]{V. Pavlidou}
\affiliation{Department of Physics and Institute of Theoretical and Computational Physics, University of Crete, 71003 Heraklion, Greece}
\affiliation{Institute of Astrophysics, Foundation for Research and Technology-Hellas, GR-71110 Heraklion, Greece}
\author[0000-0001-7470-3321]{J. A. Zensus}
\affiliation{Max-Planck-Institut~f{\"u}r~Radioastronomie,~Auf~dem H{\"u}gel~69,~53121~Bonn,~Germany}
\author[0000-0003-2483-2103]{M.F. Aller}
\affiliation{Department of Astronomy, University of Michigan, 323 West Hall, 1085 S. University Avenue, Ann Arbor, MI 48109, USA}
\author[0000-0001-5957-1412]{P. V. de la Parra}
\affiliation{CePIA, Astronomy Department, Universidad de Concepci\'on,  Casilla 160-C, Concepci\'on, Chile}

\author{M.~Hodges}
\affiliation{Owens Valley Radio Observatory, California Institute of Technology,  Pasadena, CA 91125, USA}
\author[0000-0002-2024-8199]{T. Hovatta}
\affiliation{Finnish Centre for Astronomy with ESO (FINCA), University of Turku, FI-20014 University of Turku, Finland}
\affiliation{Aalto University Mets\"ahovi Radio Observatory,  Mets\"ahovintie 114, 02540 Kylm\"al\"a, Finland}
\author[0000-0001-6314-9177]{S.~Kiehlmann}
\affiliation{Institute of Astrophysics, Foundation for Research and Technology-Hellas, GR-71110 Heraklion, Greece}
\author[0000-0001-9200-4006]{I. Liodakis}
\affiliation{Institute of Astrophysics, Foundation for Research and Technology-Hellas, GR-71110 Heraklion, Greece}
\author[0000-0002-5491-5244]{W. Max-Moerbeck}
\affiliation{Departamento de Astronomía, Universidad de Chile, Camino El Observatorio 1515, Las Condes, Santiago, Chile}
\author[0009-0000-9963-6874]{B. Molina}
\affiliation{CePIA, Astronomy Department, Universidad de Concepci\'on,  Casilla 160-C, Concepci\'on, Chile}
\author[0000-0001-5213-6231]{T. J. Pearson}
\affiliation{Owens Valley Radio Observatory, California Institute of Technology,  Pasadena, CA 91125, USA}
\author[0000-0002-7252-5485]{V. Ravi}
\affiliation{Owens Valley Radio Observatory, California Institute of Technology,  Pasadena, CA 91125, USA}
\author[0000-0001-5704-271X]{R.A. Reeves}
\affiliation{CePIA, Astronomy Department, Universidad de Concepci\'on,  Casilla 160-C, Concepci\'on, Chile}
\author[0009-0004-2614-830X]{A. Synani}
\affiliation{Department of Physics and Institute of Theoretical and Computational Physics, University of Crete, 71003 Heraklion, Greece}
\affiliation{Institute of Astrophysics, Foundation for Research and Technology-Hellas, GR-71110 Heraklion, Greece}
\author[0000-0002-8831-2038]{K. Tassis}
\affiliation{Department of Physics and Institute of Theoretical and Computational Physics, University of Crete, 71003 Heraklion, Greece}
\affiliation{Institute of Astrophysics, Foundation for Research and Technology-Hellas, GR-71110 Heraklion, Greece}

\collaboration{1000}{Owens Valley Radio Observatory 40-m Telescope Group}

\author{A. Foisseau}
\affiliation{Universit{\'e} Paris Cit{\'e}, CNRS, Astroparticule et Cosmologie, F-75013 Paris, France}
\author{A. Coleiro}
\affiliation{Universit{\'e} Paris Cit{\'e}, CNRS, Astroparticule et Cosmologie, F-75013 Paris, France}
\author{F. Cangemi}
\affiliation{Universit{\'e} Paris Cit{\'e}, CNRS, Astroparticule et Cosmologie, F-75013 Paris, France}

\author{D.~Dornic}
\affiliation{Aix~Marseille~Univ,~CNRS/IN2P3,~CPPM,~Marseille,~France}

\author{C. Lachaud}
\affiliation{Universit{\'e} Paris Cit{\'e}, CNRS, Astroparticule et Cosmologie, F-75013 Paris, France}
\author{P. Maggi }
\affiliation{Observatoire Astronomique de Strasbourg, Universit\'e de Strasbourg, CNRS, 11 rue de l'Universit\'e, F-67000 Strasbourg, France}
\author{D. Gotz}
\affiliation{Universit\'e Paris-Saclay, Universit\'e Paris Cit\'e, CEA, CNRS, AIM, 91191 Gif-sur-Yvette, France}
\author{L. Xin}
\affiliation{Key Laboratory of Space Astronomy and Technology, National Astronomical Observatories, Chinese Academy of Sciences, Beijing 100101, China}
\author{B. Cordier}
\affiliation{Universit\'e Paris-Saclay, Universit\'e Paris Cit\'e, CEA, CNRS, AIM, 91191 Gif-sur-Yvette, France}
\author{O. Godet}
\affiliation{IRAP, Universit\'e de Toulouse, CNRS, CNES, UPS, Toulouse, France}
\author{A. Goldwurm}
\affiliation{Universit\'e Paris Cit\'e, CNRS, CEA, Astroparticule et Cosmologie, F-75013 Paris, France}
\author{H. Goto}
\affiliation{Universit\'e Paris-Saclay, Universit\'e Paris Cit\'e, CEA, CNRS, AIM, 91191 Gif-sur-Yvette, France}
\author{X. Han}
\affiliation{Key Laboratory of Space Astronomy and Technology, National Astronomical Observatories, Chinese Academy of Sciences, Beijing 100101, China}
\author{N. Leroy}
\affiliation{Universit\'e Paris-Saclay, CNRS/IN2P3, IJCLab, 91405 Orsay, France}
\author{C. Plasse}
\affiliation{Universit\'e Paris-Saclay, Universit\'e Paris Cit\'e, CEA, CNRS, AIM, 91191 Gif-sur-Yvette, France}
\author{Y. Qiu}
\affiliation{Key Laboratory of Space Astronomy and Technology, National Astronomical Observatories, Chinese Academy of Sciences, Beijing 100101, China}
\author{J. Rodriguez}
\affiliation{Universit\'e Paris-Saclay, Universit\'e Paris Cit\'e, CEA, CNRS, AIM, 91191 Gif-sur-Yvette, France}
\author{J. Wang}
\affiliation{Key Laboratory of Space Astronomy and Technology, National Astronomical Observatories, Chinese Academy of Sciences, Beijing 100101, China}
\author{J. Wei}
\affiliation{Key Laboratory of Space Astronomy and Technology, National Astronomical Observatories, Chinese Academy of Sciences, Beijing 100101, China}

\collaboration{1000}{\emph{SVOM} Collaboration}

\author{P. Baldini}
\affiliation{Max-Planck-Institut f{\"u}r extraterrestrische Physik, Giessenbachstrasse 1, 85748 Garching, Germany}

\author[0000-0003-0426-6634]{J. Buchner}
\affiliation{Max-Planck-Institut f{\"u}r extraterrestrische Physik, Giessenbachstrasse 1, 85748 Garching, Germany}


\author{A. K. Erkenov}
\altaffiliation{The shown author affiliations reflect their job contracts; the KM3NeT collaboration has currently suspended all institutional relations with Russian science organisations.}
\affiliation{Special Astrophysical Observatory of the Russian Academy of Sciences, Nizhny Arkhyz 369167, Russia}

\author[0000-0001-9011-0737]{N. Globus}
\affiliation{Instituto de Astronomía, Universidad Nacional Autónoma de México Campus Ensenada, A.P. 106, Ensenada, BC 22800, México}
\affiliation{Kavli Institute for Particle Astrophysics and Cosmology, Stanford University, Stanford, CA 94305, USA}
\affiliation{Astrophysical Big Bang Laboratory, RIKEN, Wako, Saitama, Japan}

\author[0000-0002-0761-0130]{A. Merloni}
\affiliation{Max-Planck-Institut f{\"u}r extraterrestrische Physik, Giessenbachstrasse 1, 85748 Garching, Germany}

\author{A. Paggi}
\affiliation{Institute of Astrophysics, Foundation for Research and Technology-Hellas, GR-71110 Heraklion, Greece}

\author[0000-0002-0739-700X]{A. V. Popkov}
\altaffiliation{The shown author affiliations reflect their job contracts; the KM3NeT collaboration has currently suspended all institutional relations with Russian science organisations.}
\affiliation{Moscow Center for Advanced Studies, Kulakova str. 20, Moscow, 123592, Russia}
\affiliation{Institute for Nuclear Research, Russian Academy of Sciences, 60th October Anniversary Prospect 7a, Moscow 117312, Russia}

\author[0000-0001-9731-0352]{D. Porquet}
\affiliation{Aix Marseille Univ, CNRS, CNES, LAM, Marseille, France}

\author[0000-0001-7116-9303]{M. Salvato}
\affiliation{Max-Planck-Institut f{\"u}r extraterrestrische Physik, Giessenbachstrasse 1, 85748 Garching, Germany}


\author[0000-0001-9172-7237]{Y. V. Sotnikova}
\altaffiliation{The shown author affiliations reflect their job contracts; the KM3NeT collaboration has currently suspended all institutional relations with Russian science organisations.}
\affiliation{Special Astrophysical Observatory of the Russian Academy of Sciences, Nizhny Arkhyz 369167, Russia}
\affiliation{Institute for Nuclear Research, Russian Academy of Sciences, 60th October Anniversary Prospect 7a, Moscow 117312, Russia}
\affiliation{Kazan Federal University, Institute of Physics, 18 Kremlyovskaya Street, Kazan 420008, Russia}

\author[0000-0002-1290-1629]{P. A. Voitsik}
\affiliation{Independent researcher} 

\nocollaboration{1000}

\correspondingauthor{M. Lincetto}
\email{massimiliano.lincetto@uni-wuerzburg.de}
\correspondingauthor{L. Pfeiffer}
\email{leonard.pfeiffer@uni-wuerzburg.de}
\correspondingauthor{A. Plavin}
\email{alexander@plav.in}

\begin{abstract}
High-energy astrophysical neutrinos serve as crucial messengers for understanding hadronic acceleration processes and identifying the origins of cosmic rays, with blazars among the most promising neutrino sources. The KM3NeT experiment reported the detection of an ultra-high-energy neutrino with an energy estimate of $\sim 220\,\rm{PeV}$, the most energetic yet observed. The neutrino arrival direction has a 99\% confidence region of $3\degree$ radius centered at RA \qty{94.3}{\degree}, Dec \SI{-7.8}{\degree} (J2000).
In this work, seventeen candidate blazars located within this region are identified. Comprehensive new observations and archival data analysis for these sources are presented. The study provides a complete multiwavelength coverage across radio, optical, X-ray, and gamma-ray bands, including proprietary data and dedicated follow-up observations. This study highlights flaring behavior in several candidate counterparts. One object exhibits a radio flare coinciding with the neutrino arrival time, with a pre-trial chance probability of $0.26$\%. Another candidate displays a rising trend in X-ray flux in a one-year window around the neutrino arrival time, while a third undergoes a \gammaray flare during the same period. Based on the observational findings here presented, while none of these candidates can conclusively be linked to the neutrino, the implications of a possible blazar origin for the KM3NeT event are discussed.

\end{abstract}

\section{Introduction}
\label{s:intro}

The detection of an ultra-high-energy (hereafter UHE; in the range $\sim\qty{50}{PeV}$ to $\unit{EeV}$) neutrino by KM3NeT \citep{KM3NeT-2025-Nature}, the deep sea neutrino observatory in the Mediterranean, marks a significant milestone in neutrino astrophysics. This event is far more energetic than any previously observed astrophysical neutrino, and challenges the current understanding of the cosmic neutrino spectrum, which exhibits a steep decline at the highest energies \citep{IceCube_diffuse:2022, KM3NeT-landscape}. 
The observation of astrophysical neutrinos in the high-energy range (hereafter HE; in the range $\unit{TeV}$--$\unit{PeV}$) and UHE range  represents important progress in unraveling the origins of cosmic rays. As cosmic rays accelerate and propagate through space, they may interact with ambient matter or radiation fields, triggering hadronic processes that produce high-energy neutrinos. These neutrinos serve as unambiguous tracers of cosmic ray acceleration, as their production is exclusively tied to hadronic interactions.

While the detection of an astrophysical diffuse neutrino flux is well established, the nature of the astrophysical sources dominating this emission remains a major open question \cite{KM3NeT-landscape, KM3NeT:2025zmb:GRBCompanion}. The detection of the KM3NeT event could indicate the existence of a new class of extreme cosmic accelerators, distinct from those responsible for the bulk of the high-energy neutrino flux. It may also represent the first observation of a cosmogenic neutrino \citep{Berezinsky:1970, KM3NeT:2025vut:CosmogenicCompanion}, produced in interactions between ultra-high-energy cosmic rays (UHECR) and the cosmic radiation background. The observation of such an extreme event can be used to constrain fundamental physical quantities, such as the constancy of the speed of light~\citep{KM3NeT:2025mfl:SuperluminalCompanion}.

Given the typical error region of neutrino events, chance spatial coincidences with one or more astrophysical objects may occur, requiring additional evidence for a definitive neutrino-source association. For this reason, previous research has focused on population studies or variable behavior of candidate sources, in synergy with multiwavelength and time-domain astronomical observations, to identify potential associations.

Among the candidate sources of HE and UHE cosmic rays are blazars. Blazars are a subclass of active galactic nuclei (AGN) with highly-collimated jets pointing almost directly towards Earth. They emit across the whole electromagnetic spectrum from radio frequencies to very-high-energy ($\gtrsim \qty{100}{GeV}$) \gammarays~\citep{2019ARA&A..57..467B, 2019Galax...7...20B, 2023A&A...678A.140J} and can display strong variability on timescales ranging from minutes to years.
The first promising association between a high-energy neutrino event  and a blazar was reported in 2017 with the detection of neutrino IceCube-170922A from the direction of the blazar TXS\,0506+056, which was flaring in \gammarays~\citep{IceCube:2018cha}. 
Since then, correlations between blazars and neutrino detections have been examined in many studies. They suggest a potential connection between the extreme high-energy processes in or near blazars and the production of neutrinos~\citep{Krauss:2014tna, Padovani_TXS_notBLLac:2019, Giommi:2020hbx, Franckowiak:2020qrq, OVRO_IC_21,Das:2021cdf,Buson:2022fyf, Buson:2023irp, 2023MNRAS.523.1799P,  Bellenghi:2023yza, IceCube:2023htm:AlertStacking, Garrappa:2024, OVRO_IC_24,  AntaresBlazars}. 
This evidence supports the hypothesis that blazars play a major role in the production of high-energy cosmic rays and neutrinos. However, definitive proof remains elusive.

Following this motivation, in this paper a possible blazar origin for the KM3NeT event is investigated. Combining archival data, public and proprietary, and new observations, an improved identification of candidate counterparts is presented together with a detailed characterization of their multiwavelength properties. The candidate blazars positionally consistent with \uhevent are studied over the full electromagnetic spectrum, from radio to gamma-rays.

The paper is organized as follows. In \autoref{sec:km} the properties of the event are briefly presented. In \autoref{s:obsdata}, a comprehensive summary of the archival data and new dedicated observations is provided, realizing the most complete observational coverage of these sources to date. In \autoref{s:selection}, the methodology adopted to pinpoint objects of likely blazar nature in the region of \uhevent is introduced. Further developing the initial search for counterparts reported in the previous study \citep{KM3NeT-2025-Nature}, a sample of seventeen potential blazar associations in the 99\% confidence region of KM3-230213A is presented. In \autoref{s:results}, the properties and behavior of candidates exhibiting electromagnetic flaring in correspondence of the neutrino arrival time are highlighted. Finally in \autoref{s:discussion}, the consistency between these observational findings and the blazar--neutrino association hypothesis is discussed in light of current models of the high-energy emission in blazars. \autoref{s:summary} summarizes our observational results and discusses potential strategies to narrow the source selection in future neutrino events.

\section{The ultra-high-energy event KM3-230213A}
\label{sec:km}
The KM3NeT Collaboration is building two large-volume neutrino detectors in the depths of the Mediterranean Sea~\citep{KM3Net:2016zxf}. 
The KM3NeT/ORCA detector, located near Toulon (France), will be equipped with 115 detector lines, with a geometry optimized for the detection of GeV--scale neutrinos and the study of atmospheric neutrino oscillations. 
The KM3NeT/ARCA detector is located near Portopalo di Capo Passero in Sicily (Italy) and will consist of two blocks of 115 detector lines, with larger spacings optimized for TeV--PeV astrophysical neutrinos~\citep{KM3NeT:2024paj}. 
Each detection line, anchored on the sea bed, comprises 18 digital optical modules ~\citep[DOM,][]{KM3NeT:2022pnv}, containing 31 $\SI{80}{mm}$ photomultiplier tubes \cite{KM3NeT:2025qao:PMT}. 

On 2023 February 13, KM3NeT/ARCA detected an exceptionally high-energy muon, originating most likely in the interaction of a cosmic neutrino. The energy of the detected muon is estimated at $120^{+110}_{-60}\,\rm{PeV}$. Assuming an astrophysical $E^{-2}$ flux, the best fit neutrino energy is $\sim \qty{220}{PeV}$ with a 90\% confidence interval of $\qty{72}{PeV}-\qty{2.6}{EeV}$.
The event is designated \uhevent~\citep{KM3NeT-2025-Nature}. This event was detected with the ARCA21 detector configuration (21 active lines). The J2000 equatorial coordinates of \uhevent are $\textrm{RA} = \qty{94.3}{\degree}$, 
$\textrm{Dec} = \qty{-7.8}{\degree}$
($06^\mathrm{h}17^\mathrm{m}12^\mathrm{s}$,$-07\degr48\arcmin00\arcsec$); 
the UTC date and time of the detection was 2023-02-13T01:16:47.703Z (MJD = 59988.0533299). 
The directional error region is largely dominated by the current systematic uncertainty in the absolute orientation of the detector. 
The containment radii for different confidence levels are: R(68\%) = 1.5\degr, R(90\%) = 2.2\degr, and R(99\%) = 3.0\degr. In the near future, a recalibration of the detector elements will provide a significant improvement in the localization accuracy of \uhevent.

\section{Observational data}
\label{s:obsdata}
The majority of the electromagnetic emission from blazars is commonly attributed to leptonic processes, such as synchrotron emission of accelerated electrons and inverse Compton scattering. The two processes may be coupled, with emission resulting from a synchrotron self-Compton process.  In lepto-hadronic models, accelerated ions interact with photons or protons producing neutral pions, decaying into high-energy \gammarays, and charged pions, decaying into muons and neutrinos \citep[][]{1989A&A...221..211M, 1992A&A...260L...1M, Hovatta:2019ulp}.  Protons can also produce electromagnetic radiation through other means, such as Bethe-Heitler pair production or synchrotron processes. Furthermore, depending on the density of the emission region, \gammarays may cascade down to lower energies, in the MeV and/or X-ray bands. Different electromagnetic signatures can, therefore, be exploited to identify and characterize candidate blazar sources.

The purpose of this study is two-fold. First, to obtain a complete survey of the region-of-interest of \uhevent, aimed at blazar-like objects, across the wavelengths where such a task is achievable. Second, to improve the selection of blazar candidates introduced in \cite{KM3NeT-2025-Nature} and present a comprehensive multiwavelength characterization of such objects. The collection of observational data analyzed to this aim are presented in this section.

\subsection{Radio}
\paragraph{Very Long Baseline Interferometry}
For the purpose of this work, the Very Long Baseline Array (VLBA) has been used to observe all radio objects detected by the VLA Sky Survey \citep[VLASS;][]{VLASS} with a (2$-$4)\,GHz flux density greater than 50\,mJy within the region of interest of KM3-230213A. The dedicated observation was conducted on November 25, 2024, achieving a deeper and complete survey of the parsec-scale properties of the radio objects in the region, complementing the existing VLBI archival data. The selection results in forty-two objects. Notably, the observation did not target the previously identified object \sourcename{5}, which is a likely blazar candidate. Almost all of its radio emission comes from parsec-scales and has a rising spectrum, leading to a VLASS (2$-$4~GHz) flux density $< 50$~mJy while the VLBI (8~GHz) flux density is $\geq 100$~mJy. Lacking clear indications of strong variability, the difference between the two measurements is likely caused by the source emission spectrum. For the forty-two observed objects, measurements at two center frequencies of 4.4~GHz and 7.6~GHz were obtained using the wide C-band receiver, with 4~Gbps total bandwidth, 128-MHz-wide frequency channels, and two 6-minute scans per target. In the VLBA archive, this experiment is listed as as BK257. Source positions for observations and correlations are taken from the Radio Fundamental Catalog \citep[RFC;][]{RFC} when available, or from VLASS if no RFC counterpart is present. 

Results for sources observed within this VLBA program are presented in \autoref{t:vlba_sources}. The calibrated visibility and image FITS files of these
observations are published in an electronic format.
These measurements allow to select highly-luminous AGN jets with strong beaming. Sources where the parsec-scale radio emission (VLBI) is bright, dominates over larger scales (VLA), and has a flat radio spectrum at parsec scales due to synchrotron opacity are identified as likely blazars, with the radio emission coming from jets aligned with the line of sight.

Sources are marked as primary candidate blazars in \autoref{t:vlba_sources} and highlighted in \autoref{fig:roi-map} and \hyperref[f:vlba_sources]{Figure A3} when they meet all the following criteria:
(i) detected on parsec scales with compactness $S^\mathrm{5\,GHz}_\mathrm{VLBA}/S_\mathrm{VLASS}>0.5$,
(ii) parsec scale (5$-$8)\,GHz spectral index $\alpha_\mathrm{VLBA}>-0.5$,
(iii) core brightness temperature $T_\mathrm{b}>10^{10}$~K at either 5\,GHz or 8\,GHz.
The observed flux densities at the end of 2024 are similar to historical averages (\autoref{t:vlba_sources}). In particular, no major flare was detected during this time.

\begin{table*}[h]
    \scriptsize
  \caption{Properties of sources observed with VLBA within $3\degree$ from the \uhevent position.}
  \label{t:vlba_sources}
  \centering
  \begin{tabular}{@{}c@{\,\,}ll@{\,\,}c@{\,\,}c@{\,\,}c@{\,\,}c@{\,\,}c@{\,\,}c@{\,\,}r@{\,\,\,\,}l@{\,\,\,\,}l@{}}
  \hline\hline
     \noalign{\medskip}
    &&&&&\tcenter{$\mathbf{S_\mathrm{VLASS}}$}&\multicolumn{3}{c}{\bfseries $\mathbf{S_\mathrm{VLBA}}$}&\thead{$\alpha_\mathrm{VLBA}$}&\multicolumn{2}{c}{\bfseries Core $\mathbf{T_\mathrm{b}}$} \\
    Blaz & \tcenter{\bfseries Name} & \thead{Position}     & \thead{Sep}         & \tcenter{$\mathbf z$} & \tcenter{2-4\,GHz} & \tcenter{Hist\,8\,GHz} & \tcenter{5\,GHz} & \tcenter{8\,GHz} & \tcenter{5-8\,GHz} & \tcenter{5\,GHz} & \tcenter{8\,GHz} \\
    & \tcenter{(J2000)} & \tcenter{J2000 RA and Dec.} & \tcenter{($\degree$)} &          & \tcenter{(mJy)}       & \tcenter{(mJy)}       & \tcenter{(mJy)}       & \tcenter{(mJy)} & & \tcenter{(K)} & \tcenter{(K)} \\
    \multicolumn{1}{@{}c@{}}{(1)} & \multicolumn{1}{@{}c@{}}{(2)} & \multicolumn{1}{@{}c@{}}{(3)} & \multicolumn{1}{@{}c@{}}{(4)} & \multicolumn{1}{@{}c@{}}{(5)} & \multicolumn{1}{@{}c@{}}{(6)} & \multicolumn{1}{@{}c@{}}{(7)} & \multicolumn{1}{@{}c@{}}{(8)} & \multicolumn{1}{@{}c@{}}{(9)} & \multicolumn{1}{@{}c@{}}{(10)} & \multicolumn{1}{@{}c@{}}{(11)} & \multicolumn{1}{@{}c@{}}{(12)} \\
       \noalign{\medskip}
    \hline
       \noalign{\medskip}
 x & J0607$-$0834 & 06:07:59.699 $-$08:34:49.978  $^*$ & 2.4 &   0.87 &  3100 &     $2067$ &        $ 2176\pm  220$ &      $ 2240\pm  226$ &    $0.05\pm0.26$ &  $10^{11.9\pm0.1}$ &  $10^{11.3\pm0.1}$\\ 
 x & J0606$-$0724 & 06:06:43.546 $-$07:24:30.231  $^*$ & 2.6 &  1.227 &   461 &      $309$ &        $  379\pm   38$ &      $  306\pm   31$ &   $-0.38\pm0.26$ &      $> 10^{12.0}$ &  $10^{10.4\pm0.1}$\\ 
 x & J0616$-$1041 & 06:16:41.808 $-$10:41:08.456  $^*$ & 2.9 &        &   197 &      $185$ &        $  215\pm   22$ &      $  248\pm   25$ &    $0.26\pm0.26$ &      $> 10^{10.4}$ &      $> 10^{12.1}$\\ 
 x & J0622$-$0656 & 06:22:58.046 $-$06:56:51.971  $^*$ & 1.7 &        &   101 &      $103$ &        $  111\pm   11$ &      $   87\pm    9$ &   $-0.42\pm0.26$ &      $> 10^{12.1}$ &      $> 10^{10.9}$\\ 
 x & J0610$-$0959 & 06:10:50.711 $-$09:59:33.958  $^*$ & 2.7 &        &    97 &       $78$ &        $  100\pm   10$ &      $   94\pm   10$ &   $-0.11\pm0.26$ &      $> 10^{12.1}$ &      $> 10^{11.4}$\\ 
 x & J0622$-$0846 & 06:22:37.999 $-$08:46:18.263  $^*$ & 1.7 &        &    61 &       $74$ &        $   93\pm   10$ &      $  109\pm   12$ &    $0.28\pm0.27$ &      $> 10^{12.2}$ &      $> 10^{12.0}$\\ 
 x & J0605$-$0759 & 06:05:26.187 $-$07:59:31.786  $^*$ & 2.9 &        &   149 &       $74$ &        $   76\pm    8$ &      $   67\pm    7$ &   $-0.24\pm0.27$ &   $10^{9.9\pm0.1}$ &  $10^{10.2\pm0.1}$\\ 
 x & J0614$-$0918 & 06:14:06.042 $-$09:18:37.911  $^*$ & 1.7 &        &    95 &       $63$ &        $   72\pm    7$ &      $   55\pm    6$ &   $-0.46\pm0.27$ &  $10^{10.5\pm0.2}$ &       $> 10^{9.8}$\\ 
 x & J0612$-$0700 & 06:12:04.843 $-$07:00:22.853       & 1.5 &        &    63 &            &        $   54\pm    6$ &      $   62\pm    7$ &    $0.24\pm0.27$ &  $10^{10.2\pm0.1}$ &      $> 10^{11.0}$\\ 
 x & J0612$-$0552 & 06:12:37.360 $-$05:52:44.418  $^*$ & 2.2 &        &    65 &       $62$ &        $   53\pm    6$ &      $   45\pm    5$ &   $-0.29\pm0.27$ &  $10^{10.5\pm0.3}$ &      $> 10^{10.0}$\\ 
   & J0609$-$0615 & 06:09:39.974 $-$06:15:05.858  $^*$ & 2.4 &  2.219 &    83 &       $39$ &        $   70\pm    7$ &      $   48\pm    5$ &   $-0.66\pm0.27$ &   $10^{9.8\pm0.1}$ &   $10^{9.5\pm0.1}$\\ 
   & J0615$-$0934 & 06:15:35.497 $-$09:34:27.996  $^*$ & 1.8 &        &   105 &       $30$ &        $   67\pm    7$ &      $   39\pm    4$ &   $-0.97\pm0.27$ &   $10^{9.9\pm0.1}$ &   $10^{9.0\pm0.1}$\\ 
   & J0620$-$1036 & 06:20:27.542 $-$10:36:22.106  $^*$ & 2.9 &        &   172 &       $39$ &        $   63\pm    7$ &      $   43\pm    5$ &   $-0.66\pm0.27$ &   $10^{9.6\pm0.1}$ &   $10^{9.2\pm0.1}$\\ 
   & J0624$-$0735 & 06:24:55.342 $-$07:35:37.194       & 1.9 &        &   110 &            &        $   58\pm    6$ &      $   50\pm    5$ &   $-0.25\pm0.27$ &   $10^{9.9\pm0.1}$ &   $10^{9.2\pm0.1}$\\ 
   & J0622$-$0723 & 06:22:44.911 $-$07:23:34.839       & 1.4 &        &    94 &            &        $   53\pm    6$ &      $   32\pm    4$ &   $-0.86\pm0.28$ &   $10^{9.1\pm0.1}$ &       $> 10^{9.8}$\\ 
   & J0621$-$0609 & 06:21:10.364 $-$06:09:54.206  $^*$ & 1.9 &        &   132 &       $65$ &        $   49\pm    7$ &      $   44\pm    5$ &   $-0.18\pm0.31$ &       $> 10^{9.9}$ &      $> 10^{10.6}$\\ 
   & J0620$-$0638 & 06:20:42.818 $-$06:38:24.334  $^*$ & 1.5 &        &   101 &     $26^\mathrm{c}$ &        $   25\pm    3$ &      $   18\pm    2$ &   $-0.55\pm0.30$ &      $> 10^{10.4}$ &       $> 10^{9.8}$\\ 
   & J0608$-$0727 & 06:08:04.370 $-$07:27:32.822  $^*$ & 2.3 &        &   356 &       $19$ &        $   23\pm    3$ &      $   22\pm    3$ &   $-0.09\pm0.30$ &      $> 10^{10.9}$ &      $> 10^{10.6}$\\ 
   & J0623$-$0609 & 06:23:23.723 $-$06:09:22.673       & 2.3 &        &    53 &            &        $   20\pm    2$ &      $   23\pm    3$ &    $0.23\pm0.31$ &      $> 10^{10.7}$ &      $> 10^{10.3}$\\ 
   & J0616$-$0648 & 06:16:33.667 $-$06:48:03.289       & 1.0 &        &    52 &            &        $   19\pm    2$ &      $   18\pm    2$ &   $-0.04\pm0.31$ &       $> 10^{9.7}$ &       $> 10^{9.9}$\\ 
   & J0611$-$0800 & 06:11:58.357 $-$08:00:11.217  $^*$ & 1.3 &        &    88 &       $14$ &        $   17\pm    2$ &      $   18\pm    2$ &    $0.14\pm0.28$ &      $> 10^{10.3}$ &      $> 10^{10.1}$\\ 
   & J0624$-$0826 & 06:24:30.354 $-$08:26:38.949       & 1.9 &        &    56 &            &        $   13\pm    2$ &      $   17\pm    2$ &    $0.41\pm0.37$ &       $> 10^{9.5}$ &       $> 10^{8.4}$\\ 
   & J0617$-$0515 & 06:17:50.307 $-$05:15:07.396       & 2.6 &        &    98 &            &        $   28\pm    3$ &               $> 16$ &                  &   $10^{7.6\pm0.1}$ &   $10^{6.7\pm0.4}$\\ 
   & J0605$-$0757 & 06:05:10.018 $-$07:57:50.331       & 3.0 &        &    53 &            &                 $> 26$ &                      &                  &       $> 10^{6.7}$ &                   \\ 
   & J0616$-$0525 & 06:16:31.146 $-$05:25:32.225       & 2.4 &        &    77 &            &                  $> 9$ &                $> 9$ &                  &       $> 10^{8.6}$ &       $> 10^{7.3}$\\ 
   & J0610$-$0725 & 06:10:02.513 $-$07:25:12.715       & 1.8 &        &    57 &            &                  $> 7$ &               $> 10$ &                  &       $> 10^{6.6}$ &       $> 10^{7.6}$\\ 
   & J0613$-$0955 & 06:13:39.450 $-$09:55:34.713       & 2.3 &        &    75 &            &                        &                $> 9$ &                  &                    &       $> 10^{7.7}$\\ 
   & J0608$-$0843 & 06:08:28.480 $-$08:43:53.539       & 2.4 &        &    69 &            &                     \multicolumn{2}{c}{$< 6$} &                  &                    &                   \\ 
   & J0613$-$0655 & 06:13:40.138 $-$06:55:34.607       & 1.2 &        &   123 &      $<12$ &                     \multicolumn{2}{c}{$< 6$} &                  &                    &                   \\ 
   & J0619$-$0731 & 06:19:06.739 $-$07:31:07.183       & 0.6 &        &    51 &            &                     \multicolumn{2}{c}{$< 6$} &                  &                    &                   \\
 & J0605$-$0716 & 06:05:27.091 $-$07:16:23.639       & 3.0 &        &    50 &            & \multicolumn{2}{c}{$< 6$} & & &\\ 
 & J0607$-$0610 & 06:07:16.721 $-$06:10:29.815       & 2.9 &        &    71 &      $<20$ & \multicolumn{2}{c}{$< 6$} & & &\\ 
 & J0608$-$0738 & 06:08:03.661 $-$07:38:31.374       & 2.3 &        &   192 &      $<12$ & \multicolumn{2}{c}{$< 6$} & & &\\ 
 & J0608$-$0700 & 06:08:36.934 $-$07:00:01.723       & 2.3 &        &    93 &      $<20$ & \multicolumn{2}{c}{$< 6$} & & &\\ 
 & J0610$-$0559 & 06:10:09.582 $-$05:59:51.932       & 2.5 &        &    67 &      $<20$ & \multicolumn{2}{c}{$< 6$} & & &\\ 
 & J0610$-$0611 & 06:10:50.604 $-$06:11:49.999       & 2.2 &        &   111 &      $<20$ & \multicolumn{2}{c}{$< 6$} & & &\\ 
 & J0614$-$1038 & 06:14:45.393 $-$10:38:04.475       & 2.9 &        &    81 &            & \multicolumn{2}{c}{$< 6$} & & &\\ 
 & J0615$-$0644 & 06:15:47.557 $-$06:44:47.652       & 1.1 &        &    78 &            & \multicolumn{2}{c}{$< 6$} & & &\\ 
 & J0618$-$0719 & 06:18:47.839 $-$07:19:15.503       & 0.6 &        &    59 &            & \multicolumn{2}{c}{$< 6$} & & &\\ 
 & J0624$-$0721 & 06:24:14.681 $-$07:21:05.721       & 1.8 &        &    57 &            & \multicolumn{2}{c}{$< 6$} & & &\\ 
 & J0625$-$0904 & 06:25:18.204 $-$09:04:35.619       & 2.4 &        &    64 &            & \multicolumn{2}{c}{$< 6$} & & &\\ 
 & J0625$-$0703 & 06:25:29.624 $-$07:03:37.547       & 2.2 &        &    60 &            & \multicolumn{2}{c}{$< 6$} & & &\\
     \noalign{\medskip}
   \hline
   \noalign{\medskip}
\multicolumn{12}{@{}p{18cm}@{}}{\scriptsize
Note: 
(1) -- Primary blazar candidate mark, see details in \hyperref[s:vlbi_sources]{Appendix A.1.1};
(2) -- J2000 source name;
(3) -- J2000 source positions, Right Ascension and Declination; 
asterisk ``$^*$'' denotes VLBI-astrometry positions from RFC, the rest of coordinates are taken from the VLASS;
(4) -- Separation between the \uhevent position and the source position in degrees;
(5) -- Redshift as reported by the NASA Extragalactic Database;
(6) -- Flux density at (2$-$4)\,GHz measured within the VLASS;
(7) -- Historic total VLBI flux density from the RFC database at 8\,GHz; in case the 5\,GHz VLBI flux density is available instead, it is marked by the note ``$^\mathrm{c}$'';
(8) and (9) -- Total VLBI flux density at 5\,GHz and 8\,GHz with its $1\sigma$ uncertainty, measured by the VLBA observations on 25~November~2024, respectively;
(10) -- VLBA total spectral index between 5\,GHz and 8\,GHz $\alpha$ ($S\propto\nu^{\alpha}$);
(11) and (12) -- brightness temperature of the core at 5\,Ghz and 8\,GHz with is $1\sigma$ uncertainty, respectively.
Uncertainties of the VLBI flux scale are about or less than 10\,\%; this uncertainty is accounted for in the measurements presented in this table.
}  
\end{tabular}
\end{table*}

\paragraph{Radio light curves}
Light curves at 15~GHz obtained from the Owens Valley Radio Observatory (OVRO; \citealt{2011ApJS..194...29R}) and RATAN-600 observational campaigns (\hyperref[s:ovro_data]{Appendix A.1.2}, \hyperref[s:ratan_data]{Appendix A.1.3}) are exploited for the time-dependent study of the radio emission.

The calibrator source catalog of the Atacama Large Millimeter/submillimeter Array (ALMA\footnote{\url{https://almascience.nrao.edu/alma-data/calibrator-catalogue}}) is used for complementary variability information.

\subsection{Infrared and optical}
Infrared observations are retrieved from public archival data of the \WISE \citep{WISE} and the Near-Earth Object Wide-field Infrared Survey Explorer \cite[\NEOWISE;][]{NEOWISE} programs. Public archival optical data are collected from the Asteroid Terrestrial-impact Last Alert System  \citep[ATLAS;][]{ATLAS1,ATLAS2,ATLAS3}, the Catalina Real-Time Transient Survey \cite[CRTS;][]{CRTS}, the Zwicky Transient Facility \citep[ZTF;][]{ZTF} ground-based telescopes and the space-based \GAIA mission \citep{Gaia1,Gaia2,Gaia3}. See \hyperref[s:ir-optical]{Appendix A.2}.

\subsection{X-rays}
For the purpose of this work, MXT and ECLAIR instruments onboard the \textit{Space Variable Objects Monitor} \cite[\textit{SVOM};][]{2016arXiv161006892W} were used to observe the 90\% containment region in order to identify bright X-ray sources potentially associated with \uhevent. This survey complements the proprietary four-epoch data obtained from the \eROSITA instrument onboard of the Spektrum-Roentgen-Gamma Observatory \citep[SRG;][]{Predehl_2021}, covering all AGN-like objects in the region. Target-of-opportunity observations were conducted in December 2024 with the \SwiftXRT telescope onboard the \textit{Neil Gehrels Swift Observatory} \citep[\Swift;][]{2005SSRv..120..165B}, for two sources identified as privileged candidates for a dedicated follow-up (in the \Swift archive, these observations are listed as \texttt{ObsId} 00018990001 and 00036371010).

For all the sources identified as candidate blazars in this work, public archival data were collected from \SwiftXRT, the \textit{Chandra X-ray Observatory} \citep[\Chandra;][]{Weisskopf:2000tx}, and the \textit{ROentgen SATellite} \citep[\ROSAT;][]{1991PoAst..39...15C}. All the X-ray data were analyzed to estimate the intrinsic flux in the $\qtyrange{0.2}{2.3}{keV}$ band. A detailed summary of the estimated fluxes is provided in \autoref{tab:xray_flux}. The historical average of the available measurement for each source is further reported in \autoref{tab:candidate-blazars}.

\begin{figure}
    \centering
    \includegraphics[width=\linewidth]{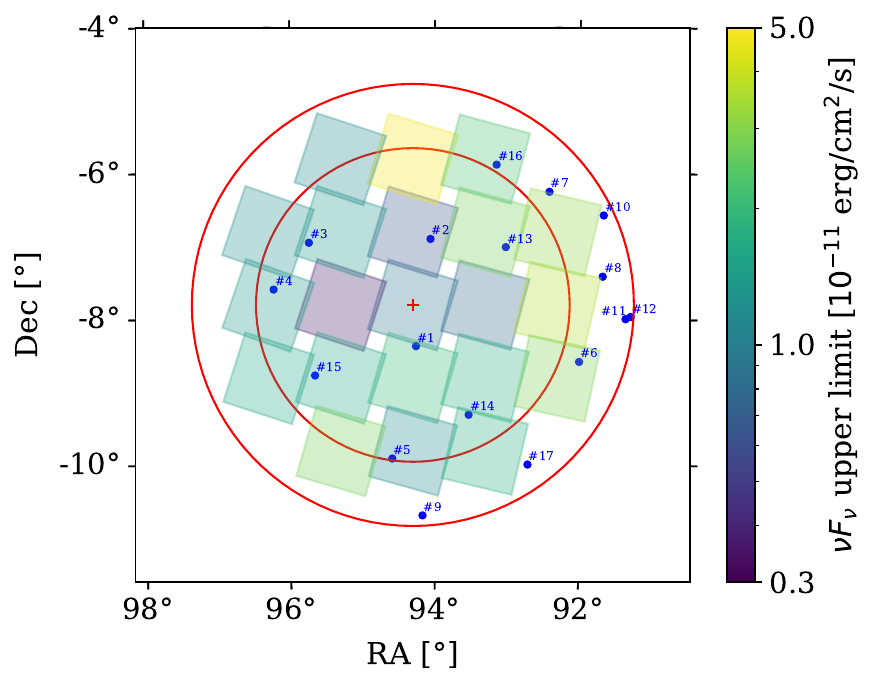}
    \caption{3$\,\sigma$ upper limits derived from the tiling of the neutrino error region with \SVOM-MXT over the 0.5--10\,keV energy band. The concentric red circles indicate the 90\% and 99\% error regions of \uhe, respectively. The numbers refer to the candidate blazars listed in \autoref{tab:candidate-blazars}.}
    \label{f:svom-mxt-tiling}
\end{figure}

\paragraph{SVOM MXT and ECLAIR survey}
The SVOM satellite was launched on June 22, 2024. At the time of the observations, between 2024-12-16T16:52:25.535Z~UTC and 2024-12-17T13:53:34.693Z~UTC (spanning fourteen orbits), \SVOM was in the commissioning and verification phase. \SVOM observed the 90\% neutrino error region using a Multi-Messenger Target of Opportunity (ToO-MM) program. ToO-MM allowed the narrow field-of-view (FoV) instruments onboard \SVOM, the MXT \citep[Microchannel X-ray Telescope;][]{2023ExA....55..487G}, operating in the soft 0.5--10\,keV X-ray band with a 58$\times$58\, arcmin sq. FoV, and the VT \citep[Visible Telescope;][]{2020SPIE11443E..0QF}, operating in the $VT_B$ (400--650\,nm) and $VT_R$ (650--1000\,nm) bands, with a FoV of \qty{26}{\arcmin} diameter, to perform tiled observations of the error region with exposure times ranging from 400 to 2350\,s.
A total of 21 tiles were observed with an average exposure time of 1218.9\,s. \SVOM-MXT data were processed with its dedicated analysis pipeline. No soft X-ray source was detected in the 90\% neutrino error region. Consequently, $3\sigma$~upper limits on the 0.5--10 keV energy flux were computed at the center of each tile following \cite{1991ApJ...374..344K}. The variability of upper limit across the tiled region is primarily driven by differences in exposure time for each tile. The tiling pattern and the corresponding MXT energy flux upper limits for each tile are shown in \autoref{f:svom-mxt-tiling}. None of the candidate blazars listed in \autoref{tab:candidate-blazars} and lying in the 90\% neutrino error region were located in the VT FoV.

\begin{figure}
    \centering
    \includegraphics[width=\linewidth]{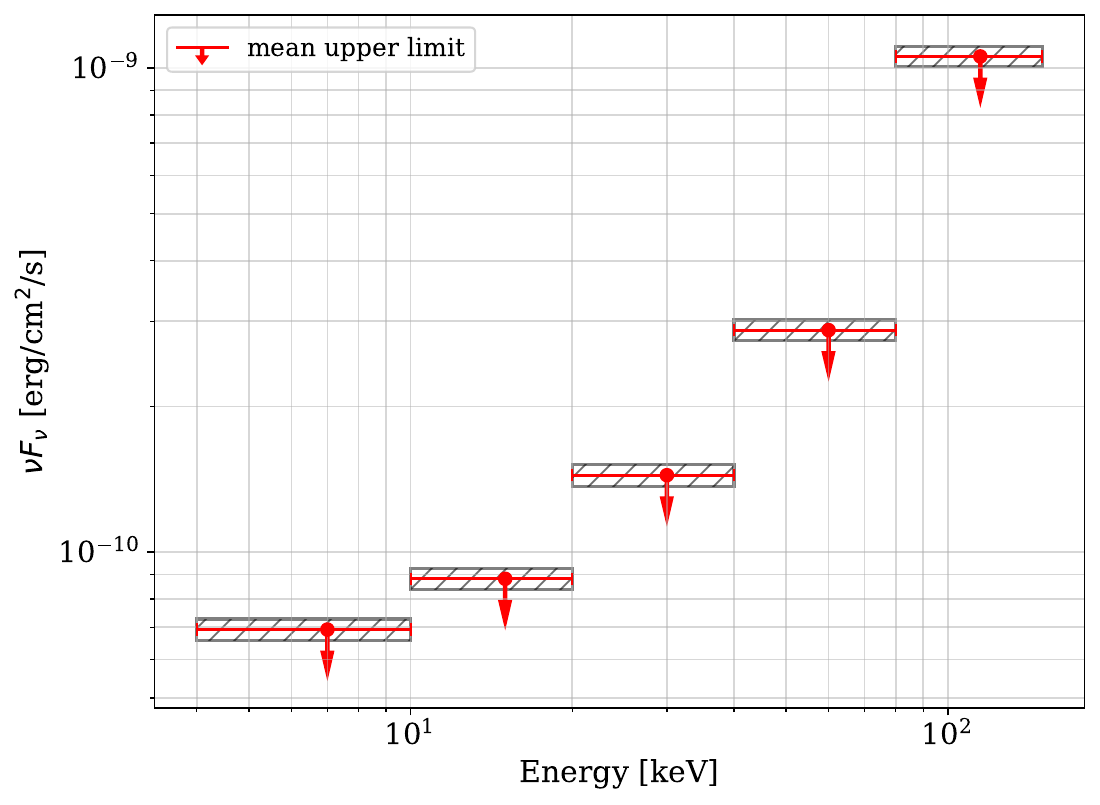}
    \caption{\SVOM-ECLAIRs $3\,\sigma$ upper limits on the energy flux in different energy ranges for the candidate blazars listed in \autoref{tab:candidate-blazars}. The average upper limits for the seventeen targets are shown with red markers. The horizontal hatched bands cover the interval between the minimum and maximum upper limits for the candidate sample.}
   \label{f:svom-eclair}
\end{figure}

In addition, the coded-mask instrument \SVOM-ECLAIRs \citep{2014SPIE.9144E..24G}, with its wide FoV of \qty{2}{\steradian}, observed the entire error region over the 4--150\,keV energy range during the total ToO exposure of 25.6\,ks. No source was detected. For each of the seventeen blazars listed in \autoref{tab:candidate-blazars}, $3\,\sigma$ upper limits over different energy bands were derived from the ECLAIRs variance maps produced by the ECPI data processing pipeline v1.14.4 and assuming a power law spectral model with $\Gamma=2$. They are summarized in \autoref{f:svom-eclair}.

\begin{table*}[h!]
\centering
\caption{Fluxes derived from the available X-ray data in the \qtyrange{0.2}{2.3}{keV} band, given in units of $\SI{E-13}{erg.cm^{-2}.s^{-1}}$. Uncertainties are reported at the $1\,\sigma$ confidence level. The following notes are used: $^\mathrm{a}$, derived flux in the \qtyrange{0.1}{2.4}{keV} energy range; $^\mathrm{b}$, derived flux in the \qtyrange{0.2}{2.0}{keV} energy range. Those marked with a \textdagger~symbol denote upper limits.}
\label{tab:xray_flux}
\renewcommand*{\arraystretch}{0.8}
\scriptsize\begin{tabular}{lccc}
\hline \hline
\noalign{\smallskip}
\textbf{Source}     & \textbf{Instrument} & \textbf{MJD} & \textbf{Flux}                    \\ \noalign{\smallskip}
\hline
\multirow{6}{*}{\#1 \sourcename{1}}  & \ROSAT$^\mathrm{a}$               & 47893        & $1.39^{+0.47}_{-0.48}$           \\
                    & \eROSITA             & 58945        & $3.56^{+0.98}_{-0.79}$           \\
                    & \eROSITA             & 59132        & $3.94^{+1.49}_{-1.06}$           \\
                    & \eROSITA             & 59308        & $6.43^{+1.15}_{-1.05}$           \\
                    & \eROSITA             & 59492        & $7.97^{+1.27}_{-1.11}$           \\
                    & \Swift-XRT                 & 60666        & $10.2^{+1.61}_{-1.43}$            \\ \noalign{\smallskip}
\hline
\multirow{4}{*}{\#3 \sourcename{3}}  & \eROSITA             & 58945        & $\leq 0.69^{\text{\textdagger}}$           \\
                    & \eROSITA             & 59132        & $1.02^{+1.94}_{-0.75}$           \\
                    & \eROSITA             & 59308        & $1.46^{+2.62}_{-0.79}$           \\
                    & \eROSITA             & 59492        & $\leq 0.65^{\text{\textdagger}}$            \\ \noalign{\smallskip}
\hline
\multirow{4}{*}{\#4 \sourcename{4}}  & \eROSITA             & 58945        & $\leq 2.50^{\text{\textdagger}}$            \\
                    & \eROSITA             & 59132        & $1.49^{+1.98}_{-0.82}$           \\
                    & \eROSITA             & 59308        & $\leq 2.00^{\text{\textdagger}}$           \\
                    & \eROSITA             & 59492        & $\leq 0.81^{\text{\textdagger}}$           \\ \noalign{\smallskip}
\hline
\multirow{17}{*}{\#6 \sourcename{6}} & \ROSAT$^\mathrm{b}$               & 47893        & $2.42^{+0.95}_{-0.95}$            \\
                    & \Chandra-ACIS-S               & 52031        & $9.04^{+0.70}_{-0.64}$            \\
                    & \Swift-XRT                 & 54214        & $8.32^{+13.52}_{-7.26}$           \\
                    & \Swift-XRT                 & 54235        & $10.40^{+3.02}_{-1.85}$              \\
                    & \Swift-XRT                 & 54321        & $12.50^{+5.10}_{-3.26}$            \\
                    & \Swift-XRT                 & 54323        & $24.60^{+102.13}_{-13.90}$            \\
                    & \Swift-XRT                 & 54682        & $16.30^{+8.80}_{-5.24}$               \\
                    & \Swift-XRT                 & 54684        & $19.00^{+5.50}_{-3.90}$               \\
                    & \Swift-XRT                 & 54687        & $17.40^{+2.01}_{-1.73}$               \\
                    & \Swift-XRT                 & 54787        & $16.10^{+7.81}_{-4.42}$             \\
                    & \Chandra-ACIS-S               & 55179        & $18.70^{+0.72}_{-0.63}$             \\
                    & \Chandra-ACIS-S               & 55182        & $18.20^{+0.70}_{-0.63}$             \\
                    & \eROSITA             & 58945        & $5.20^{+1.26}_{-1.06}$            \\
                    & \eROSITA             & 59132        & $4.24^{+1.12}_{-0.97}$           \\
                    & \eROSITA             & 59308        & $2.82^{+1.17}_{-0.83}$           \\
                    & \eROSITA             & 59492        & $4.97^{+1.24}_{-1.07}$           \\ 
                    & \Swift-XRT                 & 60670        & $7.52^{+1.39}_{-1.19}$           \\
\noalign{\smallskip}
\hline
\#7 \sourcename{7} & \Swift-XRT                 & 55571        & $\leq 25.00^{\text{\textdagger}}$ \\ \noalign{\smallskip}
\hline
\multirow{4}{*}{\#8 \sourcename{8}}  & \eROSITA             & 58945        & $\leq 0.71^{\text{\textdagger}}$             \\
                    & \eROSITA             & 59132        & $\leq 0.63^{\text{\textdagger}}$           \\
                    & \eROSITA             & 59308        & $0.36^{+0.43}_{-0.24}$           \\
                    & \eROSITA             & 59492        & $\leq 0.35^{\text{\textdagger}}$           \\ \noalign{\smallskip}
\hline
\multirow{4}{*}{\#9 \sourcename{9}}  & \eROSITA             & 58945        & $4.40^{+1.09}_{-0.93}$            \\
                    & \eROSITA             & 59132        & $1.68^{+0.64}_{-0.51}$           \\
                    & \eROSITA             & 59308        & $5.18^{+4.02}_{-2.14}$           \\
                    & \eROSITA             & 59492        & $\leq 0.26^{\text{\textdagger}}$                 \\ \noalign{\smallskip}
\hline
\multirow{5}{*}{\#10 \sourcename{10}} & \ROSAT               & 49244        & $0.77^{+0.23}_{-0.18}$           \\
                    & \eROSITA             & 58945        & $\leq 0.72^{\text{\textdagger}}$           \\
                    & \eROSITA             & 59132        & $0.36^{+0.34}_{-0.26}$           \\
                    & \eROSITA             & 59308        & $\leq 1.32^{\text{\textdagger}}$           \\
                    & \eROSITA             & 59492        & $\leq 0.65^{\text{\textdagger}}$           \\  \noalign{\smallskip}
\hline
\multirow{4}{*}{\#11 \sourcename{11}} & \eROSITA             & 58945        & $1.47^{+1.22}_{-0.66}$            \\
                    & \eROSITA             & 59132        & $0.75^{+0.77}_{-0.40}$            \\
                    & \eROSITA             & 59308        & $2.13^{+0.86}_{-0.68}$           \\
                    & \eROSITA             & 59492        & $3.62^{+1.33}_{-0.95}$           \\ \noalign{\smallskip}
\hline
\multirow{4}{*}{\#12 \sourcename{12}} & \eROSITA             & 58945        & $3.11^{+3.48}_{-1.48}$           \\
                    & \eROSITA             & 59132        & $0.92^{+1.55}_{-0.58}$           \\
                    & \eROSITA             & 59308        & $1.29^{+1.62}_{-0.72}$            \\
                    & \eROSITA             & 59492        & $2.38^{+1.13}_{-0.75}$           \\ \noalign{\smallskip}
\hline
\multirow{4}{*}{\#13 \sourcename{13}} & \eROSITA             & 58945        & $\leq 0.29^{\text{\textdagger}}$            \\
                    & \eROSITA             & 59132        & $\leq 0.72^{\text{\textdagger}}$           \\
                    & \eROSITA             & 59308        & $\leq 1.81^{\text{\textdagger}}$           \\
                    & \eROSITA             & 59492        & $\leq 1.10^{\text{\textdagger}}$            \\ \noalign{\smallskip}
\hline
\multirow{4}{*}{\#17 \sourcename{17}} & \eROSITA             & 58945        & $\leq 1.35^{\text{\textdagger}}$            \\
                    & \eROSITA             & 59132        & $\leq 0.25^{\text{\textdagger}}$           \\
                    & \eROSITA             & 59308        & $\leq 0.76^{\text{\textdagger}}$           \\
                    & \eROSITA             & 59492        & $0.77^{+0.64}_{-0.38}$            \\ \noalign{\smallskip}
\hline
\end{tabular}
\end{table*}

\subsection{Gamma rays}
The data from the Large Area Telescope (LAT) onboard the \textit{Fermi Gamma-ray Space Telescope}, providing continuous observations of the full sky in the $0.1-1000$~GeV energy range, are analyzed \citep[\FermiLAT;][]{Fermi-LAT:2009ihh}. The aim is to characterize the \gammaray emission at the best fit neutrino position of the KM3NeT event, search for potential new point sources in the uncertainty region, and characterize all sources, whether previously cataloged or not, positionally consistent with \uhevent.

To check for previously undetected sources, an iterative procedure using a test statistic (TS) map is applied. The TS is defined as $2\log(L/L_0)$, where \textit{$L$} is the likelihood of the model with a point source at a given position and \textit{$L_0$} is the likelihood without the source. A $TS$ value of 25 corresponds to a statistical significance of $\gtrsim4.0\,\sigma$~\citep[as adopted in][]{4FGL,Mattox:1996}. The details of the adopted maximum-likelihood analysis are given in Appendix A. A TS map is generated by inserting a putative point source at each pixel and evaluating its significance relative to the current best fit model. The test source is modeled with a power-law spectrum, allowing only the normalization to vary while keeping the photon index fixed at 2.0. A search for peaks with TS $>25$ and a minimum separation of \qty{0.5}{\degree} from existing sources is performed. When a significant peak is found, a new point source is added to the model at that position, re-fitting the RoI, and generating a new TS map. This process is iterated until no more significant excesses are found.

To search for transient and variable \gammaray emission on intermediate timescales in the LAT data, the same method is applied to 1-day and 1-month intervals prior to the detection of \uhevent. No new significant ($>5\,\sigma$) excess emission is found within the 90\% confidence localization of \uhevent. A 95\% confidence upper limit for the \qty{0.1}{GeV} flux is determined at the neutrino position: $< 2.6 \times 10^{-10}$\,ph\,cm$^{-2}$ s$^{-1}$ for the $\sim$16-year integration, and $< 1.75 \times 10^{-8}$ ($<1.7 \times 10^{-6}$)\,ph\,cm$^{-2}$ s$^{-1}$ for the 1-month (1-day) integration time. Over these timescales, none of the candidate objects exhibit significant flux enhancements that would suggest flaring behavior or spectral change at \gammarays.

Since the target is close to the Galactic Plane, the analysis is repeated using a more restrictive lower energy threshold of $\geq 1$~GeV to reduce contamination from bright galactic emission. This approach also takes advantage of the improved LAT point spread function at higher energies. The results remain consistent with those obtained from the full energy range analysis. 

The TS maps of the sky region obtained in the analysis are shown in \hyperref[fig:Fermi_tsmap]{Figure A2}.

\section{Candidate blazar counterparts}
\label{s:selection}\label{s:selection-methods}

As part of the initial study of \uhevent~\citep{KM3NeT-2025-Nature}, multiwavelength archival data were explored to compile a census of potential blazar companions within the 99\% confidence region of the event. In this work, while maintaining the original four selection methods, the sample is expanded from twelve to seventeen candidates thanks to Very Long Baseline Interferometry (VLBI) observations conducted with the Very Long Baseline Array (VLBA) (see \hyperref[s:vlbi_sources]{Appendix A.1.1}). The updated selection methods are the following: 
\begin{itemize}
\item Method 1: this strategy relies on the distinctive X-ray, radio and infrared properties of blazars. By cross matching the first \eROSITA catalog \citep[eRASS1;][]{merloni2024} with the NRAO VLA Sky Survey catalog~\citep[NVSS, at 1.4~GHz;][]{1998AJ....115.1693C}, eighteen sources with radio fluxes exceeding \SI{10}{mJy} are identified. 
The sources are positionally cross-matched with the Wide-field Infrared Survey Explorer \citep[\WISE;][]{2014yCat.2328....0C}. The measured \WISE colors of these objects are used to select those that populate the ``\WISE blazar strip''~\citep{2012ApJ...750..138M}, narrowing the list of likely blazar candidates to seven objects (see \hyperref[as:wise-strip]{Appendix A.2.1}). Consistent results are found when cross-matching the eROSITA data set with the DESI Legacy Survey DR10 \citep{DESI:2018ymu} using the Bayesian tool NWAY \citep{Salvato:2017tyv} combined with machine learning, as described in \cite{Salvato:2021lxu}.

\item Method 2: radio blazars are selected on the basis of their parsec-scale flux measured by VLBI, as motivated by population studies \citep{MOJAVEangles} and previous works \citep[e.g., blazars/ANTARES studies;][]{AntaresBlazars}.
Following the approach suggested in \cite{2023MNRAS.523.1799P}, VLBI measurements from the Radio Fundamental Catalog \citep[RFC, version 2024с;][]{RFC} are used. The catalog is complete down to a  cutoff level of \qty{100}{mJy} at \qty{8}{GHz}. Objects above this threshold are selected, resulting in five objects with median VLBI flux density ranging from $\qty{0.1}{Jy}$ to $\qty{2}{Jy}$. In addition to RFC cataloged sources, six objects are pinpointed on the basis of their VLBI properties from the dedicated VLBA observation (see \hyperref[s:vlbi_sources]{Appendix A.1.1}). These are marked with the indication "2a" under "method" in \autoref{tab:candidate-blazars}. 

\item Method 3: the 5th Roma BZ Catalog~\citep[5BZCAT;][]{2015Ap&SS.357...75M}, a compilation of 3561 blazars, has been previously used to investigate the possible neutrino-blazar connection~\citep{KM3NeT-2025-Nature,Buson:2022fyf,Buson_erratum:2022,Buson:2023irp}. Three objects in this catalog are located within the 99\% confidence region.
\item Method 4: The \FermiLAT fourth source catalog \cite[4FGL-DR4;][]{Fermi_4FGL:2020,4FGL_DR4} provides the most recent census of \gammaray sources in the 50\,MeV--1\,TeV energy range. Four 4FGL-DR4 objects are found within the 99\% confidence region of \uhe.
The \Fermi\ Long-Term Transient Sources Catalog \citep[1FLT,][]{Fermi_1FLT:2021}, and the incremental version iFLT\footnote{\href{dd}{https://www.ssdc.asi.it/fermi\_iflt/} } are also considered. No iFLT sources are found within the uncertainty of the neutrino direction.
\end{itemize}

\label{s:main-list}

\begin{figure*}
\centering
\includegraphics[width=\linewidth]{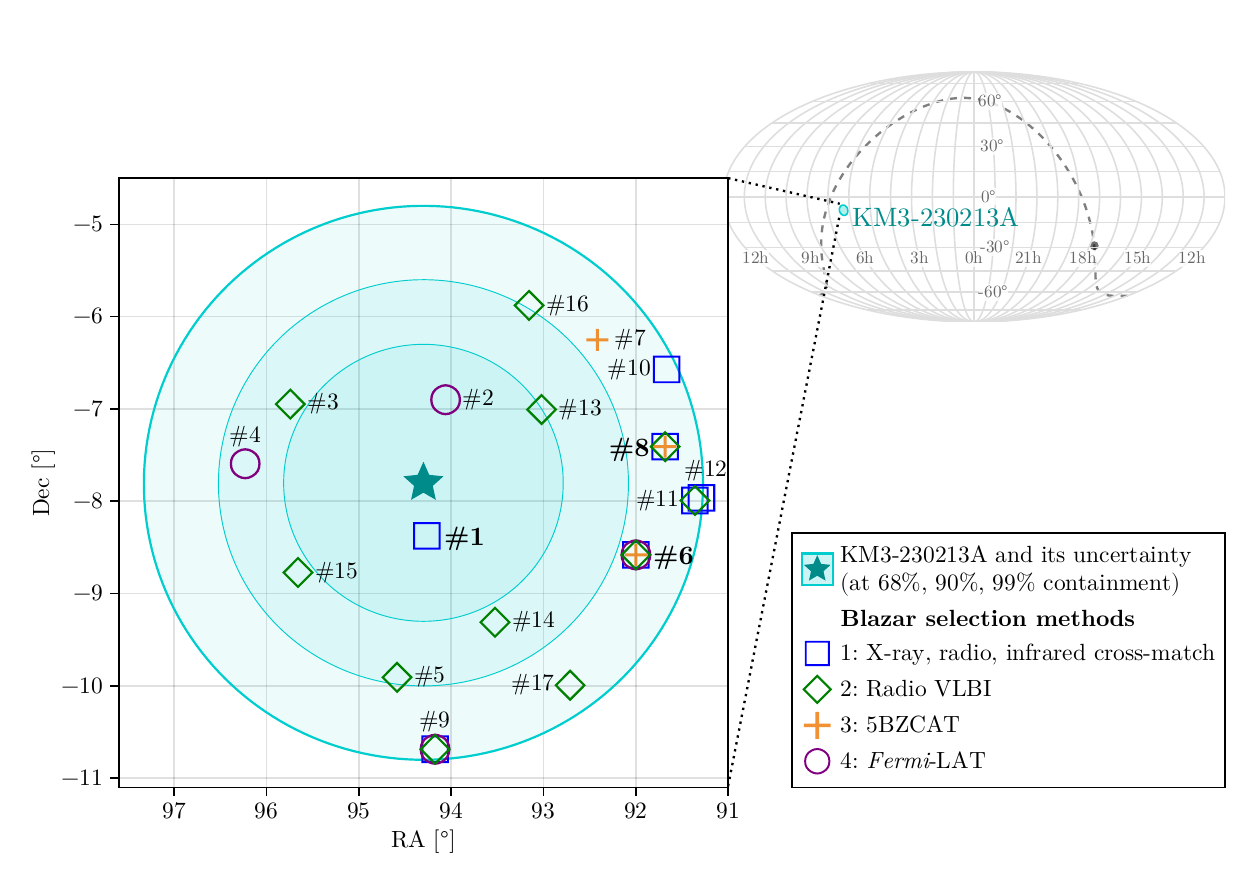}
\caption{
Summary of \uhe{} and its candidate blazar counterparts selected in \autoref{s:selection}. Both the location of the KM3NeT event in equatorial coordinates (J2000) and its uncertainty regions are shown. The markers indicate the criteria used for the inclusion in the list of candidates as presented in \autoref{s:selection}. Source numbers refer to \autoref{tab:candidate-blazars}; the three blazars discussed in \autoref{s:discussion} are labeled in bold here: \#1 (\sourcename{1}), \#6 (\sourcename{6}), \#8 (\sourcename{8}).
}
\label{fig:roi-map}
\end{figure*}

In total, \numberOfCandidates blazar candidates were pinpointed through the methods described above, with four sources selected through more than one method. The final list of objects is provided in~\autoref{tab:candidate-blazars}, which includes their spatial distance from the \uhe best fit position, redshift (when available), associated electromagnetic counterparts, VLBI radio and X-ray flux measurements. The positions of these sources relative to the \uhevent confidence region are shown in Figure~\ref{fig:roi-map}. With the exception of \sourcename{2} (\#2), that is only detected through \gammarays, further discussed in \autoref{s:gamma-nature}, all the other sources are detected in radio.

Spectroscopic classifications and redshift determinations are available only for the three sources reported in the 5BZCAT. They are classified as flat spectrum radio quasars (FSRQs) based on the presence of broad emission lines in their optical spectra \citep{2015Ap&SS.357...75M}.
Object \sourcename{6} (\#6, 5BZQ\,J0607-0834) is one of the fifty brightest radio blazars in the sky.  
Observations with the Hobby-Eberley Telescope (HET) detect the H$\beta$ and \ion{Mg}{2} emission lines, enabling a redshift estimate of $z = 0.870$ \citep{Shaw_2012}.  
\sourcename{7} (\#7, 5BZQ\,J0609-0615) is observed with the New Technology Telescope (NTT), revealing a broad \ion{C}{4} emission line corresponding to a redshift of $z = 2.219$ \citep{Shaw_2012}.  
\sourcename{8} (\#8, 5BZQ\,J0606-0724) appears in the Candidate Gamma-Ray Blazar Survey Catalog \citep[CGRaBS, ][]{2008ApJS..175...97H} and has a spectroscopically measured redshift of $z = 1.227$.

Some of the candidate sources are recorded in existing catalogs of candidate blazars. Three objects (\#6, \#8 and \#9) are reported in CGRaBS. Similarly, three objects (\#1, \#9 and \#11) appear in WIBRaLS
\citep{DAbrusco:2014nij:WIBRaLS}, a catalog of radio-loud candidate gamma-ray emitting blazars selected through the WISE mid-infrared colors, with a strategy similar to the blazar strip selection adopted in Method 1 of \autoref{s:selection}. The corresponding identifiers are indicated as associations in \autoref{tab:candidate-blazars}.




\startlongtable

\begin{deluxetable*}{llrrrclrr}


\tabletypesize{\small}


\tablecaption{Candidate blazars identified using the methods described in Section~\ref{s:selection}, located within the 99\% confidence region of \uhevent.}

\tablenum{1}

\tablehead{\colhead{Source} & \colhead{Names} & \colhead{Sep.} & \colhead{RA} & \colhead{Dec} & \colhead{z} & \colhead{Methods} & \colhead{VLBI} & \colhead{Xray} \\ 
\colhead{} & \colhead{} & \colhead{[$\degree$]} & \colhead{[$\degree$] (J2000)} & \colhead{[$\degree$] (J2000)} & \colhead{} & \colhead{} & \colhead{mJy} & \colhead{[$\qty{E-13}{\frac{erg}{cm^{2}s}}$]} } 

\startdata
\#1 & \textbf{\object{MRC 0614-083}} & 0.6 & 94.2623 & -8.3749 & -- & 1 & -- & $5.58^{+1.16}_{-0.99}$ \\
 & NVSS J061703-082225 &  &  &  &  &  &  & (R,e,S) \\
 & VLASS1QLCIR J061702.97-082229.2 &  &  &  &  &  &  &  \\
 & WISE J061702.95-082229.5 &  &  &  &  &  &  &  \\
 & 1eRASS J061702.8-082230 &  &  &  &  &  &  &  \\
 & WIBRaLS J0617-0822 &  &  &  &  &  &  &  \\
\hline
\#2 & \textbf{\object{4FGL J0616.2-0653}} & 0.9 & 94.06 & -6.90 & -- & 4 & -- & $\leq 0.89$ \\
 &  &  &  &  &  &  &  & (e) \\
\tablebreak
\#3 & \textbf{\object{PMN J0622-0657}} & 1.7 & 95.7419 & -6.9478 & -- & 2 & $   87\pm    9$ & $0.62^{+1.48}_{-0.39}$ \\
 & NVSS J062258-065652 &  &  &  &  &  &  & (e) \\
 & VLASS1QLCIR J062258.04-065651.9 &  &  &  &  &  &  &  \\
 & RFC J0622-0656 &  &  &  &  &  &  &  \\
 & WISEA J062258.02-065652.0 &  &  &  &  &  &  &  \\
 & eRASSU J062257.8-065654 &  &  &  &  &  &  &  \\
\hline
\#4 & \textbf{\object{NVSS J062455-073536}} & 1.9 & 96.2306 & -7.5936 & -- & 4 & $   50\pm    5$ & $0.37^{+1.82}_{-0.21}$ \\
 & VLASS1QLCIR J062455.34-073536.9 &  &  &  &  &  &  & (e) \\
 & eRASSU J062454.9-073541 &  &  &  &  &  &  &  \\
 & 4FGL J0624.8-0735 &  &  &  &  &  &  &  \\
\hline
\#5 & \textbf{\object{PMN J0618-0954}} & 2.1 & 94.5861 & -9.9071 & -- & 2 & $ 100 $ & $\leq 0.93$ \\
 & NVSS J061820-095426 &  &  &  &  &  &  & (e) \\
 & VLASS1QLCIR J061820.67-095425.3 &  &  &  &  &  &  &  \\
 & RFC J0618-0954  &  &  &  &  &  &  &  \\
\hline
\#6 & \textbf{\object{0605-085}} & 2.4 & 91.9987 & -8.5805 & 0.87 & 1, 2, 3, 4 & $ 2240\pm  226$ & $11.6^{+9.25}_{-2.89}$ \\
 & NVSS J060759-083450 &  &  &  &  &  &  &  (R,C,S,e) \\
 & VLASS1QLCIR J060759.69-083450.3 &  &  &  &  &  &  &  \\
 & RFC J0607-0834 &  &  &  &  &  &  &  \\
 & WISE J060759.61-083451.6 &  &  &  &  &  &  &  \\
 & 1eRASS J060759.7-083448 (H) &  &  &  &  &  &  &  \\
 & 4FGL J0608.0-0835 &  &  &  &  &  &  &  \\
 & CGRaBS J0607-0834 &  &  &  &  &  &  &  \\
 & BZQ J0607-0834 &  &  &  &  &  &  &  \\
\hline
\#7 & \textbf{\object{PMN J0609-0615}} & 2.4 & 92.41656 & -6.25163 & 2.219 & 3 & $   48\pm    5$ & $\leq 1.47$ \\
 & NVSS J060940-061505 &  &  &  &  &  &  &  (e) \\
 & VLASS1QLCIR J060939.98-061505.6 &  &  &  &  &  &  &  \\
 & RFC J0609-0615 &  &  &  &  &  &  &  \\
 & WISEA J060939.96-061506.0 &  &  &  &  &  &  &  \\
 & BZQ J0609-0615 &  &  &  &  &  &  &  \\
\hline
\#8 & \textbf{\object{PMN J0606-0724}} & 2.6 & 91.68144 & -7.40840 & 1.277 & 1, 2, 3 & $  306\pm   31$ & $0.09^{+0.53}_{-0.06}$ \\
 & NVSS J060643-072430 &  &  &  &  &  &  & (e) \\
 & VLASS1QLCIR J060643.54-072430.3 &  &  &  &  &  &  &  \\
 & RFC J0606-0724 &  &  &  &  &  &  &  \\
 & WISEA J060643.53-072430.1 &  &  &  &  &  &  &  \\
 & 1eRASS J060643.3-072429 &  &  &  &  &  &  &  \\
 & CGRaBS J0606-0724 &  &  &  &  &  &  &  \\
 & BZQ J0606-0724 &  &  &  &  &  &  &  \\
\hline
\#9 & \textbf{\object{PMN J0616-1040}} & 2.9 & 94.17420 & -10.68568 & -- & 1, 2, 4 & $  248\pm   25$ & $2.82^{+1.5}_{-0.89}$ \\
 & NVSS J061641-104108 &  &  &  &  &  &  & (e) \\
 & VLASS1QLCIR J061641.80-104108.5 &  &  &  &  &  &  &  \\
 & RFC J0616-1041 &  &  &  &  &  &  &  \\
 & WISE J061641.80-104108.4 &  &  &  &  &  &  &  \\
 & 1eRASS J061641.5-104107 &  &  &  &  &  &  &  \\
 & 4FGL J0616.7-1049 &  &  &  &  &  &  &  \\
 & CGRaBS J0616-1041 &  &  &  &  &  &  &  \\
 & WIBRaLS J061641.80-104108.4 &  &  &  &  &  &  &  \\
\hline
\#10 & \textbf{\object{NVSS J060639-063421}} & 2.9 & 91.6667 & -6.5721 & -- & 1 & -- & $0.23^{+0.65}_{-0.09}$ \\
 & VLASS1QLCIR J060640.01-063419.3 &  &  &  &  &  &  & (R, e) \\
 & WISEA J060640.00-063419.6 &  &  &  &  &  &  &  \\
 & 1eRASS J060640.2-063424 &  &  &  &  &  &  &  \\
\hline
\#11 & \textbf{\object{PMN J0605-0759}} & 2.9 & 91.35911 & -7.99216 & -- & 1, 2a & $   67\pm    7$ & $1.99^{+1.04}_{-0.67}$ \\
 & NVSS J060526-075928 &  &  &  &  &  &  & (e) \\
 & VLASS1QLCIR J060526.19-075931.9 &  &  &  &  &  &  &  \\
 & RFC J0605-0759 &  &  &  &  &  &  &  \\
 & WISE J060526.16-075931.7 &  &  &  &  &  &  &  \\
 & 1eRASS J060527.0-075925 &  &  &  &  &  &  &  \\
 & WIBRaLS J0605-0759 &  &  &  &  &  &  &  \\
\hline
\#12 & \textbf{\object{NVSS J060509-075747}}  & 3.0 & 91.28843 & -7.96308 & -- & 1 & -- & $1.93^{+1.94}_{-0.88}$ \\
 & VLASS1QLCIR J060508.92-075747.4 &  &  &  &  &  &  & (e) \\
 & WISEA J060509.22-075747.1 &  &  &  &  &  &  &  \\
 & 1eRASS J060509.0-075739 &  &  &  &  &  &  &  \\
\hline
\#13 & \textbf{\object{PMN J0612-0700}} & 1.5 & 93.02018 & -7.00635 & -- & 2a & $   62\pm    7$ & $\leq 0.98$ \\
 & NVSS J061204-070022 &  &  &  &  &  &  & (e) \\
 & VLASS1QLCIR J061204.84-070022.8 &  &  &  &  &  &  &  \\
 & eRASSU J930117.21-070019 &  &  &  &  &  &  &  \\
\hline
\#14 & \textbf{\object{PMN J0614-0918}} & 1.7 & 93.52517 & -9.31053 & -- & 2a & $   55\pm    6$ & $\leq 0.85$ \\
 & NVSS J061406-091837 &  &  &  &  &  &  & (e) \\
 & VLASS1QLCIR J061406.04-091837.8 &  &  &  &  &  &  &  \\
 & RFC J0614-0918 &  &  &  &  &  &  &  \\
\hline
\#15 & \textbf{\object{PMN J0622-0846}} & 1.7 & 95.65833 & -8.77174 & -- & 2a & $  109\pm   12$ & $\leq 1.13$ \\
 & NVSS J062237-084617 &  &  &  &  &  &  & (e) \\
 & VLASS1QLCIR J062238.00-084618.2 &  &  &  &  &  &  &  \\
 & RFC J0622-0846 &  &  &  &  &  &  &  \\
\hline
\#16 & \textbf{\object{NVSS J061237-055244}} & 2.2 & 93.15567 & -5.87900 & -- & 2a & $   45\pm    5$ & $\leq 1.17$ \\
 & VLASS1QLCIR J061237.35-055244.4 &  &  &  &  &  &  & (e) \\
 & RFC J0612-0552 &  &  &  &  &  &  &  \\
 & SSTSL2 J061237.36-055244.4 &  &  &  &  &  &  &  \\
\hline
\#17 & \textbf{\object{NVSS J061050-095934}} & 2.7 & 92.71130 & -9.99277 & -- & 2a & $   94\pm   10$ & $0.19^{+0.75}_{-0.09}$ \\
 & VLASS1QLCIR J061050.71-095933.8 &  &  &  &  &  &  & (e) \\
 & RFC J0610-0959 &  &  &  &  &  &  &  \\
 & WISEA J061050.70-095933.8 &  &  &  &  &  &  &  \\
 & eRASSU J061051.1-095936 &  &  &  &  &  &  &  \\
\enddata


\tablecomments{Positions are provided in equatorial (J2000) coordinates, with the primary name (in bold) of each source corresponding to either a common name or the name in the first survey reporting the object. Associated sources across different wavelengths are reported, emphasising the surveys and catalogues relevant to each selection method. For completeness, the counterparts from catalogues considered in this work are indicated regardless of the selection method. The listed coordinates prioritize counterparts in the following order: Radio Fundamental Catalog (VLBI), infrared surveys, VLASS, and \FermiLAT\ analysis for the single gamma-ray-only detection. In the Methods column the selection methods are reported (see \autoref{s:selection}): 1 (X-ray, radio, infrared crossmatch), 2 (VLBI from RFC), 2a (VLBI from VLBA observation), 3 (5BZCAT), 4 (4FGL). Radio fluxes ($\qty{8}{GHz}$) and intrinsic, time-averaged X-ray fluxes in the $\qtyrange{0.2}{2.3}{keV}$ band are reported in the respective columns. The X-ray instruments from which the average has been calculated are indicated with a single letter: R (\ROSAT), e (\eROSITA), S (\SwiftXRT), C (\Chandra). Details on the data origin are provided in \hyperref[s:vlbi_sources]{Appendix A.1.1} and \hyperref[s:xray]{Appendix A.3}. For one object, not detected by WISE, the infrared association in the Spitzer (SEIP) source list (SSTSL2; \cite{SSTSL2}) is reported.}


\end{deluxetable*}
\label{tab:candidate-blazars}


\section{Multiwavelength properties of \uhe\ candidate counterparts}
The multiwavelength investigation reported in the following sections has two main goals. First, to characterize the electromagnetic properties of the seventeen candidate counterparts. Second, to reduce the degeneracy in the potential associations by exploiting multiwavelength and time-domain information.

\subsection{Variable or flaring behaviors}\label{s:results}
The remaining sample of candidate counterparts does not provide compelling evidence for a specific association due to the number of blazars within the \qty{3}{\degree} footprint. Brightness, spectrum, and time variability information are used to identify additional observational signatures that can reasonably be expected from a powerful neutrino source.

The \qty{3}{\degree} radius localization region of \uhevent contains \numberOfCandidates astronomical sources that are likely blazars (\autoref{s:selection}). Ten of them show strong radio emission on the parsec and sub-parsec scales (\autoref{s:selection-methods}, \autoref{t:vlba_sources}), indicating strong jet beaming effects.

In addition, the temporal correlation between major flares and the neutrino arrival time has been analyzed as a key observable that can help selecting source associations. 
This observational approach was used in earlier studies, comparing radio flares with neutrino detections from IceCube, ANTARES, and Baikal-GVD \citep{IceCube:2018cha,Plavin20,OVRO_IC_21,AntaresBlazars,OVRO_IC_24,2024MNRAS.527.8784A}. There is growing evidence for neutrinos being produced during large electromagnetic flares.

The most important results of the multiwavelength variability study are discussed in the following sections. While not unequivocal proof, highly variable electromagnetic emission could indicate a higher probability that the neutrino is associated to a specific source. In the following sections, three sources exhibiting flaring behavior are examined. Their properties in the electromagnetic band involved are summarized in \autoref{f:featured-blazars}.

\begin{figure*}
\gridline{
    \fig{lightcurves/Highlight0605m085_LC.pdf}{0.75\linewidth}{(a) The \textit{Fermi}-LAT light curve and a VLBI image of \sourcename{6} (\#6): the brightest radio source in the neutrino localization region that experiences a gamma-ray flaring activity around the neutrino arrival (\autoref{s:radiobright}). $\Delta\mathrm{t}_{\mathrm{peak}}$ highlights the time difference of $\sim181\ \mathrm{days}$ between the flare peak and the neutrino arrival time.}
    \fig{vlbi_images/MOJAVE_0605-085.pdf}{0.2\linewidth}{}
}
\gridline{
    \fig{lightcurves/Highlight0606m0724_LC.pdf}{0.75\linewidth}{(b) The radio light curve for \sourcename{8} (\#8) that experiences a major flare in close coincidence to the neutrino arrival (\autoref{s:radioflare}). $\Delta\mathrm{t}_{\mathrm{peak}}$ highlights the time difference of $\sim5\ \mathrm{days}$ between the flare peak and the neutrino arrival time.}
    \hspace{0.22\linewidth}
}
\gridline{
    \fig{lightcurves/Highlight0614m083_LC.pdf}{0.75\linewidth}{(c) The X-ray light curve for \sourcename{1} (\#1) that indicates a flaring activity around the neutrino arrival (\autoref{s:xrayflare}).}
    \hspace{0.22\linewidth}
}
\caption{
Summary of observational properties related to the three most promising blazars associated with the \uhe{} event. The cyan line highlights the arrival time of \uhevent{} in each light curve.\\
The astronomical coordinates of these blazars are given in \autoref{tab:candidate-blazars} and shown in \autoref{fig:roi-map}. Details on the observational data, along with multiwavelength light curves, are reported in \autoref{s:data} and \autoref{s:light curves}.
\label{f:featured-blazars}}
\end{figure*}

\subsubsection{Source \sourcename{6}: brightest in radio, experiences \gammaray flaring activity}
\label{s:radiobright}

Four known \FermiLAT cataloged objects lie within the 90\% confidence region of \uhe. Among these, \sourcename{6} (\#6) exhibits enhanced \gammaray activity around the time of the detection of \uhe (see \hyperref[f:featured-blazars]{Figure 4a}, \hyperref[fig:LC6]{Appendix Fig.\,\,A10}), with a time difference of 181 days between the peak of the flare and the neutrino arrival time.
The observed year-long flux enhancement exhibits a gradual increase and decrease in emission. Gamma-ray flaring activity has been previously used to pinpoint promising neutrino counterparts  \citep{IceCube:2018cha}, but it may be not unusual for such trends to coincide with a neutrino event by chance.

The blazar \sourcename{6} is among the fifty brightest radio objects in the whole sky in terms of the VLBI flux density, indicating a high degree of relativistic beaming. The object's X-ray activity remains unconstrained due to a sparse observational coverage (see also \hyperref[s:xray]{Appendix A.3}). The recent \SwiftXRT ToO observation measured a flux consistent with historical values. At optical wavelengths, a high flux state is observed, temporally consistent with the \gammaray flare. However, this has to be interpreted with caution, since the blazar is about $\qty{4}{\arcsec}$ away from a foreground star. Although the optical light curve for this blazar was constructed using data from ZTF and \GAIA, which provide sufficient resolution to disentangle point sources separated by such distances, the possibility of residual contamination from the star affecting the optical light curve of the blazar should be further investigated.

\subsubsection{Major radio flare from \sourcename{8}}
\label{s:radioflare}
Among the seventeen candidate blazars there are three radio-bright sources in the OVRO CGRaBS sample with dense uniformly-sampled radio light curves (OVRO + RATAN; \hyperref[f:rskymap]{Appendix Fig.\,\,A4}). One of them, \sourcename{8} (\#8), experiences a large radio flare very close in time to the \uhe\ event, with a time difference of five days, see \hyperref[f:featured-blazars]{Figure 4b} (\hyperref[fig:LC8]{Appendix Fig.\,\,A13}). In this scenario, a neutrino such as \uhe could be produced by photohadronic interactions in the radio core of the blazar, without a significant associated gamma-ray signal \citep{Kivokurtseva:2025sui}.

Such a coincidence is relatively uncommon. Thanks to the coverage of monitoring programs at OVRO and RATAN, a statistical evaluation of its chance probability is possible. The analysis is performed in \autoref{s:flaretest}, and results in a pre-trial p-value of $p=0.26\%$. This finding can be considered a hint of a possible association, consistent with a number of studies that find such associations in blazars \citep{Plavin20,AntaresBlazars,OVRO_IC_21,OVRO_IC_24,IceCube:2023htm:AlertStacking}. Given the a posteriori nature of this analysis, the exact probability should be interpreted with caution, as discussed in \autoref{s:discussion}. This concerns the observed time delay of the radio peak coming five days after the neutrino arrival as well. 

Another source, \sourcename{9}, shows a persistent rapid radio variability in \hyperref[fig:LC9]{Appendix Fig.\,\,A14}, indicative of a very compact ($\sim \qty{10}{\micro as}$) and bright structure probed by scattering of radio waves \citep{2019MNRAS.489.5365K}.

\subsubsection{X-ray flaring activity of \sourcename{1}}
\label{s:xrayflare}

The X-ray band represents a valuable tracer of neutrino production in both jetted and non-jetted AGN. Previous studies have proposed an observable connection between neutrinos and X-ray emission in uniform samples of blazars and in individual sources such as TXS~0506+056 and NGC~1068 \citep{Paliya:2020, 2024JCAP...05..133P,2024arXiv240606684A,2024PhRvD.110l3014K}. Likewise, the potential of the X-ray band to reveal hadronic components in the SED of blazars has been previously shown in the study of the blazar 5BZB\,J0630-2406 \citep{Fichet:2023,Zaballa:2025}. 

Of the seventeen objects listed in Table \ref{tab:candidate-blazars}, nine are detected at X-ray energies. The X-ray fluxes are integrated in the $0.2-2.3$~keV energy band, so that measurements across different facilities can be readily compared in the light curves.  The \eROSITA observations consist of four epochs in the period MJD\,$58677 - 59500$ ($\sim 2$ years). Upon inspection of the eROSITA X-ray data for the candidate counterparts, only two objects are found to be variable: \sourcename{1} (\#1) and \sourcename{9} (\#9). In addition, publicly available archival \SwiftXRT observations suggest long term variability for \sourcename{6} (\#6). 

ToO observations with \SwiftXRT are performed for object \#1 and \#6, on MJD~60666 and MJD~60669, respectively. The primary goal is to investigate X-ray year-long variability, albeit in a coarse manner, since nearly two years have passed since the observation of \uhe. The choice of the timescale is dictated, in this case, by data availability. The \eROSITA sample only extends to $\sim 488$ days before \uhevent, leaving a $\sim 3.2$ year gap between it and the new \SwiftXRT observations.

Object \#6 and object \#9 show no remarkable X-ray flux enhancements at the observation epochs closest to the KM3NeT event. \sourcename{1} (\#1), which is the closest to the best fit neutrino localization, displays X-ray activity that increases steadily in the years leading up to the KM3NeT event (see \hyperref[f:featured-blazars]{Figure 4c}, \hyperref[fig:LC1]{Appendix Fig.\,\,A5}). The recent \Swift follow-up indicates that the high X-ray state of the object persists. This suggests the intriguing possibility that \uhevent occurs while the object is undergoing an X-ray enhancement. Theoretical modeling of the multiwavelength behavior of this source across the optical, infrared and X-ray bands suggests that a super-Eddington accretion flare may be at the origin of the KM3NeT event~\citep{Yuan:2025zwe}.

\subsection{The nature of 4FGL~J0616.2-0653: point-like source or extended diffuse emission?}
\label{s:gamma-nature}

The unidentified, cataloged object \sourcename{2} (\#2) lacks a confident counterpart at other wavelengths. The \FermiLAT analysis of the region reveals other mild excess \gammaray emission, most of which appears to be extended and diffuse. The best fit coordinates of \sourcename{2} are RA\,= 94.06$^\circ$, Dec = -6.90$^\circ$ (J2000). Within the 95\% \gammaray error region of $0.09 {\degree}$, no counterparts are found.
The closest radio source, NVSS J061633-064804, is $\SI{0.12}{\degree}$ from the best fit position of the \FermiLAT object, outside of the 95\% error region. While the radio object is one of the targets of the VLBA observation (J0616-0648, \autoref{t:vlba_sources}), with indication of a blazar nature, it is not included in our selection of candidates due to its relatively low radio flux density and low estimate of the lower limit of the brightness temperature.
The 4FGL-DR4 catalog reports a tentative association of \sourcename{2} with an object that appears in the third EGRET catalog \citep{Hartman:1999}. The EGRET source, 3EG\,J0616$-$0720, is located at RA, Dec = $(94.16, -7.35)$, with a positional uncertainty of 0.91$\degree$ ($95\%$ positional error). With a separation of $0.46 \degree$, 3EG\,J0616-0720, classified as an unidentified object, is positionally consistent with \sourcename{2}.  \citet{Casandjian:2008} excluded 3EG\,J0616-0720 from the EGRET Revised (EGR) source list due to additional structure observed in the interstellar medium.
The lack of convincing counterparts at different wavelengths (radio, infrared, optical and X-ray) does not support the hypothesis of a point-like origin for \sourcename{2}. Furthermore, the excess emission does not appear to be isolated, but rather part of a larger continuous region with an arc-like morphology associated with the Orion molecular clouds \citep{Fermi_orion:2012}, as evident from the test-statistic map of the ROI shown in \hyperref[fig:Fermi_tsmap]{Figure A2}. 
Based on these results, mismodeled diffuse emission may explain the observed \gammaray excess.

\section{Discussion}
\label{s:discussion}

Due to the size of the neutrino direction uncertainty region, comprising \numberOfCandidates candidate blazars, it is not possible to conclusively associate the neutrino with a specific source. The temporal correlations reported in \autoref{s:results} and \autoref{f:featured-blazars} could arise by chance as well. While blazars are here considered the most promising sources for neutrinos at these energies, the broader population of AGN may harbor other candidate counterparts. The event may also originate from a cosmological population of transients, such as gamma-ray bursts \cite{KM3NeT:2025zmb:GRBCompanion}. A Galactic origin for \uhe is, on the other hand, unlikely \citep{GalacticCompanion}. 

Among the temporal correlations that have been examined, only the chance probability for the major radio flare in \sourcename{8} is quantifiable in a robust way (\autoref{s:radioflare}). The observed close coincidence of such a flare with the neutrino (within five days) can arise by chance with the probability of $p=0.26\%$. This correlation can be considered a hint of a possible association, especially in light of earlier studies on the blazar flare--neutrino connection. 
Flaring activity at gamma-ray and X-ray bands is observed in 0605-08 and MRC~0614-083 respectively; these coincidences are more common by chance, and their probabilities are not quantified here. Considering the a posteriori nature of the flare investigation, and the look-elsewhere effect arising from the exploration of time variability across multiple wavelengths, the p-value for the radio flare is interpreted as pre-trial. It serves the purpose of demonstrating that such a close coincidence by chance is rare, but far from impossible. These findings are not claimed as a statistically significant association.

The three cases of flaring sources discussed above do not exhibit strong and unequivocal correlations across different electromagnetic bands. However, if the neutrino was associated with one of such flares, the lack of correlated activity in other wavelengths would still be consistent with previous findings. As introduced in \autoref{s:selection}, the multiwavelength signatures of hadronic processes in blazars can be diverse. In the case of TXS~0506+056, while a single neutrino above \qty{100}{TeV} was detected in the direction of the source while it was flaring in gamma rays, an additional neutrino excess at lower energies was observed during a gamma-ray quiet state \citep{IceCube:2018cha, Fermi-LAT:2019hte}. Theoretical efforts have addressed neutrino production under both flaring and non-flaring gamma-ray states \citep{Padovani_TXS_notBLLac:2019, Winter:2019hee, Petropoulou:2019zqp}.

From a phenomenological standpoint, the conditions required for a blazar to be the source of a muon neutrino with energy $72\,{\rm PeV} \lesssim E_\nu \lesssim 2600\,{\rm PeV}$ can be evaluated. These neutrinos originate in the collisions of ultra-relativistic ions with protons at rest, or photons of sufficient energy. In both cases, the minimum energy required for a single proton is $E_p \sim 20 E_\nu \gtrsim \qty{1.5}{EeV}$ \citep{2021JCAP...03..050R}. Such an energy is still significantly lower than the highest energies observed for cosmic-ray particles, $\sim \qty{300}{EeV}$ \citep{FlysEyeEvent,TelescopeArray:Amaterasu}, for which the jets of active galactic nuclei have already been proposed as possible sources \citep{Rieger:2022qhs}. Assuming a rate of detected neutrino events from a single source at these energies is $T^{-1}$, the luminosity and the minimum jet power can be estimated. The effective area of ARCA with 21 lines at the most likely neutrino energy $E_\nu \simeq \qty{220}{PeV}$ is $A_{\mathrm{eff}} \sim \qty{400}{m^2}$ \citep{KM3NeT-2025-Nature}. The likely counterparts with known redshift are located at $z \approx 1$ (\autoref{tab:candidate-blazars}; luminosity distance $d_L \approx \qty{7}{Gpc}$), and these distances are also expected for neutrino associations from pure volume arguments. The emission can be enhanced with beaming effects by $f_{\rm beam}$ --- the beaming factor. For example, $f_{\rm beam}=1$ for isotropic emission and $f_{\rm beam} \approx 10^3$ when the emission is concentrated within $\qty{4}{\degree}$ around the jet direction. The latter is common for electromagnetic emission of blazars as shown by population studies \citep{MOJAVEangles}. Together, this implies the source luminosity of
\begin{equation}
L_\nu = \frac{\Phi \cdot 4\pi d_L^2}{f_{\rm beam}} = \frac{E_\nu \cdot 4\pi d_L^2}{A \cdot T \cdot f_{\rm beam}} \approx \frac{1.5\cdot10^{49} {\rm erg\,s^{-1}}}{\frac{T}{yr} \cdot f_{\rm beam}}.
\end{equation}
Following the energetics evaluations of \cite{Gottlieb:2021pzr}, the required power of the jet is at least
\begin{equation}
L_p \approx 5 L_\nu \approx \frac{7.5\cdot10^{49} {\rm erg\,s^{-1}}}{\frac{T}{\unit{yr}} \cdot f_{\rm beam}}.
\end{equation}

The maximum feasible neutrino flux and rate for the power available in blazar jets can be estimated. Bright blazars are typically associated with the total jet power of $\sim 10^{45}$~erg\,s$^{-1}$. If most of this power is carried by protons, $L_p \sim 10^{45}$~erg\,s$^{-1}$, and $f_{\rm beam} = 10^3$, a neutrino of this energy could be detected by KM3NeT every $T = \qty{75}{yr}$. For less pronounced beaming, the minimal $T$ estimate would be higher. Note that this analysis is very optimistic: it only accounts for protons with this specific energy, ignoring the contribution of the remaining spectrum to $L_p$. Furthermore, not every proton at these energies produces a neutrino, as the efficiency depends on the target photon or proton field density. %
The observation of a single event has a very limited constraining power. If more neutrinos with these energies are detected from a single source in a time frame of years or decades, even the minimum power requirements would exceed those normally associated with blazars.

Alternatively, such neutrinos can come from many individual sources over the sky; these sources are likely to be at redshift $z\sim1$ by the same volume arguments. The proton luminosity density associated with them can be estimated as 
\begin{equation}
{{\cal L}_p \sim 2\cdot10^{39}{\rm erg\,s^{-1} Mpc^{-3}} \tau^{-1}(\unit{yr})}
\end{equation}
where $\tau^{-1}$ is the total detection rate of neutrinos with these energies at KM3NeT. The isotropic flux of $\Phi_\mathrm{sky}^{1f} = 7.5\times10^{-10}$~GeV\,cm$^{-2}$\,s$^{-1}\,sr^{-1}$ estimated in \cite{KM3NeT-landscape} corresponds to one event every $\tau = 62$~yr. This rate requires the luminosity density of ${\cal L}_p \sim 3\cdot10^{37}{\rm erg\,s^{-1} Mpc^{-3}}$. Such an ${\cal L}_p$ constitutes a fraction of a percent of the total AGN luminosity density, consistent with blazars being the sources of UHE neutrinos.

Another consideration is whether a blazar can accelerate particles to such high energies at all. A generic argument can be used to place another lower bound on the jet power required to accelerate a proton to an energy $E_p$ \citep[][Globus \& Blandford, ARAA submitted]{2000PhST...85..191B}. Before creating the neutrino, the proton must have traversed a potential difference ${\gtrsim E_p/q = E_p \,\unit{(eV)^{-1}}\,\unit{V}}$ where $q$ is its charge.
In most models of acceleration, the effective impedance $Z$ is a fraction of the impedance of vacuum, or $Z \sim \qty{30}{\ohm}$. This implies a minimum power requirement of $L_{\rm jet} > (E_p/q)^2 Z^{-1}$; for the neutrino energy in the hundreds of PeV, the requirement is $L_{\rm jet} > 10^{42}\,\unit{erg.s^{-1}}$.
Accelerating a small number of protons that can produce UHE neutrinos is not as challenging as accelerating to the highest observed cosmic-ray energy, $\sim300\,{\rm EeV}$. The acceleration timescale at these high energies will always be shorter than the light crossing time across the jet and, in some scenarios, like relativistic magnetic reconnection, could be much shorter. Definite association of UHE neutrino events with fast flares in future observations would be more prescriptive.


\section{Summary}
\label{s:summary}

In this paper, we present a multiwavelength observational study of blazar candidates as possible associations of the KM3NeT event \uhevent \citep{KM3NeT-2025-Nature}. Our work combines new dedicated observations with archival data across the electromagnetic spectrum to provide the most complete dataset available for these sources. In total, seventeen likely blazar candidates sources are found within the 99\% uncertainty region. Among them, the three most interesting candidates are highlighted in \autoref{fig:roi-map} and \autoref{f:featured-blazars}:

\begin{enumerate}
\item Object \sourcename{1} is the closest object, located 0.6\degree away from the best fit neutrino position, with indications of X-ray activity during the relevant period.
\item Object \sourcename{6} is among the fifty brightest blazars in the sky on parsec scales. Our data reveal a long-term \gammaray flare peaking before the neutrino arrival.
\item Object \sourcename{8} exhibits a major radio flare, peaking at 15\,GHz within 5~days of the neutrino arrival time. The pre-trial chance coincidence probability is estimated as 0.26\,\%.
\end{enumerate}

Based on our findings, the KM3NeT event KM3-230213A cannot be conclusively associated with enhanced multiwavelength emission from any single blazar. However, the hypothesis that the event originates from a blazar remains viable, with several blazar-like cataloged objects identified as potential counterparts in the neutrino field. Additionally, the event could potentially be associated with a population of faint and/or remote sources not individually resolved.

Improvements in the calibration of the KM3NeT telescope are expected to decrease systematic uncertainties in the directional reconstruction of KM3-230213A, thereby reducing the degeneracy in identifying possible counterparts. The comprehensive multiwavelength observational data presented in this work will be crucial for reliable candidate selection in future analyses, providing an essential baseline for understanding the electromagnetic behavior of these sources.

\section*{Acknowledgments}
This work was supported by the European Research Council, ERC Starting grant \emph{MessMapp}, S.B. Principal Investigator, under contract no. 949555.
This work was funded by the European Union (ERC MuSES project No 101142396).
The authors acknowledge the financial support of:
KM3NeT-INFRADEV2 project, funded by the European Union Horizon Europe Research and Innovation Programme under grant agreement No 101079679;
Funds for Scientific Research (FRS-FNRS), Francqui foundation, BAEF foundation.
Czech Science Foundation (GAČR 24-12702S);
Agence Nationale de la Recherche (contract ANR-15-CE31-0020), Centre National de la Recherche Scientifique (CNRS), Commission Europ\'eenne (FEDER fund and Marie Curie Program), LabEx UnivEarthS (ANR-10-LABX-0023 and ANR-18-IDEX-0001), Paris \^Ile-de-France Region, Normandy Region (Alpha, Blue-waves and Neptune), France,
The Provence-Alpes-Côte d'Azur Delegation for Research and Innovation (DRARI), the Provence-Alpes-Côte d'Azur region, the Bouches-du-Rhône Departmental Council, the Metropolis of Aix-Marseille Provence and the City of Marseille through the CPER 2021-2027 NEUMED project,
The CNRS Institut National de Physique Nucléaire et de Physique des Particules (IN2P3);
Shota Rustaveli National Science Foundation of Georgia (SRNSFG, FR-22-13708), Georgia;

The General Secretariat of Research and Innovation (GSRI), Greece;
Istituto Nazionale di Fisica Nucleare (INFN) and Ministero dell’Universit{\`a} e della Ricerca (MUR), through PRIN 2022 program (Grant PANTHEON 2022E2J4RK, Next Generation EU) and PON R\&I program (Avviso n. 424 del 28 febbraio 2018, Progetto PACK-PIR01 00021), Italy; IDMAR project Po-Fesr Sicilian Region az. 1.5.1; A. De Benedittis, W. Idrissi Ibnsalih, M. Bendahman, A. Nayerhoda, G. Papalashvili, I. C. Rea, A. Simonelli have been supported by the Italian Ministero dell'Universit{\`a} e della Ricerca (MUR), Progetto CIR01 00021 (Avviso n. 2595 del 24 dicembre 2019); KM3NeT4RR MUR Project National Recovery and Resilience Plan (NRRP), Mission 4 Component 2 Investment 3.1, Funded by the European Union – NextGenerationEU,CUP I57G21000040001, Concession Decree MUR No. n. Prot. 123 del 21/06/2022;
Ministry of Higher Education, Scientific Research and Innovation, Morocco, and the Arab Fund for Economic and Social Development, Kuwait;
Nederlandse organisatie voor Wetenschappelijk Onderzoek (NWO), the Netherlands;
The grant “AstroCeNT: Particle Astrophysics Science and Technology Centre”, carried out within the International Research Agendas programme of the Foundation for Polish Science financed by the European Union under the European Regional Development Fund; The program: “Excellence initiative-research university” for the AGH University in Krakow; The ARTIQ project: UMO-2021/01/2/ST6/00004 and ARTIQ/0004/2021;
Ministry of Research, Innovation and Digitalisation, Romania;
Slovak Research and Development Agency under Contract No. APVV-22-0413; Ministry of Education, Research, Development and Youth of the Slovak Republic;
MCIN for PID2021-124591NB-C41, -C42, -C43 and PDC2023-145913-I00 funded by MCIN/AEI/10.13039/501100011033 and by “ERDF A way of making Europe”, for ASFAE/2022/014 and ASFAE/2022 /023 with funding from the EU NextGenerationEU (PRTR-C17.I01) and Generalitat Valenciana, for Grant AST22\_6.2 with funding from Consejer\'{\i}a de Universidad, Investigaci\'on e Innovaci\'on and Gobierno de Espa\~na and European Union - NextGenerationEU, for CSIC-INFRA23013 and for CNS2023-144099, Generalitat Valenciana for CIDEGENT/2018/034, /2019/043, /2020/049, /2021/23, for CIDEIG/2023/20, for CIPROM/2023/51 and for GRISOLIAP/2021/192 and EU for MSC/101025085, Spain;
Khalifa University internal grants (ESIG-2023-008, RIG-2023-070 and RIG-2024-047), United Arab Emirates;
The European Union's Horizon 2020 Research and Innovation Programme (ChETEC-INFRA - Project no. 101008324). 
Views and opinions expressed are those of the author(s) only and do not necessarily reflect those of the European Union or the European Research Council. Neither the European Union nor the granting authority can be held responsible for them.

This work is supported by NSF grants AST2407603 and AST2407604. We thank the California Institute of Technology and the Max Planck Institute for Radio Astronomy for supporting the OVRO 40\,m program under extremely difficult circumstances over the last 8~years in the absence of agency funding. Without this private support these observations could not have been made.  We also thank all the volunteers who have enabled this work to be carried out.
Prior to~2016, the OVRO program was supported by NASA grants \hbox{NNG06GG1G}, \hbox{NNX08AW31G}, \hbox{NNX11A043G}, and \hbox{NNX13AQ89G} from~2006 to~2016 and NSF grants AST-0808050 and AST-1109911 from~2008 to~2014.\\
K.T. acknowledges support from the European Research Council (ERC) under the European Unions Horizon 2020 research and innovation programme under grant agreement No.~771282.\\
I.L, S.K. and AP were funded by the European Union ERC-2022-STG - BOOTES - 101076343.\\
W.M., R.R., B.M. and P.V.d.l.P. acknowledge support from ANID BASAL FB210003 (CATA). R.R., B.M. and P.V.d.l.P. acknowledge support from Núcleo Milenio TITANs (NCN2023\_002). R.B. and N.G. acknowledge support by a grant from the Simons Foundation (00001470, RB, NG). P.V.d.l.P. also acknowledges support by the National Agency for Research and Development
(ANID) / Scholarship Program / Doctorado Nacional/2023--21232103.\\
T.H. acknowledges support from the Academy of Finland projects 317383, 320085, 322535, and 345899.\\
The work by A.V.~Popkov, Y.A.K., S.V.T., A.K.E, Y.V.S, and P.A.V. is supported in the framework of the State project ``Science’’ by the Ministry of Science and Higher Education of the Russian Federation under the contract 075-15-2024-541.\\
A.V.~Plavin is a postdoctoral fellow at the Black Hole Initiative, which is funded by grants from the John Templeton Foundation (grants 60477, 61479, 62286) and the Gordon and Betty Moore Foundation (grant GBMF-8273).\\

This research has made use of data and/or software provided by the High Energy Astrophysics Science Archive Research Center (HEASARC), which is a service of the Astrophysics Science Division at NASA/GSFC.\\
This research has made use of the NASA/IPAC Extragalactic Database, which is funded by the National Aeronautics and Space Administration and operated by the California Institute of Technology. Part of this work is based on archival data, software or online services provided by the Space Science Data Center - ASI.\\
This work made use of the Swinburne University of Technology software correlator, developed as part of the Australian Major National Research Facilities Programme and operated under license \citep{DIFX2}.\\
This work is partly based on the data obtained with the RATAN-600 radio telescope at the Special Astrophysical Observatory of the Russian Academy of Sciences (SAO RAS).\\
The National Radio Astronomy Observatory is a facility of the National Science Foundation operated under cooperative agreement by Associated Universities, Inc.\\
This research has made use of data from the Radio Fundamental Catalog \citep[RFC, \href{https://doi.org/10.25966/dhrk-zh08}{10.25966/dhrk-zh08},][]{RFC}. \\
This research has made use of data from the MOJAVE database that is maintained by the MOJAVE team \citep{2018ApJS..234...12L}.
This paper makes use of the following ALMA data: ADS/JAO.ALMA\#2011.0.00001.CAL. ALMA is a partnership of ESO (representing its member states), NSF (USA) and NINS (Japan), together with NRC (Canada), NSTC and ASIAA (Taiwan), and KASI (Republic of Korea), in cooperation with the Republic of Chile. The Joint ALMA Observatory is operated by ESO, AUI/NRAO and NAOJ.  
This publication makes use of data products from the Wide-field Infrared Survey Explorer, which is a joint project of the University of California, Los Angeles, and the Jet Propulsion Laboratory/California Institute of Technology, funded by the National Aeronautics and Space Administration. \\
This work presents results from the European Space Agency (ESA) space mission Gaia. Gaia data are being processed by the Gaia Data Processing and Analysis Consortium (DPAC). Funding for the DPAC is provided by national institutions, in particular the institutions participating in the Gaia MultiLateral Agreement (MLA).The Gaia mission website is \url{https://www.cosmos.esa.int/gaia}. The Gaia archive website is \url{https://archives.esac.esa.int/gaia}.\\
This work has made use of data from the Asteroid Terrestrial-impact Last Alert System (ATLAS) project. The Asteroid Terrestrial-impact Last Alert System (ATLAS) project is primarily funded to search for near earth asteroids through NASA grants NN12AR55G, 80NSSC18K0284, and 80NSSC18K1575; byproducts of the NEO search include images and catalogs from the survey area. This work was partially funded by Kepler/K2 grant J1944/80NSSC19K0112 and HST GO-15889, and STFC grants ST/T000198/1 and ST/S006109/1. The ATLAS science products have been made possible through the contributions of the University of Hawaii Institute for Astronomy, the Queen’s University Belfast, the Space Telescope Science Institute, the South African Astronomical Observatory, and The Millennium Institute of Astrophysics (MAS), Chile.\\
This work is based on observations obtained with the Samuel Oschin Telescope 48-inch and the 60-inch Telescope at the Palomar Observatory as part of the Zwicky Transient Facility project. ZTF is supported by the National Science Foundation under Grant No. AST-2034437 and a collaboration including Caltech, IPAC, the Weizmann Institute for Science, the Oskar Klein Center at Stockholm University, the University of Maryland, Deutsches Elektronen-Synchrotron and Humboldt University, the TANGO Consortium of Taiwan, the University of Wisconsin at Milwaukee, Trinity College Dublin, Lawrence Livermore National Laboratories, and IN2P3, France. Operations are conducted by COO, IPAC, and UW.\\
We acknowledge the use of public data from the Catalina Real-Time Transient Survey. \\
This work is partly based on data from eROSITA, the soft X-ray instrument aboard SRG, a joint Russian-German science mission supported by the Russian Space Agency (Roskosmos), in the interests of the Russian Academy of Sciences represented by its Space Research Institute (IKI), and the Deutsches Zentrum für Luft- und Raumfahrt (DLR). The SRG spacecraft was built by Lavochkin Association (NPOL) and its subcontractors, and is operated by NPOL with support from the Max Planck Institute for Extraterrestrial Physics (MPE). The development and construction of the eROSITA X-ray instrument was led by MPE, with contributions from the Dr. Karl Remeis Observatory Bamberg \& ECAP (FAU Erlangen-Nuernberg), the University of Hamburg Observatory, the Leibniz Institute for Astrophysics Potsdam (AIP), and the Institute for Astronomy and Astrophysics of the University of Tübingen, with the support of DLR and the Max Planck Society. The Argelander Institute for Astronomy of the University of Bonn and the Ludwig Maximilians Universität Munich also participated in the science preparation for eROSITA. The eROSITA data shown here were processed using the eSASS software system developed by the German eROSITA consortium.\\
We thank the Swift PI, the Observation Duty Scientists, and the science planners for promptly approving and executing our Swift observations (Obs ID 18990, Obs ID 36371). We acknowledge the use of public data from the Neil Gehrels Swift Observatory data archive.\\
This research has made use of data obtained from the Chandra Data Archive and the Chandra Source Catalog, both provided by the Chandra X-ray Center (CXC). This paper employs a list of Chandra datasets, obtained by the Chandra X-ray Observatory, contained in the Chandra Data Collection \dataset[doi: 10.25574/cdc.447]{https://doi.org/10.25574/cdc.447}."\\
We acknowledge the use of public data from the ROSAT survey. The ROSAT project is supported by the German Bundesministerium für Bildung, Wissenschaft, Forschung und Technologie (BMBF/DARA) and the Max-Planck-Society.\\
This work was supported by CNES, focused on \SVOM. The Space Variable Objects Monitor (\SVOM) is a China-France joint mission led by the Chinese National Space Administration (CNSA), French Space Agency (CNES), and the Chinese Academy of Sciences (CAS), which is dedicated to observing \gammaray~bursts and other transient phenomena in the energetic universe.\\
The \textit{Fermi} LAT Collaboration acknowledges generous ongoing support
from a number of agencies and institutes that have supported both the
development and the operation of the LAT as well as scientific data analysis.
These include the National Aeronautics and Space Administration and the
Department of Energy in the United States, the Commissariat \`a l'Energie Atomique
and the Centre National de la Recherche Scientifique / Institut National de Physique
Nucl\'eaire et de Physique des Particules in France, the Agenzia Spaziale Italiana
and the Istituto Nazionale di Fisica Nucleare in Italy, the Ministry of Education,
Culture, Sports, Science and Technology (MEXT), High Energy Accelerator Research
Organization (KEK) and Japan Aerospace Exploration Agency (JAXA) in Japan, and
the K.~A.~Wallenberg Foundation, the Swedish Research Council and the
Swedish National Space Board in Sweden. Additional support for science analysis during the operations phase is gratefully
acknowledged from the Istituto Nazionale di Astrofisica in Italy and the Centre
National d'\'Etudes Spatiales in France. This work performed in part under DOE
Contract DE-AC02-76SF00515.\\

\vspace{5mm}
\facilities{KM3NeT, VLA, VLBA, RATAN-600, OVRO:40m, 
\Swift, \Fermi, SRG/\eROSITA, \GAIA, CRTS, ATLAS, ZTF, WISE/NEOWISE, \Chandra, \ROSAT.}


\appendix
\counterwithin{figure}{section}

\section{Multi-wavelength data} \label{s:data}

In the following sections, the analysis procedures for the different collected data are described. The corresponding multi-wavelength light curves for each source are displayed in \autoref{fig:LC1} through \autoref{fig:LC17}.
\subsection{Radio data}

\subsubsection{VLBI and parsec-scale AGN jets} \label{s:vlbi_sources}

This paper utilizes both archival and new dedicated VLBI observations. Archival measurements are used in \autoref{s:selection} in the form of the RFC \citep{RFC} that catalogs a complete uniform all-sky sample of VLBI-bright sources, dominated by blazars. Further, a VLBI image from the MOJAVE (Monitoring Of Jets in Active galactic nuclei with VLBA Experiments; \citealt{2018ApJS..234...12L}) is used for illustration of the parsec-scale structure in \autoref{f:featured-blazars}.

\textbf{VLBA data analysis procedure}
A~priori amplitude, phase, and bandpass calibration are performed in \texttt{AIPS} \citep{AIPS}.  
For comparison, an independent a~priori calibration using \texttt{rPicard/CASA} \citep{rPicard} is performed, obtaining consistent results, with \texttt{rPicard}-calibrated correlated flux densities being systematically 12\,\% lower.  
Radio frequency interference partially affects the data in the 7.6\,GHz sub-band, which is marked manually in \texttt{AIPS} and automatically in \texttt{rPicard}.  
The \texttt{PIMA} software \citep{PIMA} is used to evaluate the probability of false positives and the detection sensitivity.  
The detection limit is found to be about 6\,mJy, which is in agreement with the expectations from thermal noise.  

Automated hybrid imaging is performed, applying self-calibration and CLEAN imaging to the \texttt{AIPS}-calibrated data in \texttt{difmap} \citep{difmap, hybrid_maging}.  
The phase self-calibration solution interval is limited to $\SI{12}{s}$ to avoid creating a ``fake'' signal from noise, following \citet{2021AJ....161...88P}.  Most of the sources are slightly resolved, and the properties of their structure are measured by fitting the source structure with two circular Gaussian components directly to the self-calibrated visibility data.

\begin{figure}
    \centering
    \includegraphics[width=0.45\linewidth]{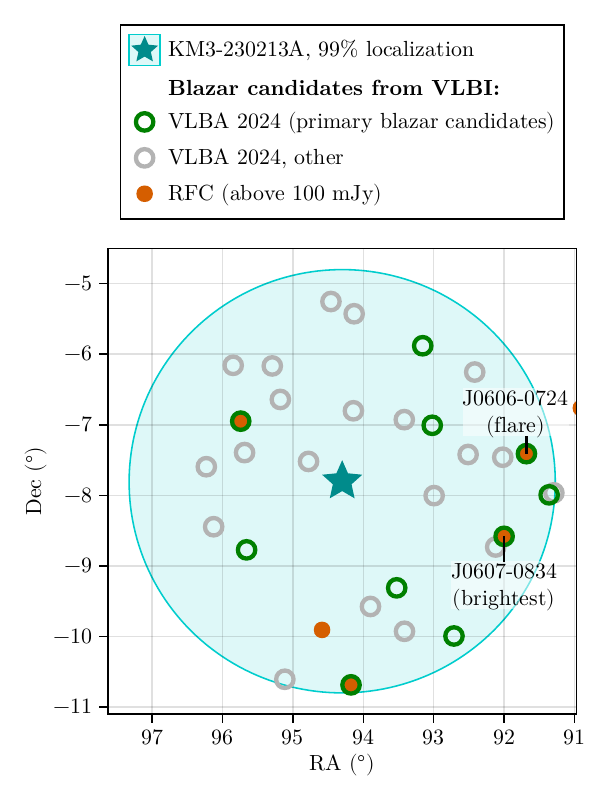}
    \caption{Likely blazar candidates within the \uhevent 99\% localization area, based on their parsec-scale radio emission properties. See \autoref{tab:candidate-blazars} and \hyperref[s:vlbi_sources]{Appendix A.1.1} for details. Two blazars are labeled in the plot: \sourcename{8} exhibits the strongest temporal correlation of radio flares to the neutrino arrival (analyzed in \autoref{s:flaretest}); \sourcename{6} is the brightest in terms of parsec-scale emission.}
    \label{f:vlba_sources}
\end{figure}

\subsubsection{OVRO 40~m}\label{s:ovro_data}

The Owens Valley Radio Observatory (OVRO) 40-meter Telescope has been dedicated to monitoring a carefully-selected sample of $1158$ blazars on a (3$-$4)-day cadence since January 2008, using symmetrical off-axis dual-beam optics in double switching mode to minimize atmospheric effects. The details of the observing method, data reduction, and observations are given in \citet{1989ApJ...346..566R} and \citet{2011ApJS..194...29R}. The cryogenic receiver is centered at 15\,GHz with a 2\,GHz equivalent noise bandwidth.

The quasar 3C\,286 is used for absolute flux density calibration, with an assumed value of 3.44\,Jy \citep{1977A&A....61...99B}. Occasionally, the molecular cloud DR\,21 is used, calibrated relative to 3C\,286 using early OVRO 40\,m data.  
A temperature stabilized noise diode is used to compensate for gain drifts.

The sample of OVRO sources used for the temporal analysis in \autoref{s:flaretest} is the Candidate Gamma-Ray Blazar Survey Source (CGRaBS) sample north of declination $\qty{-20}{\degree}$ \citep{2008ApJS..175...97H}.  
CGRaBS is a catalog of AGN with radio and X-ray properties similar to EGRET \gammaray blazars, selected from a parent population of flat-spectrum radio sources ($\alpha > -0.5$) with 4.8\,GHz flux density $> \qty{65}{mJy}$.  
After removing J1310+3233, a faint target that suffers from confusion with a bright nearby source, the sample used in this paper is reduced to 1157 objects.

\subsubsection{RATAN-600} \label{s:ratan_data}
The RATAN-600\footnote{in Russian, РАТАН-600 - радиоастрономический телескоп Академии наук, Radio Telescope of the Academy of Sciences} observations of broadband radio spectra in the meridian transit mode are used. They cover six frequency bands: 0.96/1.2, 2.3, 4.7, 7.7/8.2, 11, and 22\,GHz, quasi-simultaneously across the different bands, within \mbox{3--5}~minutes \citep{1993IAPM...35....7P}.  
Ten flux density calibrators are used to construct multi-frequency calibration curves covering the declination range between $-35^{\circ}$ and $+45^{\circ}$: J0025$-$26, J0137$+$33, J0240$-$23, J0521$+$16, J0627$-$05, J1154$-$35, J1331$+$30 (3C\,286), J1347$+$12, J2039$+$42 (DR\,21), and J2107$+$42 (NGC\,2027) \citep{1977A&A....61...99B,1994A&A...284..331O,1985AISAO..19...60A,2013ApJS..204...19P,2017ApJS..230....7P}.  
Details of the RATAN-600 antenna, radiometers and data reduction are presented in \citet{1999A&AS..139..545K,2011AstBu..66..109T,2016AstBu..71..496U,2018AstBu..73..494T,2020gbar.conf...32S}.  
RATAN-600 performs long-term monitoring of a complete sample of about 600~AGN with VLBI correlated flux density $>400$\,mJy at 8\,GHz from the RFC catalog at a cadence of about 3~months.

In this work, the quasi-simultaneous multi-frequency RATAN-600 data are interpolated to 15\,GHz for use in the time dependent analysis together with OVRO 15\,GHz measurements.  
The interpolation uses a power law between 22\,GHz and the nearest of the lower RATAN-600 frequencies (11\,GHz, 8\,GHz or 5\,GHz) for each observation epoch.  
The correction for different amplitude scales between telescopes is performed by normalizing the interpolated RATAN-600 fluxes by their median ratio to the OVRO measurements.  
The correction factor is within $\sim10\%$ for all sources. For use in the analysis, we restrict the RATAN-600 data to the time period from January 2008 to be consistent with the OVRO light curve coverage (\hyperref[s:ovro_data]{Appendix A.1.2}).

\subsubsection{ALMA}
The calibrator source catalog of the Atacama Large Millimeter/submillimeter Array (ALMA) is a comprehensive database of astronomical measurements of calibrator sources frequently observed for calibration purposes. The catalog includes observations for three sources, with data available across three receiver bands (Band 3, 6 and 7, corresponding to $84-116\,\mathrm{GHz}$, $211-275\,\mathrm{GHz}$, $275-373\,\mathrm{GHz}$), providing additional variability information.

\subsubsection{UMRAO}
Data from the University of Michigan Radio Astronomy Observatory \citep[UMRAO;][]{UMRAO1,UMRAO2} are provided by M. Aller (private communication), supplementary to the \qty{15}{GHz} data for \sourcename{6} (\#6) from OVRO and RATAN-600.

\subsection{Infrared and optical data}
\label{s:ir-optical}
Archival data in optical and infrared (IR) wavelengths are collected for the sources highlighted in Table \ref{tab:candidate-blazars}. In IR, observations performed with \citep[\WISE;][]{WISE} and the reactivated mission \textit{Near-Earth Object Wide-field Infrared Survey Explore} \citep[\NEOWISE;][]{NEOWISE}) between 2009 and 2024 are used. Both missions observe magnitudes in the W1 ($\qty{3.4}{\micro\meter}$) and W2 ($\qty{4.6}{\micro\meter}$) bands for all sources with \WISE counterparts in Table \ref{tab:candidate-blazars}. As some sources are observed multiple times per day, the weighted mean of the magnitudes is taken for further processing in each observation day. Magnitudes are then transformed to flux units using the corresponding filter curves and zero points \citep{WISEfilters}. Similarly, optical light curves are gathered for multiple sources using publicly available data from various facilities. The Asteroid Terrestrial-impact Last Alert System \citep[ATLAS;][]{ATLAS1,ATLAS2,ATLAS3}), observes the cyan (c) and orange (o) filter bands. The $5\,\sigma$ limiting magnitudes are at $m=19.7$. The filter specifications and zero points are provided by the Spanish Virtual Observatory Filter Profile Service \citep[SVO;][]{SVO1,SVO2}, in addition to the mission publications. The Catalina Realtime Transient Survey  \citep[CRTS;][]{CRTS} observed the sky region at this declination with two out of three telescopes, CSS and SSS. Both provide magnitudes close to the V-band (UBV photometric system) with magnitude limits of $m=19.5$ (CSS) and $m=19.0$ (SSS). For times between 2014 and 2017, \GAIA \citep{Gaia1,Gaia2,Gaia3} provides observations in three custom filters. The filter information and zero points for the G, $\mathrm{G}_{BP}$ and $\mathrm{G}_{RP}$ band is available in \cite{Gaiafilters}. Lastly, observations taken with the Zwicky Transient Facility \citep[ZTF;][]{ZTF} are available in the red (r) and the green (g) band with filter information available in SVO. Note that all magnitudes are corrected for galactic extinction using the source dependent $E(B-V)$ value from \cite{Schlafly_2011} and following the extinction law of \cite{Fitzpatrick_1999} before the conversion to flux units.

\subsubsection{\WISE "blazar strip" selection}
\label{as:wise-strip}
The population of \gammaray blazars has been shown to follow a specific distribution in its infrared observable properties \citep{2012ApJ...750..138M}. In particular, the infrared color-color diagrams evaluated from all four \WISE observation bands are used to build a selection of blazar-like candidates. Given $W1$, $W2$, $W3$ and $W4$ the \WISE bands at 3.4, 4.6, 12 and \qty{22}{\micro\meter}, respectively, the selection is performed in the three dimensional space where the axes are defined by the colors: $W1-W2$, $W2-W3$, $W3-W4$. This classification is statistical in nature. Since the selection adopted in this work is enhanced by the detection of an X-ray counterpart, a 40\% extension of the contours area has been applied to select the candidates reported in \autoref{tab:candidate-blazars}.

\subsection{X-ray observations}
\label{s:xray}
Publicly available archival data for the sources in the sample are collected.  
Of the \numberOfCandidates candidate sources, five (sources \#1, \#6, \#7, \#10, and \#17) have archival data from the \textit{Neil Gehrels Swift Observatory} (\Swift), \textit{Chandra X-ray Observatory} (\Chandra), and/or the \textit{ROentgen SATellite} (\ROSAT).  
In particular, \sourcename{6} \#6 has been observed most extensively, with data from 1990 to 2008 from all of the above instruments.  
In addition, proprietary data from the Spektrum-Roentgen-Gamma Observatory (SRG), in particular from the \eROSITA instrument, are collected and analyzed for the available sources.  
In addition to the archival observations, a Target of Opportunity (ToO) request with \textit{Swift} was submitted to observe sources \#1 and \#6.  
The \textit{Space Variable Objects Monitor} (\textit{SVOM}) satellite has been used to observe the 90\% uncertainty region using a tiling pattern.
The procedures used to estimate the flux for each instrument, as well as the pipeline used for data analysis, are illustrated in the following sections. For all observations, the intrinsic flux is calculated assuming a power-law model with galactic absorption, unless stated differently. The estimated fluxes are shown in \autoref{tab:xray_flux}.

\subsubsection{\ROSAT}

\ROSAT performed the first imaging X-ray survey of the entire sky between 1990 and 1991.
In the following years, until 1999, \ROSAT entered the pointed observing phase, during which more than \num{100000} sources were cataloged \citep{1999A&A...349..389V}.  
Among the sample of \numberOfCandidates blazar-like candidate sources, NVSS\,J061703$-$082225 (1RXS\,J061702.4$-$08222; $8.2^{\prime\prime}$ offset from optical coordinates) and RFC\,J0607$-$0834 (1RXS\,J060758.7$-$08344; $14.9^{\prime\prime}$ offset from optical coordinates) have been included in the \ROSAT All-Sky Survey Faint Source Catalog \citep{2000IAUC.7432....3V}.  
With count rates of \qty{1.16 \pm 0.57 E-2}{s^{-1}} and \qty{1.71 \pm 0.68 E-2}{s^{-1}}, and hardness ratios of $0. 68 \pm 0.73$ and $1.00 \pm 0.44$, for NVSS\,J061703$-$082225 and RFC\,J0607$-$0834, respectively, the intrinsic flux in the (0.1$-$2.0)\,keV band is estimated.
This is done using the count rate-to-energy flux conversion formula presented in the literature \citep{1995ApJ...450..392S}.

\sourcename{10} (\#10) has been cataloged in WGACAT, a comprehensive catalog derived from all \ROSAT PSPC (Position Sensitive Proportional Counter) pointed observations \citep{2000yCat.9031....0W}.  
The source is designated 1WGA\,J0606.6$-$0633, located at $28.9^{\prime\prime}$ from the optical coordinates, with a count rate of $\qty{2.99(0.76)E-2}{s^{-1}}$.  
WGACAT provides a variety of processed data products for each source, including smoothed intensity images, timing images, light-curve plots, and both source and background spectra.  
For this object, the available source spectrum is extracted using an optimized box size, along with an additional response file generated using the \texttt{pcarf} software \citep{1995ASPC...77..367B}.  
The corresponding response matrix, background spectrum, and ancillary response file are processed with \texttt{XSPEC} to estimate the flux in the $\qtyrange{0.2}{2.3}{keV}$ band.

\subsubsection{\Swift-XRT}
As mentioned above, some of the candidates were previously observed with the \Swift observatory \citep{2004ApJ...611.1005G} in addition to the requested observations.  
The X-ray data from \SwiftXRT \citep{2005SSRv..120..165B} are processed using \texttt{FTOOLS}, part of the \texttt{HEASoft} package (v6.33) designed for manipulating and analyzing FITS files.  
For this analysis, all \SwiftXRT observations were performed in photon counting (PC) mode.  
The event files are cleaned and calibrated using standard filtering criteria via the \texttt{xrtpipeline} task, using the latest files from the \Swift \texttt{CALDB} provided by \texttt{HEASARC}.

The spectral extraction for events in the $\SIrange{0.3}{10.0}{\kilo\electronvolt}$ energy range is performed using the \texttt{XSELECT} tool.  
The source signal was extracted from a circular region with a radius of 20 pixels (\qty{47}{\arcsec}), corresponding to about 90\% of the XRT point spread function (PSF), after visually centering the extraction region at the coordinates of the radio counterpart.  
The background spectrum was derived from an annular region centered on the optical coordinates of the source, with an inner radius of 40 pixels ($\sim1.5^\prime$) and an outer radius of 80 pixels ($\sim3.1^\prime$).  
For proper spectral modeling, the binning is set to at least one count per spectral channel, which allows the use of C-statistics \citep{1979ApJ...228..939C} in \texttt{XSPEC}.

\subsubsection{\eROSITA}
The SRG observatory \citep{Sunyaev_2021} is equipped with two main instruments: \eROSITA and ART-XC. \eROSITA is a groundbreaking instrument for surveying the entire sky in soft X-rays, operating at energies between 0.2 and 8~keV. It offers significant advancements compared to earlier facilities like \ROSAT, with improvements in parameters such as effective area, vignetting function, and point spread function \citep{Predehl_2021}.

Using the SRG/\eROSITA all-sky survey, an investigation was conducted within a 6$\times$6\,deg$^2$ error box around \uhevent, focusing only on X-ray point sources. By analyzing the list of potential high-energy counterparts to the neutrino event and matching them to within $\qty{15}{\arcsec}$ of the \eROSITA positions, eleven out of seventeen sources align. 

For these sources, spectra are extracted from the pipeline-processed \eROSITA event files (version 020). Events from all seven telescope modules of eROSITA and from all four completed \eROSITA All Sky Survey iterations (eRASS:4) are combined. The  \textsc{srctool} task of the \eROSITA Science Analysis Software System (eSASS) is used to produce source and background spectra, ancillary response files (ARFs) and response matrix files (RMFs). This is done by extracting source counts from circular regions with radii of 60$^{\prime\prime}$ and background counts from large and nearby regions. For the remaining counterparts to the neutrino event that did not match a point source with \eROSITA, the upper limits where extracted from \cite{Tubin:2024eroul}.

The eRASS:4 spectra were analyzed with the Bayesian X-ray Analysis software (BXA) version 4.1.2 (\cite{buchner2014bxa}), which connects the nested sampling algorithm UltraNest \citep{buchner2019ultranest,buchner2021ultranest} with the fitting environment CIAO/Sherpa \citep{2006SPIE.6270E..1VF}. Spectra were fit using an unbinned approach and the C-statistic. The fitting procedure included a principal component analysis based background model \citep{simmonds2018} derived from a large sample of \eROSITA background spectra. For all spectra, a power-law model (\textsc{powerlaw}) is fitted, including the contribution of Galactic absorption through the component \textsc{tbabs}, using the Galactic column density estimated by the HI4PI collaboration (\cite{Hi4pi}). 

Variability across different \eROSITA epochs is examined by comparing the cataloged (0.2$-$2.3)\,keV flux values of the individual eRASS surveys \cite[eRASS1 to 4;][]{merloni2024}. To identify variable sources, the maximum amplitude variability and its significance are evaluated following the definitions of \cite{boller2016rosat}. Two sources are identified as variable between eRASS surveys with a significance $>2\sigma$ (\sourcename{1} and \sourcename{9}). For these, a time-resolved spectral analysis is performed by applying the same prescriptions as above to the individual eRASS surveys.

\subsubsection{\Chandra}
From the sample of candidates, the blazar \sourcename{6} (\#6) stands out for its three observations with \Chandra. These were made with the Advanced CCD Imaging Spectrometer (ACIS-S) spectroscopic array between 2001 and 2009 \citep{Weisskopf:2000tx}.  
These observations are analyzed using the \texttt{chandra\_repro} script within the CIAO data analysis system \citep{2006SPIE.6270E..1VF}, version 4.16, along with the \Chandra CALDB version 4.11.5.  
The spectra are extracted using the \texttt{specextract} tool provided in the \texttt{CIAO} software suite. In particular, a visual analysis of the reprocessed imaging event file was done to select the region for both the source and the background. For the background, an annular area was utilized, centered on the given coordinates, with an inner radius of $\qty{6}{\arcsec}$ and an outer radius of $\qty{20}{\arcsec}$. Conversely, a circular region of approximately $\qty{5}{\arcsec}$ was selected for the source, deliberately positioned to exclude the nearby foreground star from being included in the selection. As with the rest of the observations, the spectra are binned to ensure one count per bin to be further modeled with \texttt{XSPEC}.

\subsection{Gamma-ray data from \FermiLAT}
\label{s:gamma-fermi}

Available \FermiLAT data are analyzed using the Python package \texttt{fermipy} \citep{Wood:2017}. The region of interest (RoI) is centered at the best fit position of \uhevent and selected photons from the \texttt{Pass 8 SOURCE} class \citep[][]{Atwood:2013, Bruel:2018} within a $10^\circ \times 10^\circ$ square RoI. The analysis covered the period from 2008 August 04 to 2025 January 11 (MJD\,$54682 - 60357$).

To minimize contamination from gamma rays produced in the Earth's upper atmosphere, a zenith angle cut of $\theta < 90^{\circ}$ is applied. The standard data quality cuts ($\rm DATA\_QUAL > 0) \&\& (LAT\_CONFIG == 1$) are applied, periods coinciding with solar flares and \gammaray bursts detected by \FermiLAT are excluded. The RoI model included all 4FGL Data Release 4 catalog sources \citep[4FGL-DR4;][]{4FGL_DR4} located within $15^{\circ}$ of the RoI center, as well as galactic and isotropic diffuse emission models\footnote{\href{https://fermi.gsfc.nasa.gov/ssc/data/access/lat/BackgroundModels.html}{LAT Background Models}} (\texttt{gll\_iem\_v07.fits} and \texttt{iso\_P8R3\_SOURCE\_V2\_v1.txt}).

A binned likelihood analysis over the (0.1$-$1000)\,GeV energy range is performed, using 10 bins per decade in energy and $0.1^{\circ}$ spatial bins. The \texttt{P8R3\_SOURCE\_V3} instrumental response functions (IRFs) are adopted. 

To model a putative point source at the \uhevent best fit position, a power-law spectrum with a fixed index of 2.0 is used. In the fit, the spectral shapes and parameters reported in the 4FGL-DR4 catalog for all sources in the RoI are adopted. First, a fit of the ROI is performed by means of the \textit{fermipy} method ``optimize'' to ensure that all spectral parameters are close to their global likelihood maxima. This is done by iteratively optimizing the components of the ROI model in sequential steps, starting from the largest components.\footnote{\url{https://fermipy.readthedocs.io/en/latest/fitting.html}}

\begin{figure}
    \centering
    \includegraphics[width=0.45\linewidth]{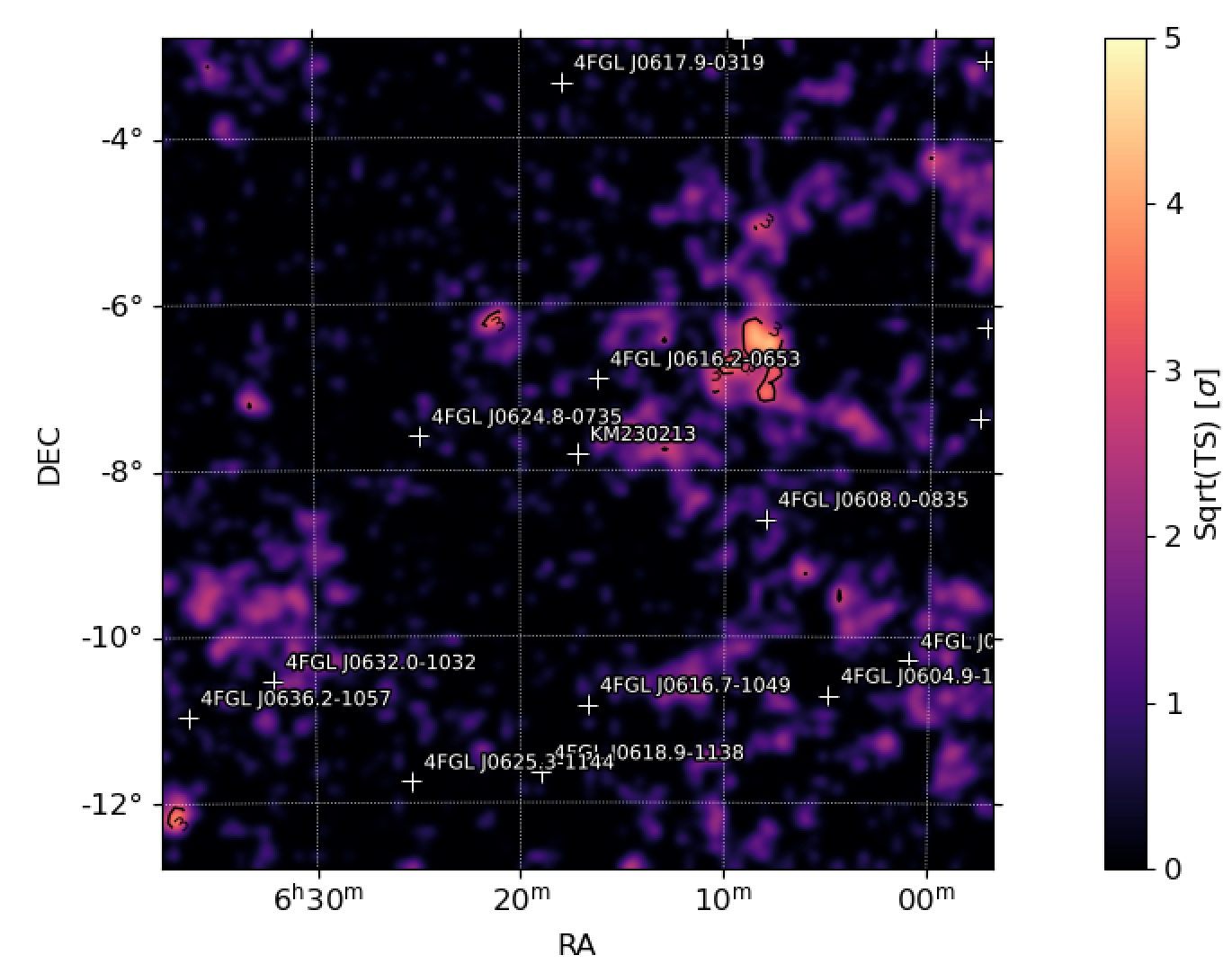}
    \includegraphics[width=.44\linewidth]{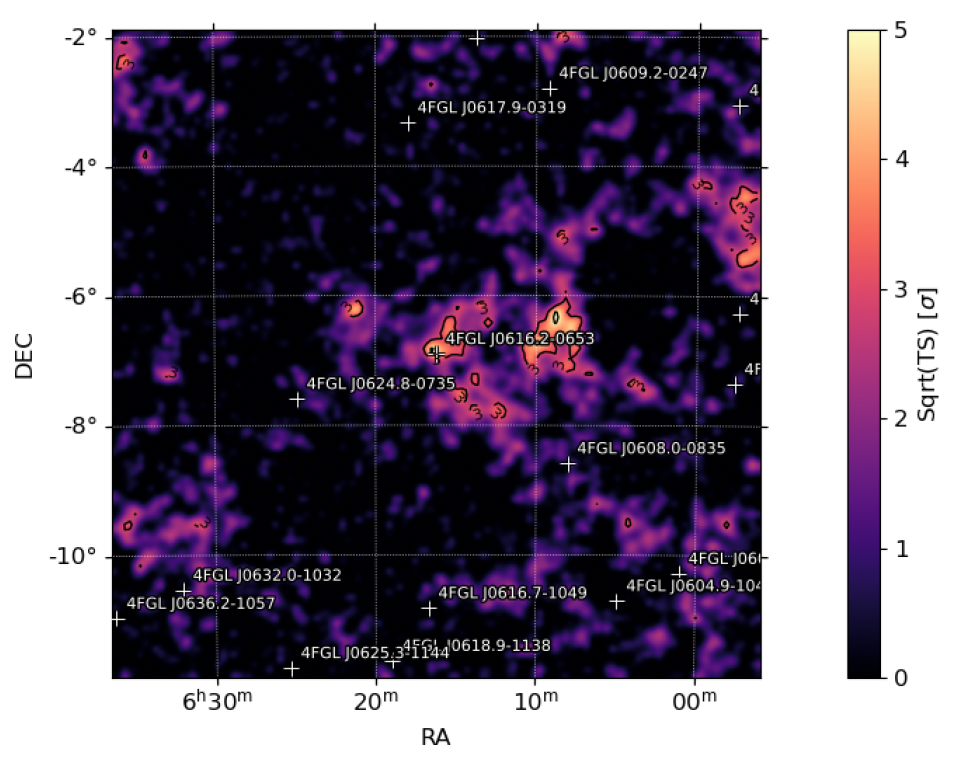}
    \caption{ 
    $Fermi$-LAT TS map ($>0.1$~GeV) of the RoI, integrating $\sim16$ years of observations.
    Left:  The RoI analysis is centered on best fit position of \uhe, and includes all 4FGL-DR4 cataloged sources present in the ROI.
    Right: The RoI analysis is centered on the position of \sourcename{2}, where the source is excluded from the model of the region, to better visualize the excess emission associated with it.
    A mild excess of gamma-ray emission is visible in the RoI, most of which appears to be extended and diffuse. The observed larger continuous region with an arc-like morphology may originate from mismodeled diffuse emission of the Orion molecular clouds. See also \autoref{s:gamma-nature}}
    \label{fig:Fermi_tsmap}
\end{figure}

\subsubsection{Gamma-ray analysis of \uhe\  candidate counterparts}

For each object listed in \autoref{tab:candidate-blazars}, the analysis described in the previous section is performed. In addition, to characterize their variability properties, \FermiLAT light curves are generated with 6-month bins. For the likelihood fits of the time bins, the best fit ROI model obtained from the full time interval analysis is adopted.

First, a fit is attempted by allowing variations in the normalization of the target source, all sources within the inner \qty{3}{\degree} of the ROI, and the diffuse components. If the fit does not converge, the number of free parameters is iteratively restricted until a successful fit is obtained. This procedure begins by fixing the sources in the ROI that are most weakly detected, starting at TS~$<4$. Next, sources with TS~$<9$ are fixed, followed by sources up to $1^\circ$ from the ROI center and those with TS~$<25$. Finally, all parameters except the normalization of the target source are fixed.

The target source is considered as detected if TS~$>9$ in the corresponding time bin. If this condition is not met, a 95\% confidence limit is reported, denoted by down pointing arrows in the \gammarays light curves.

\section{Neutrino-Radio Flare Association probability}
\label{s:flaretest}

\begin{figure*}
\centering
\includegraphics[width=\linewidth]{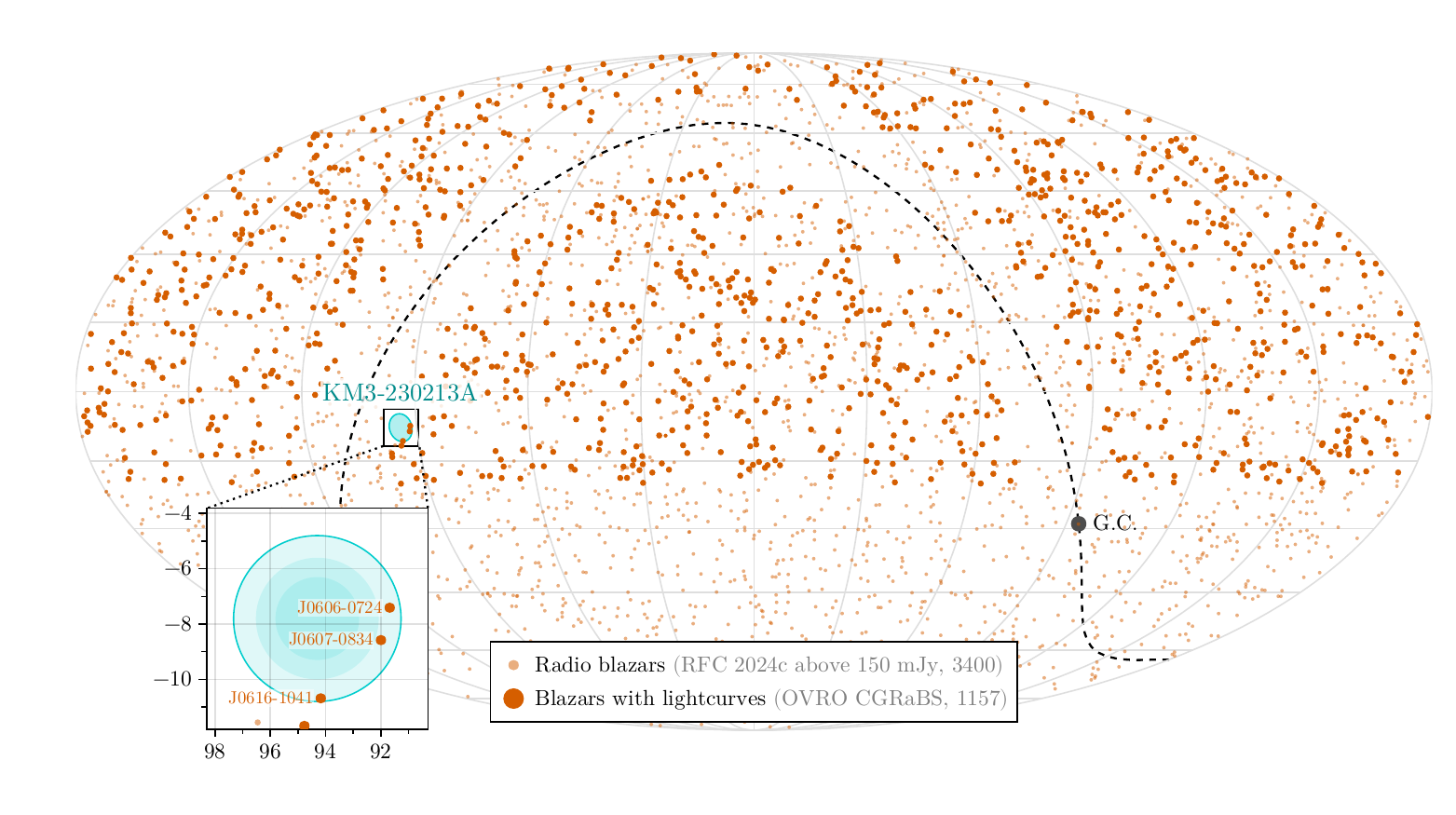}
\caption{
The \uhevent localization in the sky (68\%, 90\% and 99\% containment areas shaded in cyan) together with the RFC bright (above \qty{150}{mJy}) sources  (\hyperref[s:vlbi_sources]{Appendix A.1.1}), displayed along with the sources monitored by the OVRO program (\hyperref[s:ovro_data]{Appendix A.1.2}).
The direction of the event is close to the boundary of the OVRO monitoring area, $\pm \qty{10}{\degree}$ from the Galactic Plane (dashed black line). All bright radio blazars in this region are in the CGRaBS sample and have radio light curves available.}\label{f:rskymap}
\end{figure*}

A statistical analysis to test the association between neutrino arrival times and major blazar radio flares is performed in this section. This consists of an evaluation of the chance probability of finding, in a \qty{3}{\degree} circular region, a blazar with a time correlation as strong as the one observed in the data (\autoref{s:radioflare}). For this purpose, only blazars in the OVRO CGRaBS monitoring programs are considered: for them, a dense continuous radio coverage is available. As seen in \autoref{f:rskymap}, the KM3NeT event is relatively close to the Galactic Plane ($\qty{11}{\degree}$), at the edge of the CGRaBS coverage. Still, all bright (above 150~mJy) radio blazars within the neutrino localization region are covered by the OVRO monitoring program as seen from the lack of other radio sources from the RFC there. Combined OVRO + RATAN light curves are used, see \hyperref[s:ovro_data]{Appendix A.1.2} and \hyperref[s:ratan_data]{Appendix A.1.3} for details.

The aim of the statistical test is to evaluate two hypotheses:
\begin{itemize}
    \item null hypothesis: the neutrino is unrelated to major blazar flares, either not connected to blazars entirely or not preferring times of major flares;
    \item alternative hypothesis: the neutrino arrival time is connected to a major radio flare.
\end{itemize}

\paragraph{Simulation}
To compute the p-value, Monte Carlo simulations under the null hypothesis are used. In order to simulate the observations using minimal assumptions, $N=3$ light curves for random blazars are selected from the full CGRaBS sample. This way, every realization represents the most conservative scenario where the number of coincident blazars is always equal to the observed one.

At each of $10^6$ simulation steps, the following procedure is repeated:
\begin{enumerate}
    \item draw $N=3$ random blazars from the CGRaBS sample;
    \item compute $TS$ for their light curves according to the $TS$ definition below.
\end{enumerate}
In this procedure, the arrival time of the neutrino is fixed and individual light curves are not modified in any way.

As the result, the distribution of the test statistic under the null hypothesis is obtained. Comparing the observed $TS$ value with this distribution leads to the $p$-value.

\paragraph{Test statistic}
The test statistic adopted for the purpose of this test is the time difference between the neutrino arrival and the maximum in the radio light curve of a blazar. For multiple blazars, the shortest time difference is taken. Here, only the global all-time maximum of each light curve is considered. This is a conservative choice: there is always one flare selected for each blazar and no minimum flux threshold. A corresponding benefit is that no free parameters need to be determined or set beforehand.

The full length of the available light curves, about 17~years (\autoref{s:ovro_data}), is used for the test statistic calculation. It is to be noted that this approach makes the test statistic values dependent on the length of the radio monitoring program: the earlier the starting date, the more chances for the all-time maximum to coincide with more extreme historical flares instead of more recent ones closer to the neutrino arrival. This effect fundamentally influences both the observed and simulated coincidences in the same way, ensuring that the analysis is valid. A similar approach using the all-time normalized activity index was used for the same sample of blazars in \cite{OVRO_IC_24}. A potential alternative would be restricting the radio light curves to the operation time frame of KM3NeT. Such a restriction would involve more design choices, such as the proper accounting for the time-varying exposure of the detector. For the purpose of this work, a simple approach most similar to earlier studies was chosen in light of the \textit{a posteriori} nature of the analysis.

The calculation of this test statistic goes as follows:
\begin{itemize}
    \item for a single blazar, $TS_i =$ difference between the neutrino arrival time and the time of the maximum in the radio light curve;
    \item for multiple blazars, $TS = \min(TS_i)$, where the index $i$ runs over all the blazars.
\end{itemize}

\paragraph{Results}

\begin{figure*}
    \centering
    \includegraphics[width=0.49\columnwidth]{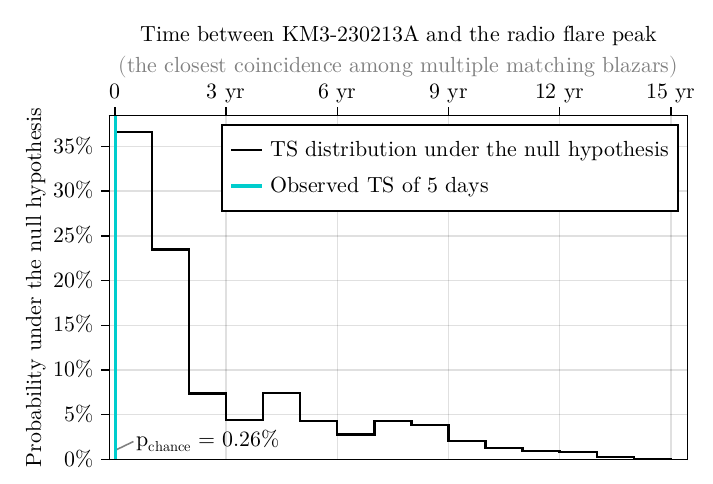}
    \caption{The visualization of the statistical test evaluating the temporal correlation between a radio flare and neutrino arrival for the \sourcename{8} blazar. As discussed in \autoref{s:radioflare}, it exhibits the strongest major flare exactly during the neutrino arrival time. The time delay between the neutrino and the flare peak is just 5 days, making the coincidence remarkable even given the \qty{3}{\degree} angular uncertainty of the neutrino event. Here, the observed value of the test statistic -- the neutrino-radio flare delay -- is shown in comparison with its distribution under the null hypothesis of no association.}
    \label{f:0604-074_flare}
\end{figure*}

\begin{figure*}[h]
\includegraphics[width=0.95\linewidth]{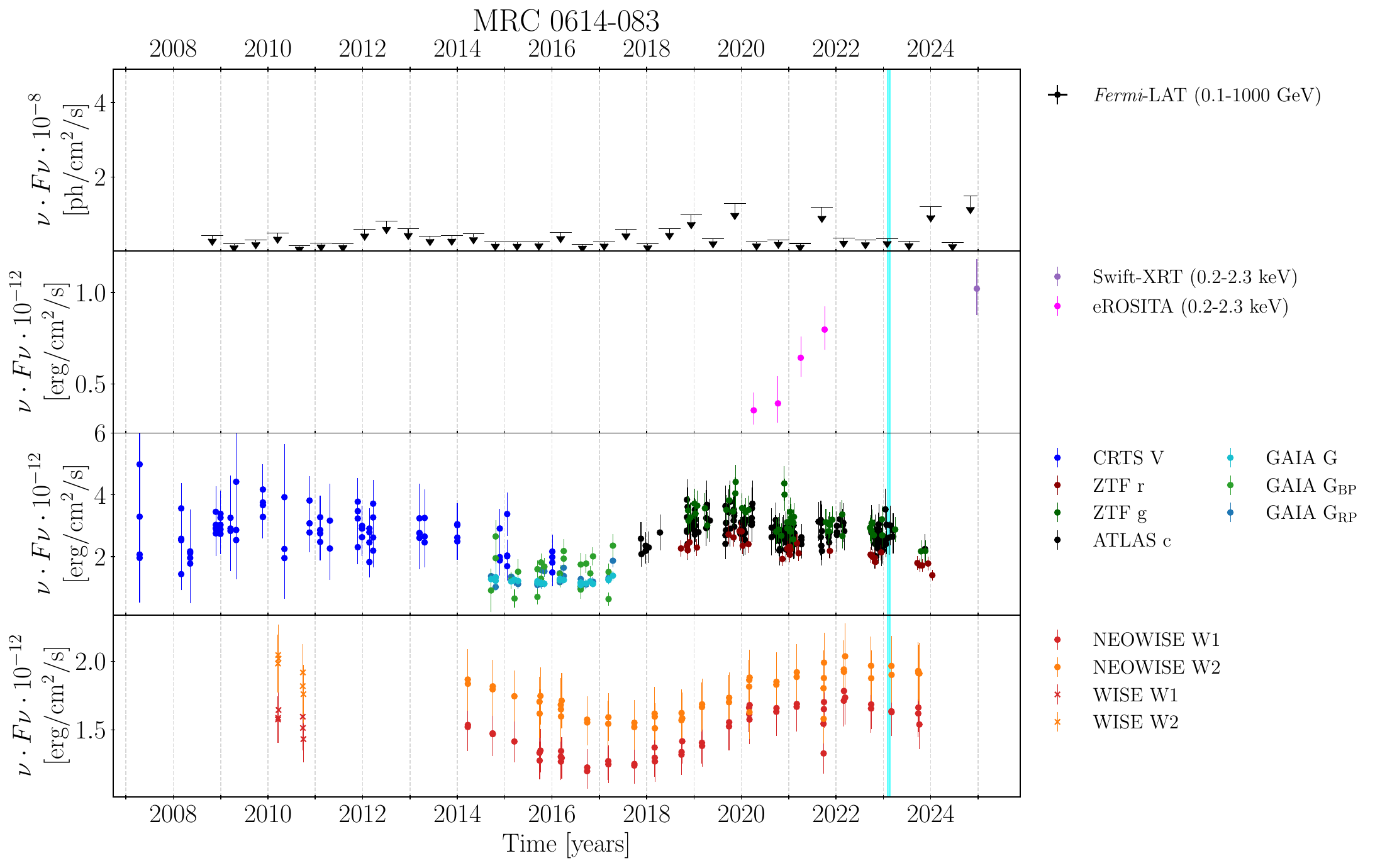}
\caption{Multi-wavelength light curves of \sourcename{1} (\#1). The panels display from top to bottom: \textit{Fermi}-LAT \gammaray light curves integrated over a 6-months time bin; X-ray data from \SwiftXRT and \eROSITA; optical data from CRTS, ZTF, ATLAS and \GAIA; IR data from WISE/NEOWISE. The cyan stripe highlights the arrival time of \uhevent.}
 \label{fig:LC1}
\end{figure*}

\begin{figure*}[h]
\includegraphics[width=0.95\linewidth]{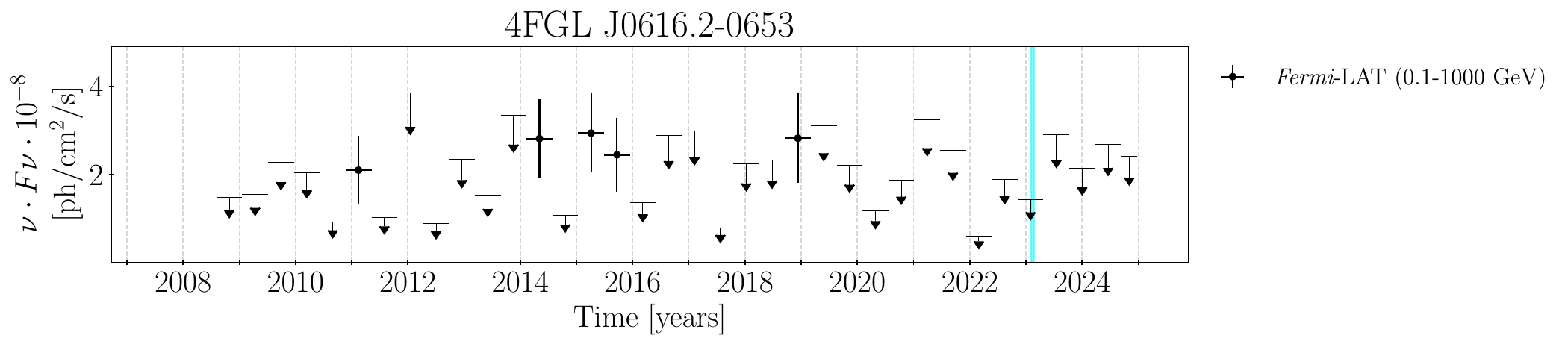}
\caption{$Fermi$-LAT light curve of \sourcename{2}($\#2$), integrated over a 6-months time bin.
The cyan stripe highlights the arrival time of \uhevent.}
 \label{fig:LC2}
\end{figure*}

\begin{figure*}[h]
\includegraphics[width=0.95\linewidth]{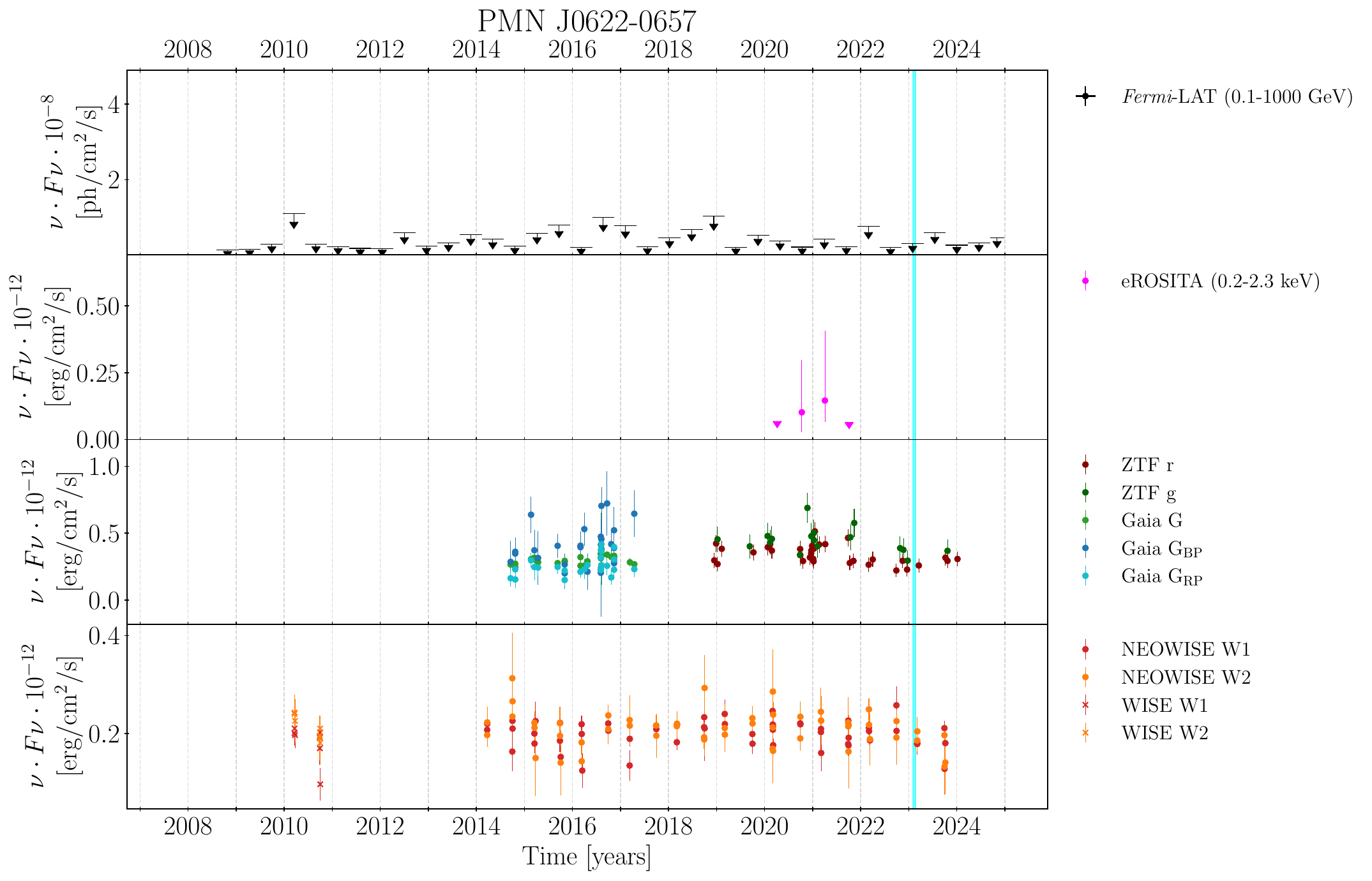}
\caption{Multi-wavelength light curves of \sourcename{3}($\#3$).
The panels display from top to bottom: \textit{Fermi}-LAT \gammaray light curves integrated over 6-months; X-ray data from  \eROSITA; optical data from ZTF and \GAIA; IR data from WISE/NEOWISE. 
The cyan stripe highlights the arrival time of \uhevent.}
 \label{fig:LC3}
\end{figure*}

\begin{figure*}[h]
\includegraphics[width=0.95\linewidth]{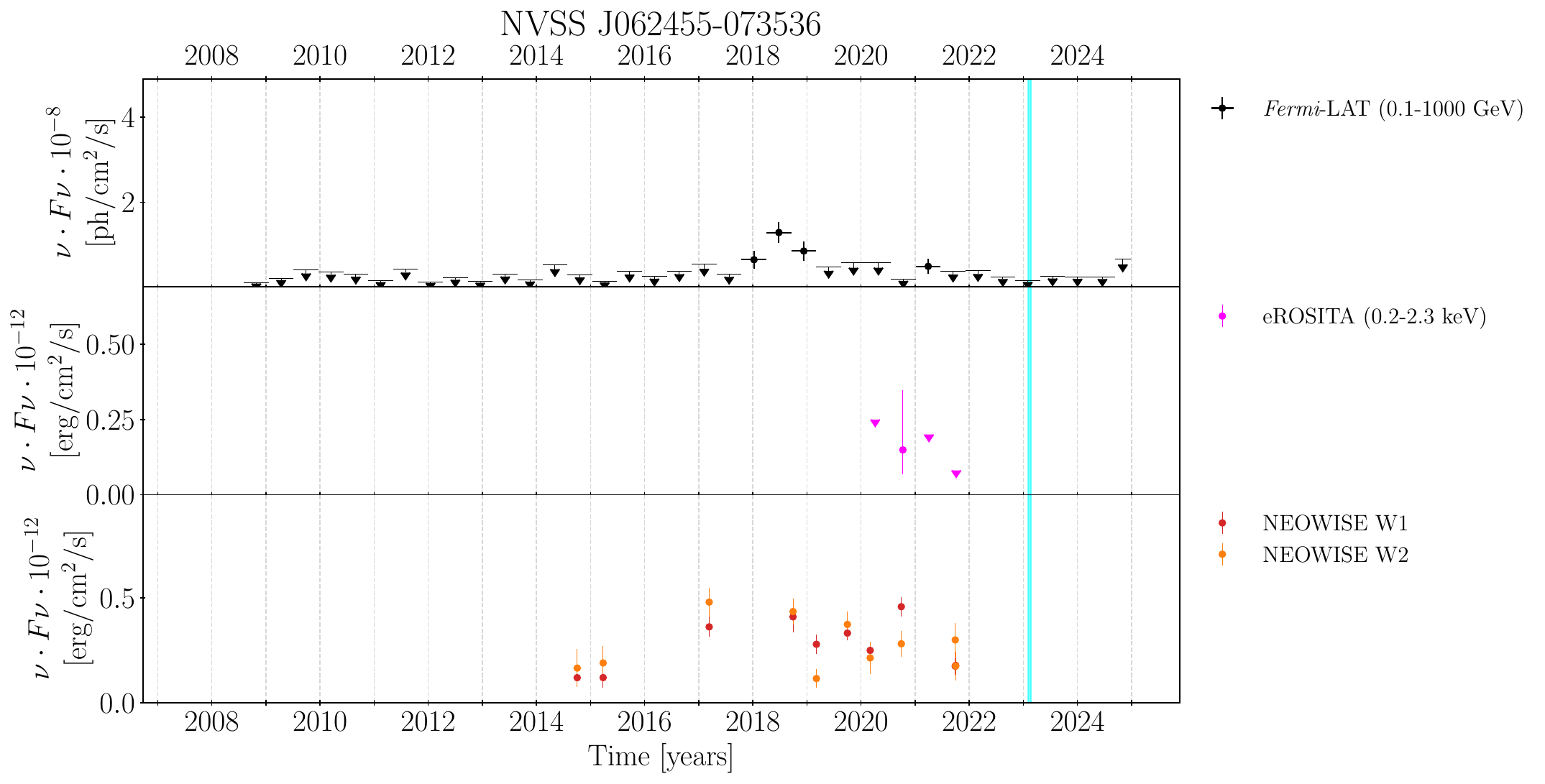}
\caption{Multi-wavelength light curves of \sourcename{4}($\#4$).
The panels display from top to bottom: \textit{Fermi}-LAT \gammaray light curves integrated over 6-months; X-ray data from  \eROSITA; IR data from WISE/NEOWISE. 
The cyan stripe highlights the arrival time of \uhevent.}
 \label{fig:LC4}
\end{figure*}

\begin{figure*}[h]
\includegraphics[width=0.95\linewidth]{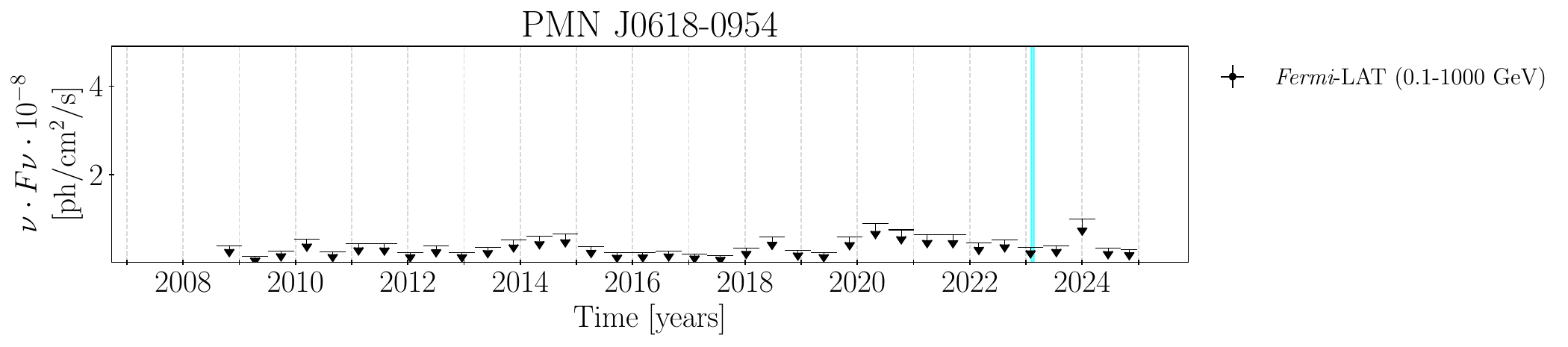}
\caption{$Fermi$-LAT light curve of \sourcename{5}($\#5$), integrated over a 6-months time bin.
The cyan stripe highlights the arrival time of \uhevent.}
 \label{fig:LC5}
\end{figure*}

\begin{figure*}[h]
\includegraphics[width=0.95\linewidth]{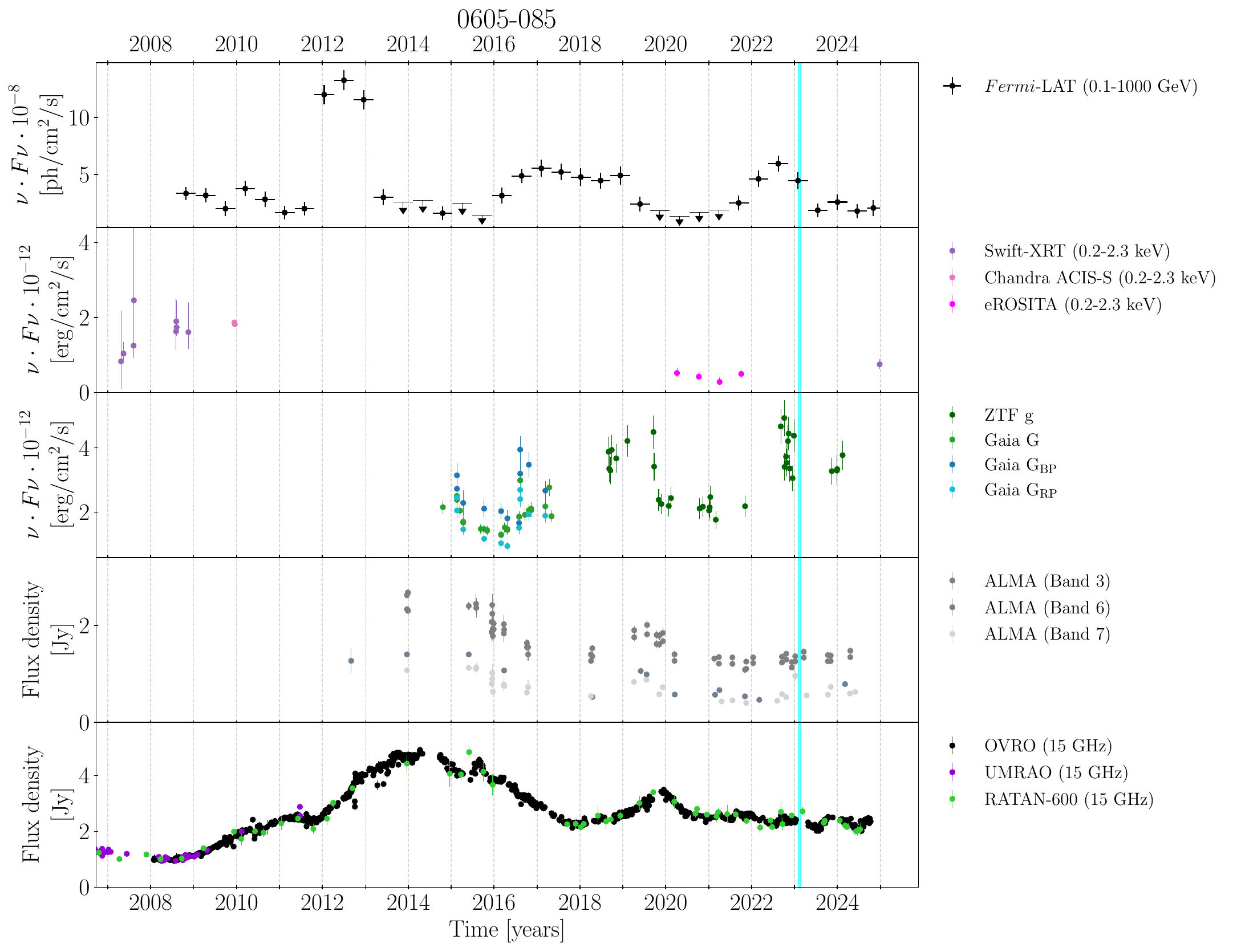}
\caption{Multi-wavelength light curves of \sourcename{6}($\#6$).
The panels display from top to bottom: \textit{Fermi}-LAT \gammaray light curves integrated over 6-months; X-ray data from  \SwiftXRT, \Chandra and \eROSITA; optical data from ZTF and \GAIA; radio data from ALMA, OVRO, UMRAO and RATAN-600. 
The cyan stripe highlights the arrival time of \uhevent.}
 \label{fig:LC6}
\end{figure*}

\begin{figure*}[h]
\includegraphics[width=0.95\linewidth]{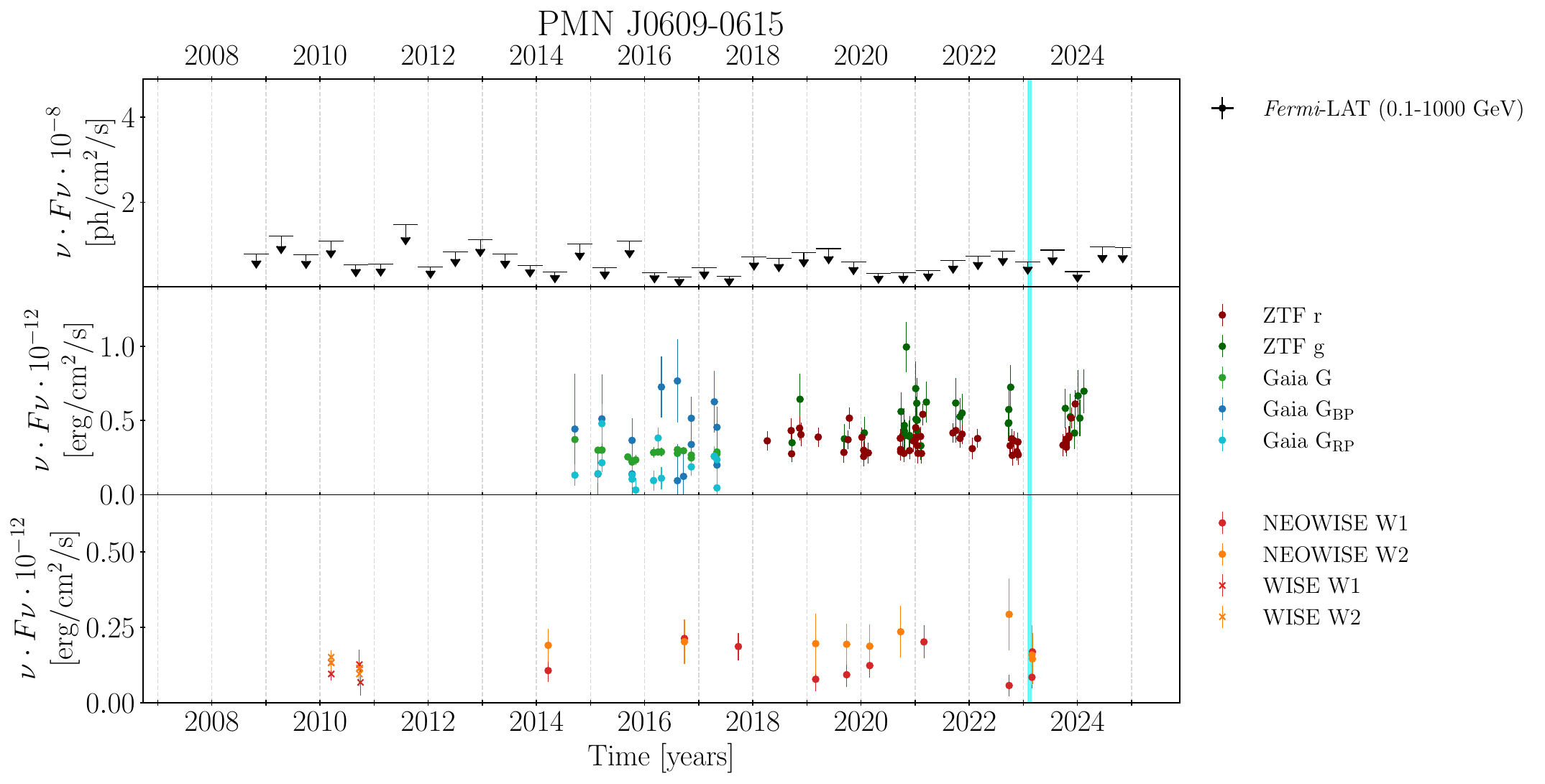}
\caption{Multi-wavelength light curves of \sourcename{7}($\#7$).
The panels display from top to bottom: \textit{Fermi}-LAT \gammaray light curves integrated over 6-months; optical data from ZTF and \GAIA; IR data from WISE/NEOWISE. 
The cyan stripe highlights the arrival time of \uhevent.}
 \label{fig:LC7}
\end{figure*}

\begin{figure*}[h]
\includegraphics[width=0.95\linewidth]{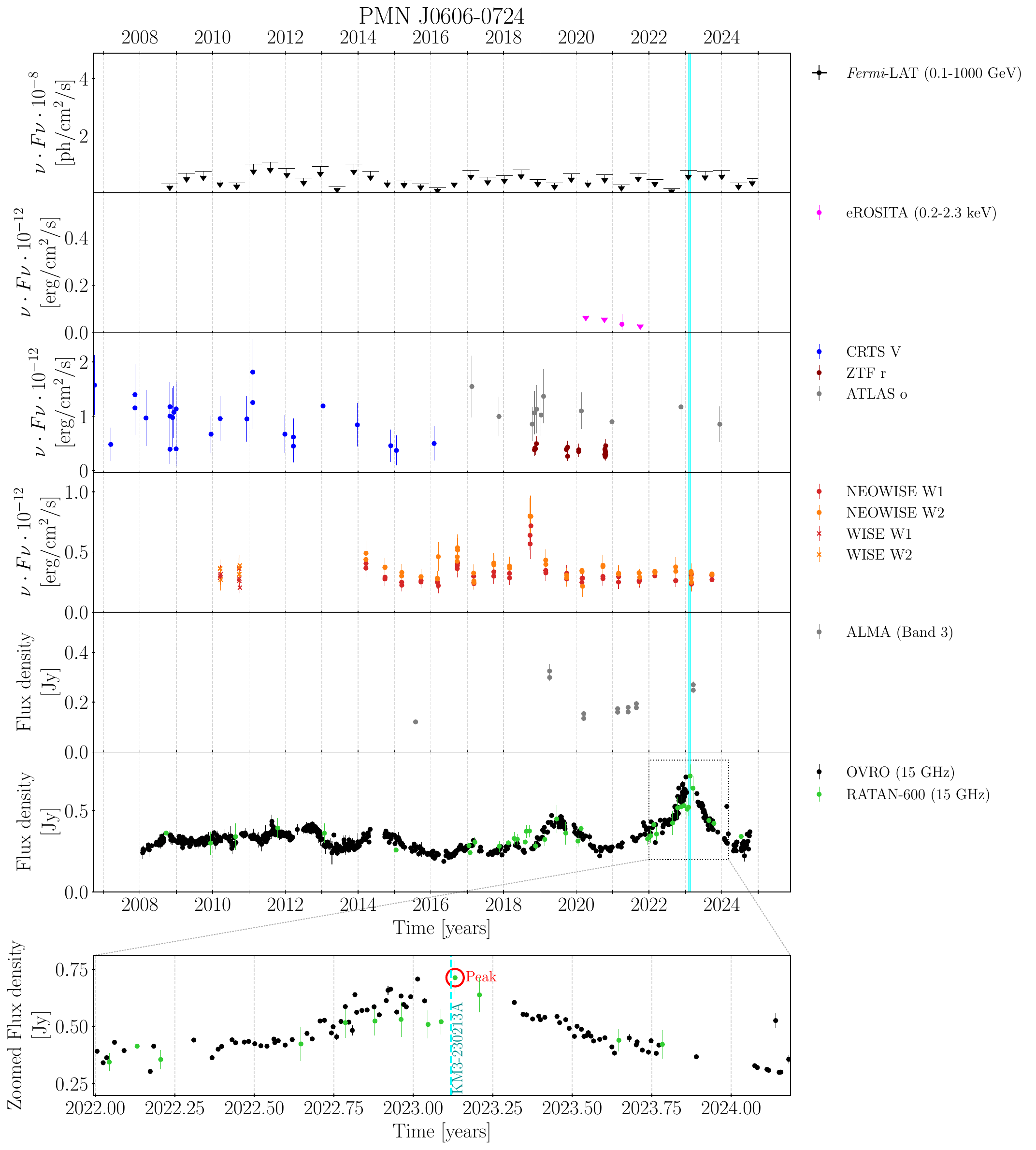}
\caption{Multi-wavelength light curves of \sourcename{8}($\#8$).
The panels display from top to bottom: \textit{Fermi}-LAT \gammaray light curves integrated over 6-months; X-ray data from  \eROSITA; optical data from CRTS, ZTF and ATLAS; IR data from WISE/NEOWISE; radio data from ALMA, OVRO and RATAN-600. The last panel displays the zoomed radio light curve centered on the neutrino arrival time. The cyan stripe highlights the arrival time of \uhevent. The red circle indicates the all-time maximum peak of the radio light curve.}
 \label{fig:LC8}
\end{figure*}

\begin{figure*}[h]
\includegraphics[width=0.95\linewidth]{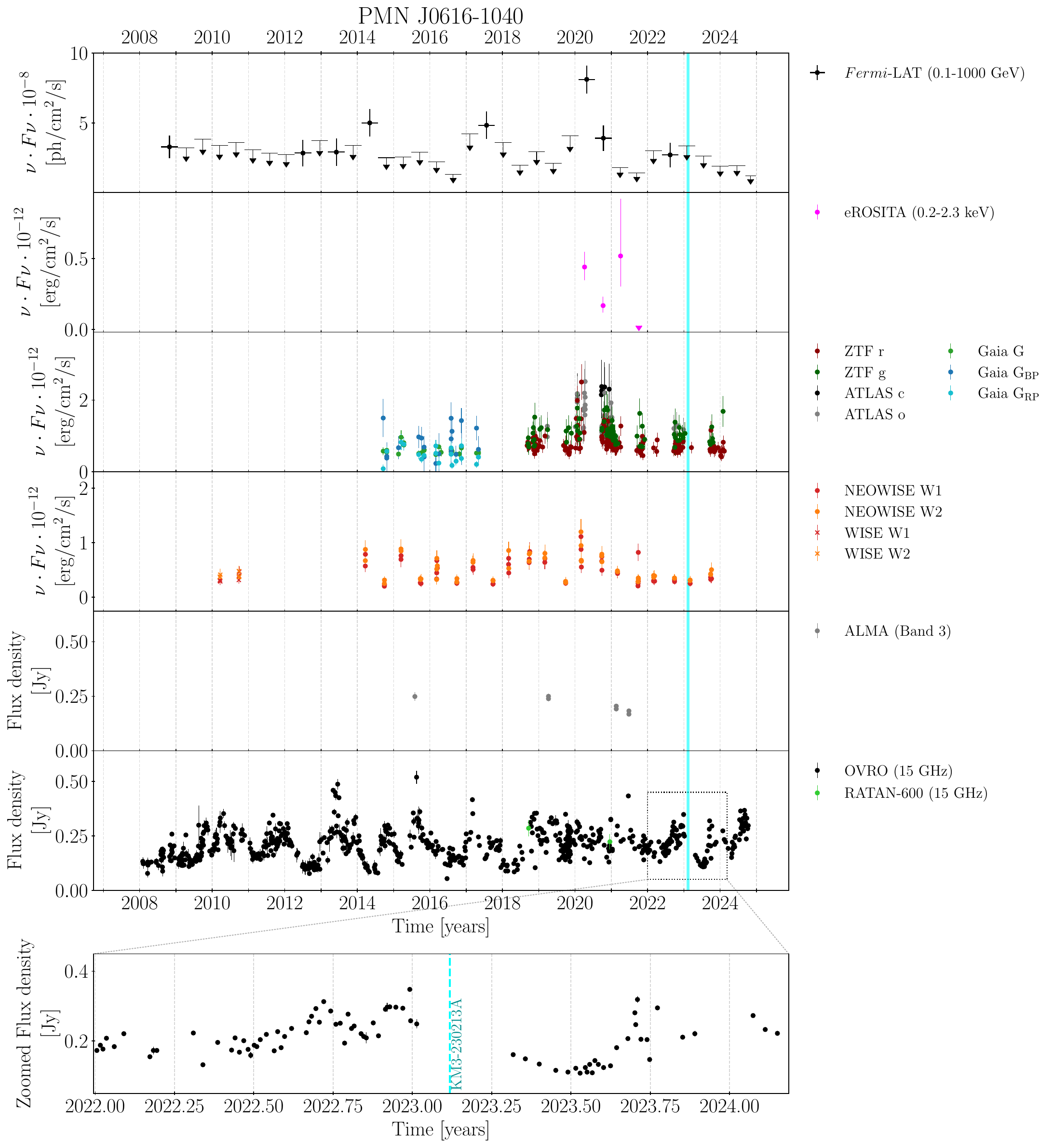}
\caption{Multi-wavelength light curves of \sourcename{9}($\#9$).
The panels display from top to bottom: \textit{Fermi}-LAT \gammaray light curves integrated over 6-months; X-ray data from  \eROSITA; optical data from ZTF, ATLAS and \GAIA; IR data from WISE/NEOWISE; radio data from ALMA, OVRO and RATAN-600. The last panel displays the zoomed radio light curve centered on the neutrino arrival time.
The cyan stripe highlights the arrival time of \uhevent.}
 \label{fig:LC9}
\end{figure*}

\begin{figure*}[h]
\includegraphics[width=0.95\linewidth]{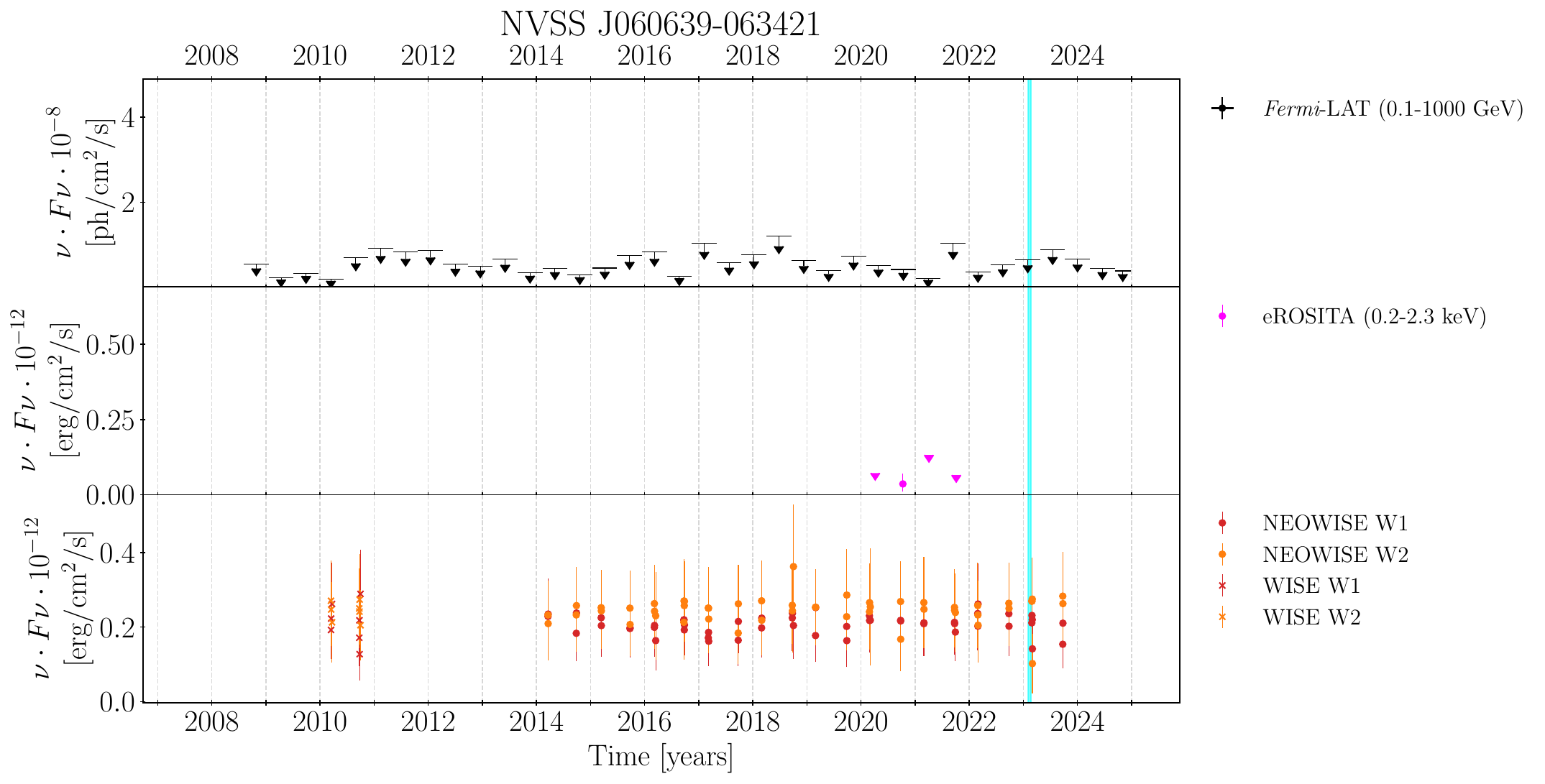}
\caption{Multi-wavelength light curves of \sourcename{10}($\#10$).
The panels display from top to bottom: \textit{Fermi}-LAT \gammaray light curves integrated over 6-months; X-ray data from  \eROSITA; IR data from WISE/NEOWISE. 
The cyan stripe highlights the arrival time of \uhevent.}
 \label{fig:LC10}
\end{figure*}

\begin{figure*}[h]
\includegraphics[width=0.95\linewidth]{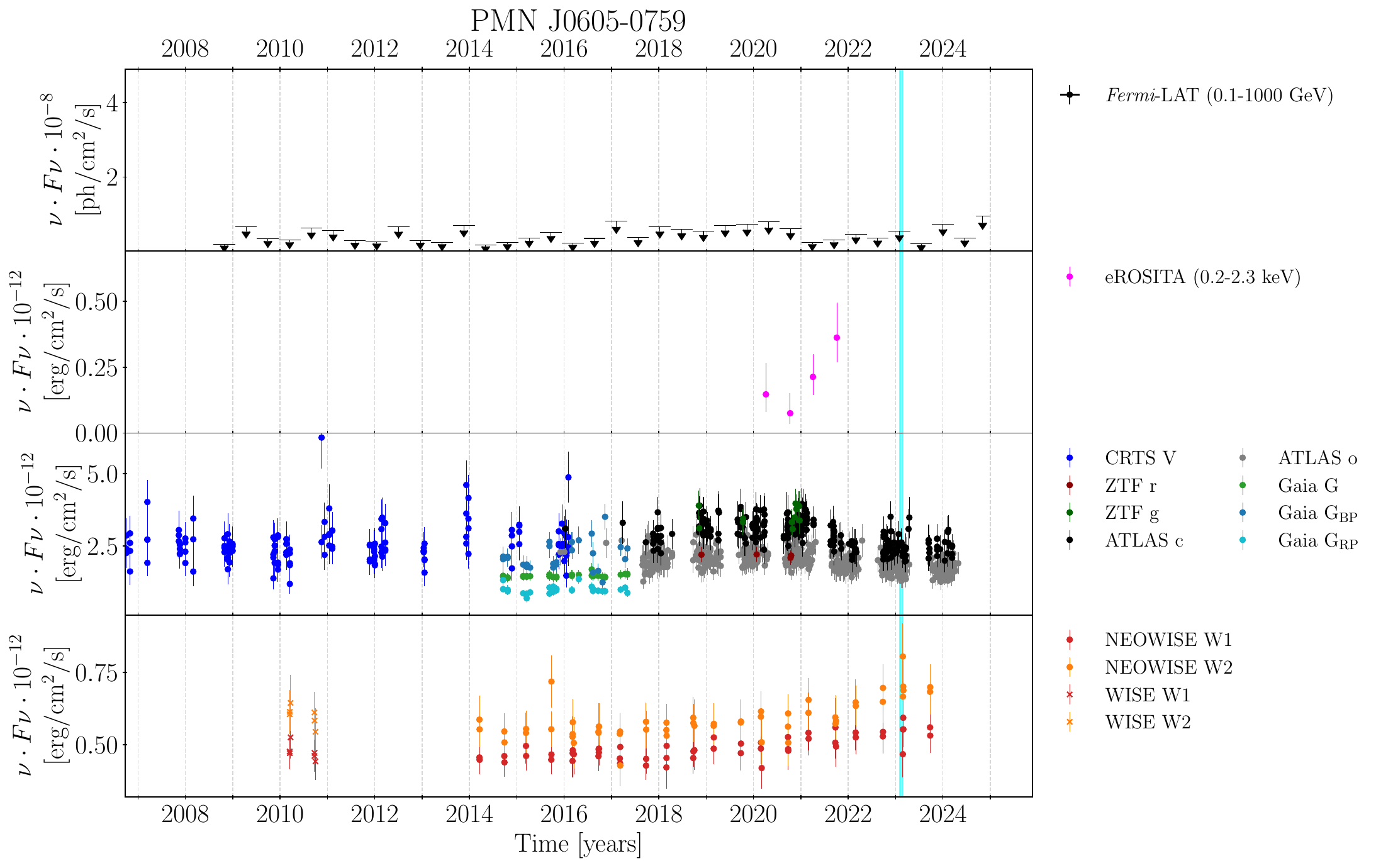}
\caption{Multi-wavelength light curves of \sourcename{11}($\#11$).
The panels display from top to bottom: \textit{Fermi}-LAT \gammaray light curves integrated over 6-months; X-ray data from  \eROSITA; optical data from CRTS, ZTF, ATLAS and \GAIA; IR data from WISE/NEOWISE. 
The cyan stripe highlights the arrival time of \uhevent.}
 \label{fig:LC11}
\end{figure*}

\begin{figure*}[h]
\includegraphics[width=0.95\linewidth]{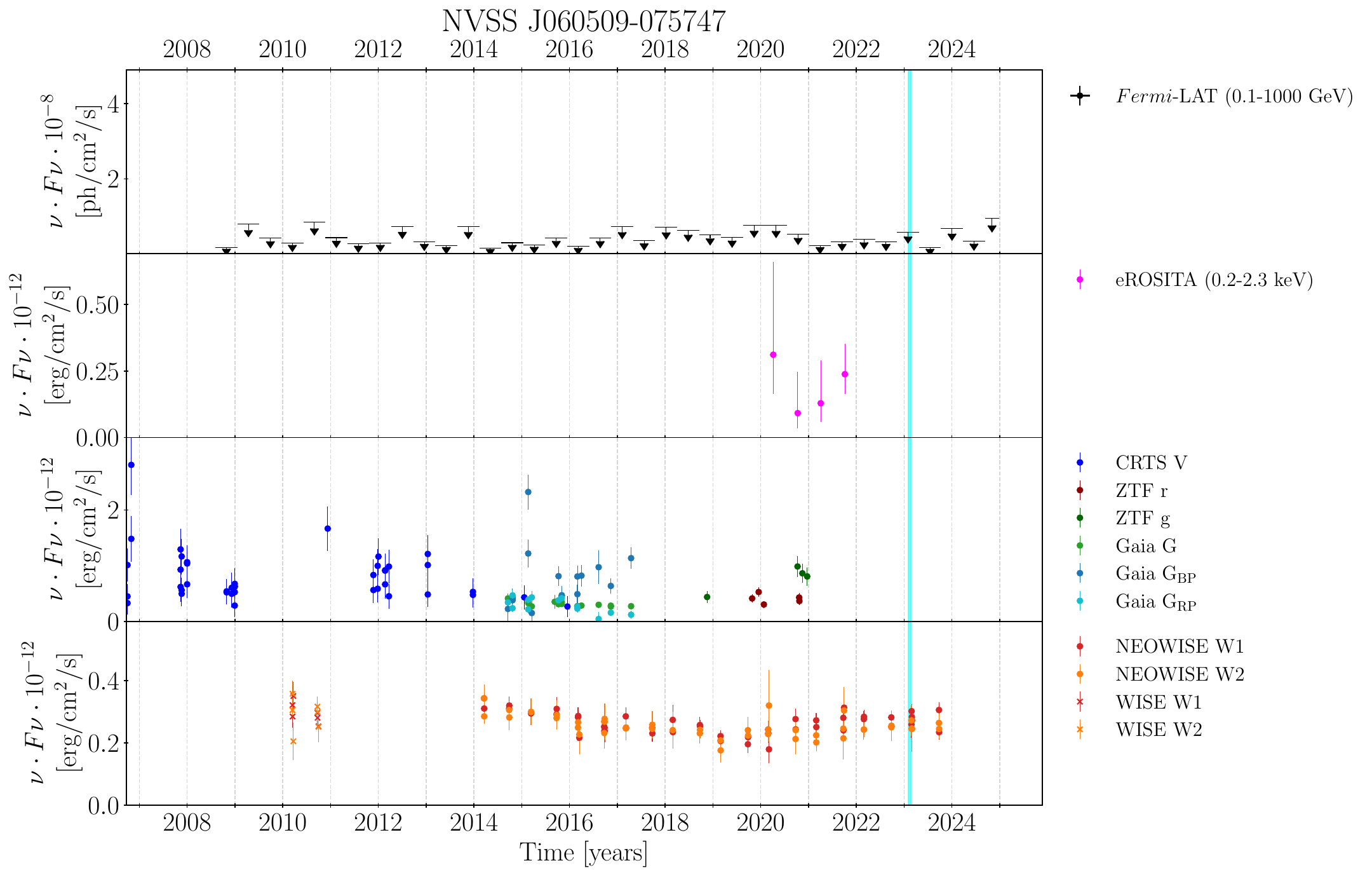}
\caption{Multi-wavelength light curves of \sourcename{12}($\#12$).
The panels display from top to bottom: \textit{Fermi}-LAT \gammaray light curves integrated over 6-months; X-ray data from \eROSITA; optical data from CRTS, ZTF and \GAIA; IR data from WISE/NEOWISE. 
The cyan stripe highlights the arrival time of \uhevent.}
 \label{fig:LC12}
\end{figure*}

\begin{figure*}[h]
\includegraphics[width=0.95\linewidth]{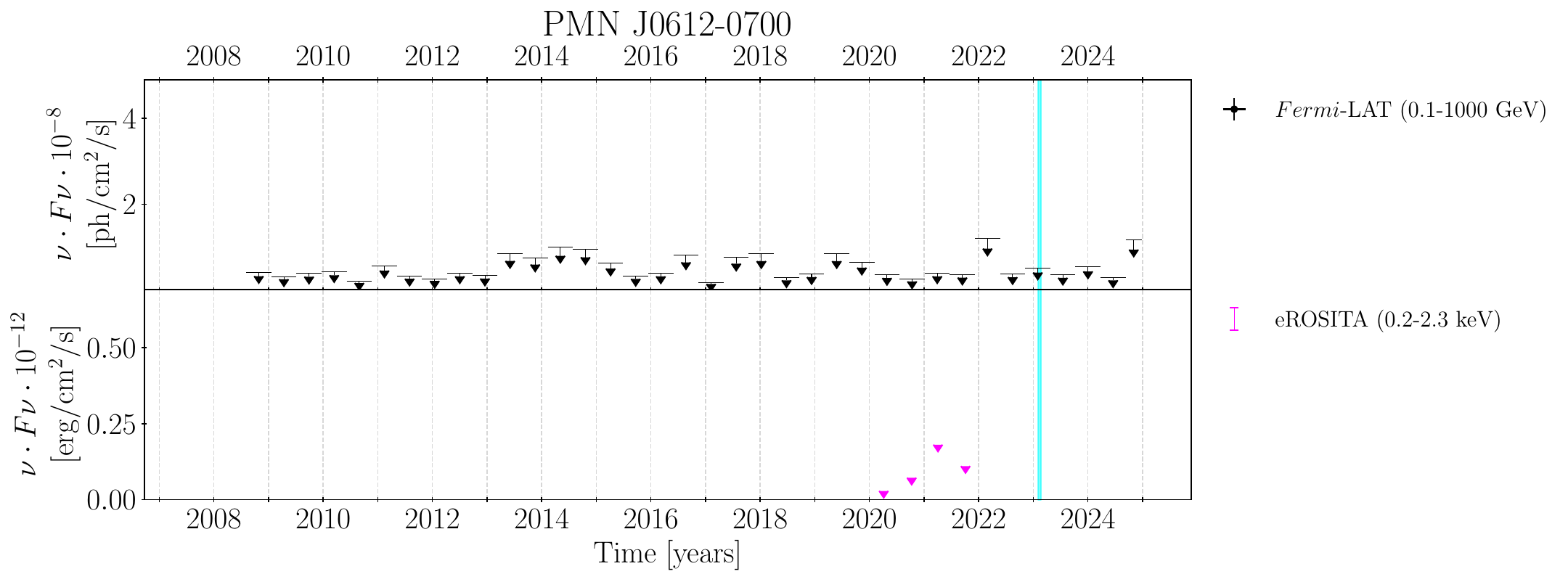}
\caption{Multi-wavelength light curves of \sourcename{13}($\#13$).
The panels display from top to bottom: \textit{Fermi}-LAT \gammaray light curves integrated over 6-months; X-ray data from \eROSITA. 
The cyan stripe highlights the arrival time of \uhevent.}
 \label{fig:LC13}
\end{figure*}

\begin{figure*}[h]
\includegraphics[width=0.95\linewidth]{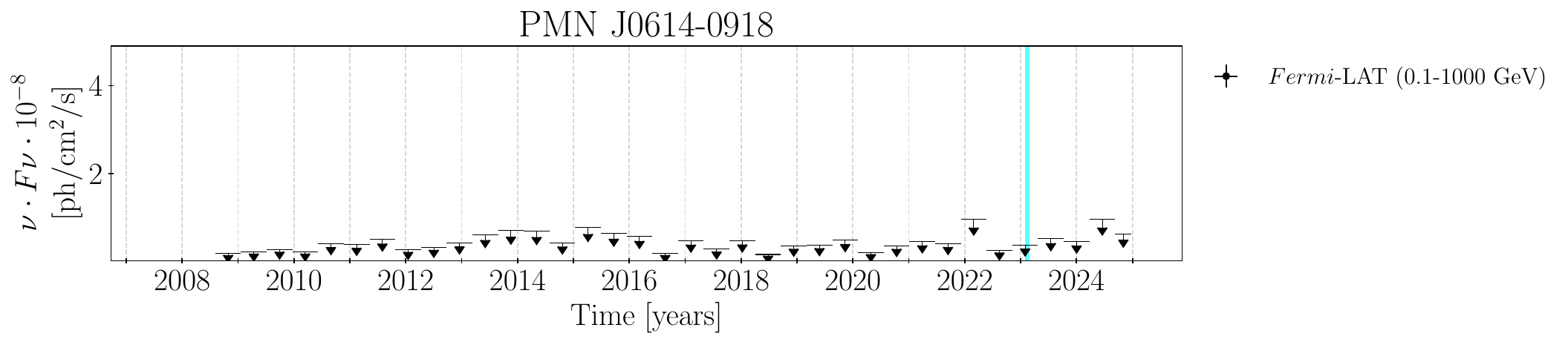}
\caption{$Fermi$-LAT light curve of \sourcename{14}($\#14$),  integrated over a 6-months time bin.
The cyan stripe highlights the arrival time of \uhevent.}
 \label{fig:LC14}
\end{figure*}

\begin{figure*}[h]
\includegraphics[width=0.95\linewidth]{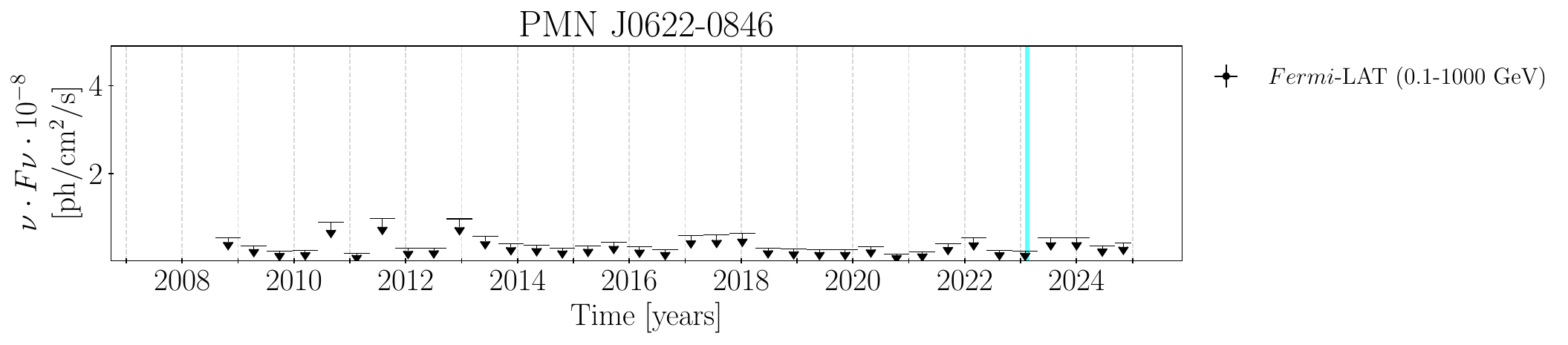}
\caption{$Fermi$-LAT light curve of \sourcename{15}($\#15$),  integrated over a 6-months time bin.
The cyan stripe highlights the arrival time of \uhevent.}
 \label{fig:LC15}
\end{figure*}

\begin{figure*}[h]
\includegraphics[width=0.95\linewidth]{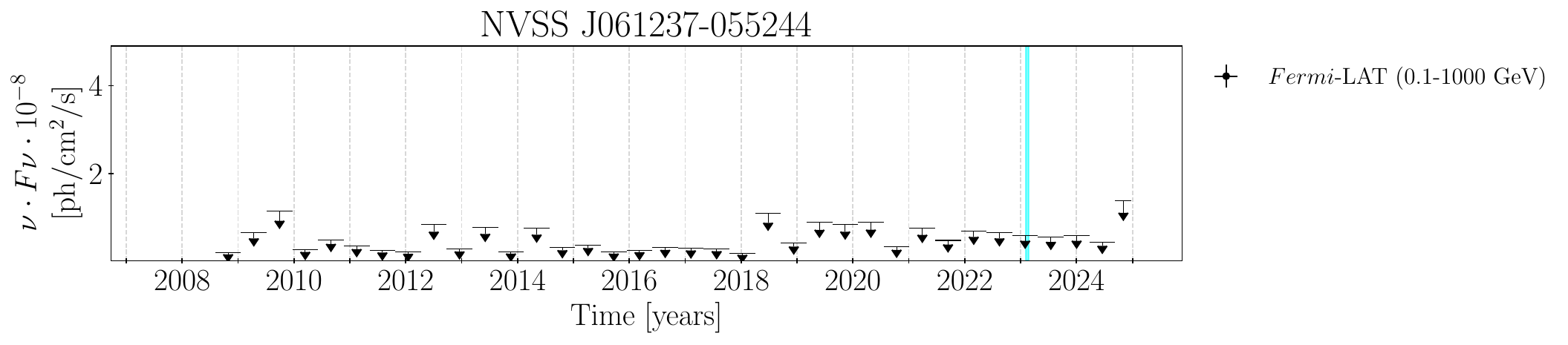}
\caption{$Fermi$-LAT light curve of \sourcename{16}($\#16$), integrated over a 6-months time bin.
The cyan stripe highlights the arrival time of \uhevent.}
 \label{fig:LC16}
\end{figure*}

\begin{figure*}[h]
\includegraphics[width=0.95\linewidth]{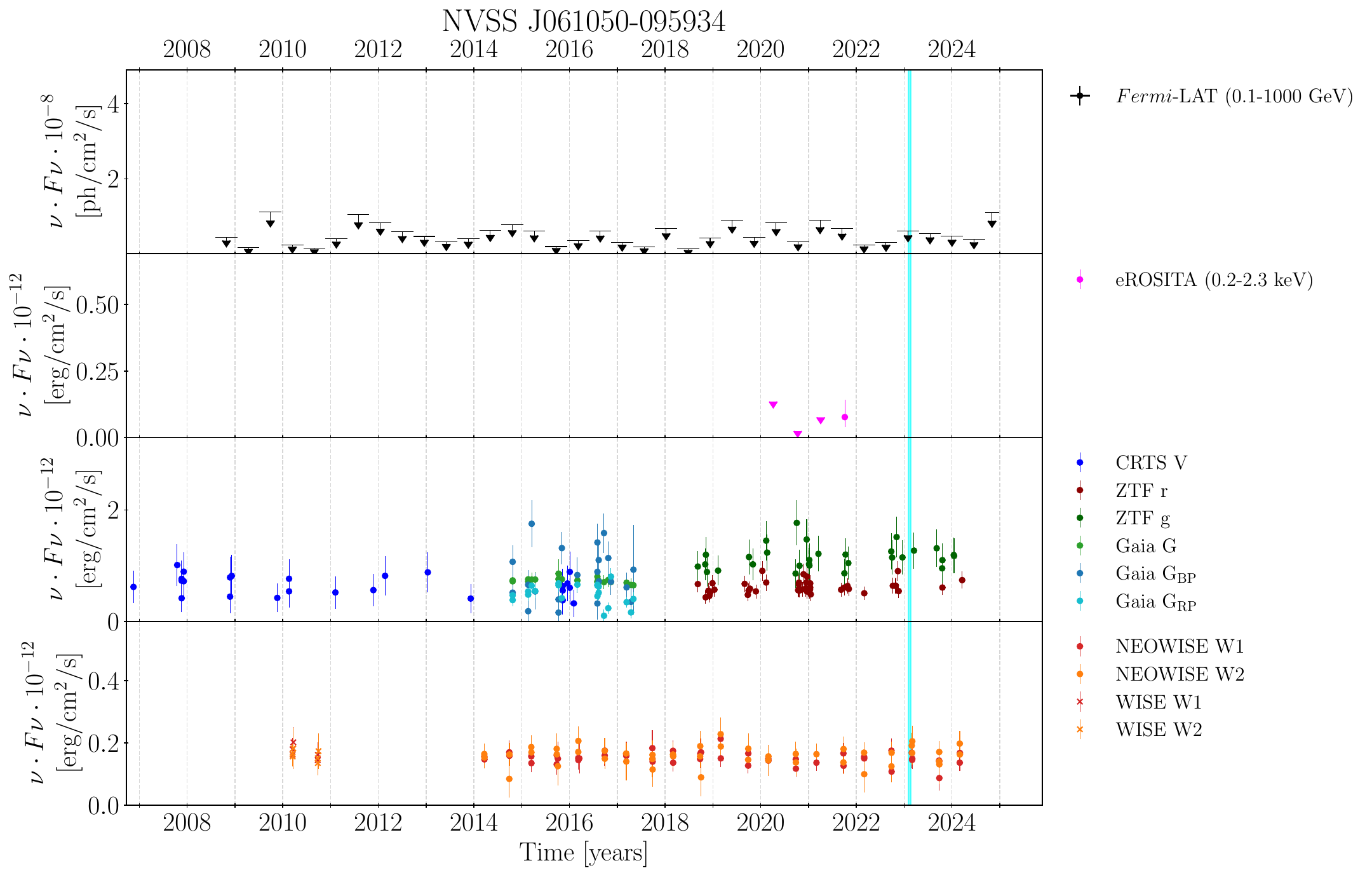}
\caption{Multi-wavelength light curves of \sourcename{17}($\#17$).
The panels display from top to bottom: \textit{Fermi}-LAT \gammaray light curves integrated over 6-months; X-ray data from \eROSITA; optical data from CRTS, ZTF and \GAIA; IR data from WISE/NEOWISE. 
The cyan stripe highlights the arrival time of \uhevent.}
 \label{fig:LC17}
\end{figure*}

The performed analysis yields the following results:

\begin{itemize}
    \item Observed $TS = 5$ days, for the blazar \sourcename{8} (\autoref{f:featured-blazars});
    \item the null distribution from the simulation is shown in \autoref{f:0604-074_flare};
    \item a p-value of $p = 0.26\%$ is obtained, indicating a 1 in 385 chance to observe a time difference of 5 days or less between the neutrino arrival time and the highest radio peak purely by random chance.
\end{itemize}

A discussion of the potential interpretation of this radio flare is given in \autoref{s:radioflare}. It should be noted that this analysis is fundamentally \emph{a posteriori} and not trial-free; in particular, it focuses solely on flares in the radio band. The band choice is not completely arbitrary, but these considerations make determining the proper correction factor for the chance probability challenging, and this correction is not attempted here.

\clearpage

\section{Multiwavelength Light Curves}\label{s:light curves}

\clearpage

\bibliography{km3net_vhe_blazars}{}
\bibliographystyle{aasjournal}



\end{document}